\RequirePackage[T1]{fontenc}
\documentclass[12pt]{article}

\usepackage[height=8.85in,width=6.45in]{geometry}

\usepackage{times}

\usepackage[utf8]{inputenc}
\usepackage{amsmath}
\usepackage{amssymb}
\usepackage{mathtools}
\numberwithin{equation}{section}
\usepackage{slashed}
\usepackage{braket}
\usepackage[svgnames,psnames]{xcolor}
\usepackage[colorlinks,citecolor=DarkGreen,linkcolor=FireBrick,linktocpage,pagebackref]{hyperref}
\usepackage{cite}
\usepackage{graphicx}
\usepackage{tikz}
\usepackage{tikz-cd}

\usepackage{courier}
\usepackage{bm}

\usepackage{upgreek}

\newcommand*{\bZ}{\mathbb{Z}}

\newcommand*{\sD}{\mathsf{D}}
\newcommand*{\sH}{\mathsf{H}}

\newcommand*{\sJ}{\mathcal{J}}
\newcommand*{\sK}{\mathcal{K}}

\def\Hom{\mathop{\mathrm{Hom}}\nolimits}

\def\ch{\mathop{\mathrm{ch}}\nolimits}
\usepackage{dsfont}

\DeclareMathOperator{\Inv}{Inv}
\DeclareMathOperator{\Ind}{Ind}
\DeclareMathOperator{\Tors}{Tors}

\begin{document}
\begin{titlepage}

    \begin{flushright}
        YITP-SB-2022-19
    \end{flushright}
    
    \vskip 3cm

    \begin{center}

    \textbf{\Large Higher Berry Phase of Fermions and Index Theorem}

    \vskip 1cm
    Yichul Choi$^{1,2}$
    and Kantaro Ohmori$^3$, 
    \vskip 1cm
    
    \begin{tabular}{ll}
    $^1$ & C. N. Yang Institute for Theoretical Physics, \\
    & SUNY, Stony Brook, NY 11794, USA\\
    $^2$ & Simons Center for Geometry and Physics, \\
    & SUNY, Stony Brook, NY 11794, USA\\
    $^3$ & Faculty of Science, University of Tokyo\\
    & Hongo 7-3-1, Bunkyo, Tokyo, Japan

    \end{tabular}

    \vskip 1cm
    
    \textbf{Abstract}

    \end{center}
    
    When a quantum field theory is trivially gapped, its infrared fixed point is an invertible field theory.
    The partition function of the invertible field theory records the response to various background fields in the long-distance limit.
    The set of background fields can include spacetime-dependent coupling constants, in which case we call the corresponding invertible theory a parameterized invertible field theory. 
    We study such parameterized invertible field theories arising from free Dirac fermions with spacetime-dependent mass parameters
    using the Atiyah-Patodi-Singer index theorem for superconnections.
    In particular, the response to an infinitesimal modulation of the mass is encoded into a higher analog of the Berry curvature, for which we provide a general formula. 
    When the Berry curvature vanishes, the invertible theory can still be nontrivial if there is a remaining torsional Berry phase, for which we list some computable examples. 
    \end{titlepage}

\tableofcontents
\section{Introduction and summary}
\subsection{Overview}\label{sec:overview}

A physical system often comes with parameters affecting the dynamics of concern.
In such cases, we can study its response to adiabatic deformations of the parameters.
In quantum mechanics, it is well-known that when one adiabatically deforms a quantum system along a cyclic path in the space of parameters, the state of the system after the process can be different from the initial state, where the difference is captured by the Berry phase \cite{Berry:1984jv}.

When the physical system considered is defined on a space with a positive number of dimensions, which can be either of a lattice or a continuous space, the parameters can be modulated both in the space and the time directions.
In particular, in a relativistic quantum field theory (QFT), it is natural to let the parameters to modulate arbitrarily in the spacetime.
The response to such general deformations is encoded into the higher Berry phase \cite{Kapustin:2020eby,Kapustin:2020mkl,Hsin:2020cgg,Kapustin:2022apy}, which in this paper we often simply call the Berry phase.
To be precise, the Berry phase is definable when the parameters vary within a locus in the parameter space along which the system is trivially gapped,\footnote{
    A system is said to be trivially gapped when it has a single gapped vacuum with no topological excitations like anyons.
}
which is a generalization of the familiar fact from quantum mechanics that the Berry phase can be defined only if there does not occur a level-crossing along the adiabatic deformation.
The IR fixed point of such a trivially gapped theory coupled to the modulated parameters is called a \emph{parameterized invertible field theory}.

An easy example of a higher Berry phase is 
that of a single free massive Dirac fermion in 1+1-dimensions.
The theory has a single complex mass parameter $m$, and the mass term preserves the vector $U(1)$ global symmetry.
We can rotate the phase of the mass parameter by the axial symmetry action on the fermion, making $m$ real.
This however generates a complex phase because of a mixed anomaly when the vector $U(1)$ background gauge field $A$ is turned on. If we write the mass parameter as $m = |m| e^{\mathrm{i}\alpha}$,
the generated phase in the fermion partition function $\mathcal{Z}$ is
\begin{equation}
    \label{eq:alphaF}
    \mathcal{Z}[m = |m| e^{\mathrm{i}\alpha}]
    = \mathcal{Z}[ m = |m|] \exp\left(-\frac{2\pi\mathrm{i}}{(2\pi)^2}\int \alpha \mathrm{d} A\right) \,.
\end{equation}
This formula applies even when $\alpha$ modulates over the spacetime up to terms suppressed by $|m|^{-1}$.
While this simple example can be thoroughly explained by the conventional 't Hooft anomaly, 
it is more direct to regard this phase as an analog of the Berry phase, because the axial symmetry does not exist when the mass is turned on.

The nontrivial Berry phase in particular means nontrivial physics on the boundary. 
In the example of the above 1+1d single Dirac fermion, this can be seen as follows.
We prepare an interface between the fermion theory with a constant positive mass $m_0$ and the fermion theory with a mass $m_\alpha = m_0 e^{\mathrm{i}\alpha}$ by linearly interpolating between the two mass values.
We regard the mass $m_0$ as the reference point and the interface as the boundary of the invertible theory with the parameter $m_\alpha$.
While the boundary is trivially gapped with a generic value of $\alpha$, it has a degeneracy when $\alpha = \pi $ because there occurs a zero-mode of a massless complex fermion localized along the boundary. 
In other words, the boundary cannot be trivially gapped uniformly over the parameter space.
This is an ``anomaly in the space of coupling constants'' on the boundary \cite{Cordova:2019jnf,Cordova:2019uob}.\footnote{This is also referred to as a ``global inconsistency.'' See \cite{Gaiotto:2014kfa,Gaiotto:2017yup,Tanizaki:2017bam,Komargodski:2017dmc,Kikuchi:2017pcp,Thorngren:2017vzn,Tanizaki:2018xto,Hidaka:2019jtv,Karasik:2019bxn,Sulejmanpasic:2020zfs,Sharon:2020doo,Unsal:2020yeh,Honda:2020txe,Tanizaki:2022ngt} for developments on this.}

From this viewpoint, the Berry phase corresponds to the ``inflow action'' in the conventional anomaly inflow language \cite{callan1985anomalies}.
In the anomaly inflow context, it is convenient to consider the formal derivative of the inflow action or the anomaly polynomial.
For an anomaly in the space of couplings,
the anomaly polynomial can be regarded as a natural generalization of the Berry curvature called the higher Berry curvature \cite{Kapustin:2020eby,Kapustin:2020mkl,Hsin:2020cgg,Kapustin:2022apy}.
In the example of \eqref{eq:alphaF}, it is 
\begin{equation}
    \mathcal{B} = -\frac{1}{(2\pi)^2} \mathrm{d}\alpha \mathrm{d}A \,.
\end{equation}

When there are multiple fermions, the form of the Berry phase becomes complicated.
The purpose of this paper is to provide new methods to calculate the Berry phases and the Berry curvatures for more general theories of massive Dirac fermions.
This will be done by first relating the Berry phase of a $d$-dimensional fermion system to a certain $(d+1)$-dimensional $\eta$-invariant, and then invoking the Atiyah-Patodi-Singer (APS) index theorem for superconnections \cite{quillen1985superconnections,kahle2011superconnections,Kanno:2021bze,Gomi:2021bhy}.
In particular, we provide an explicit formula computing the Berry curvature for such theories.
The index theorem necessary for the generalization to Majorana fermions is studied in \cite{Gomi:2021bhy}.

Some special cases of the Berry curvature of fermions with modulated mass parameters were studied in the past in \cite{Goldstone:1981kk,Abanov:1999qz}, and also in \cite{Huang:2021ubo} for the purpose of studying crystalline symmetry protected topological phases.
See also \cite{Teo:2010zb,Shiozaki:2021weu,wen2021flow,Aasen:2022cdu} for other studies of parameterized invertible theories in the condensed matter physics literature.
Fermions with modulated mass parameters were also studied in \cite{Gorantla:2020fao}.
On the phenomenological side, integrating out fermions coupled to various gauge fields as well as Higgs fields was recently discussed in \cite{Ellis:2020ivx,Angelescu:2020yzf,Quevillon:2021sfz}.
Our closed-form formula for the Berry curvature may help streamline some portion of such calculations, since the modulated mass parameters can be thought of as background Higgs fields.

\subsection{Summary of results} \label{Sec:summary_of_results}
As stated, in this paper, we examine the class of parameterized invertible theories arising from the low-energy limit of massive free Dirac fermions in $d$ spacetime dimensions.
The parameter space is the space of all possible non-degenerate mass matrices denoted as $P$, and the problem boils down to studying partition functions of free fermions with spacetime-dependent modulated mass parameters (in addition to other symmetry backgrounds). 
The phase of the (regularized) partition function is identified with the Berry phase introduced above.

Specifically, we put an arbitrary number $N_f$ of Dirac fermions on a $d$-dimensional spacetime manifold $X$ which is equipped with the various background fields, that is, the modulated mass parameters as well as background gauge fields for some global symmetries.
The parameter space $P$ is determined by $d$ and $N_f$.
Then, it turns out that it is convenient to consider an auxiliary $(d+1)$-dimensional free fermion system living on a manifold $Y_+$ which bounds $X$.
That is, $X = \partial Y_+$ and all the background fields extend into $Y_+$.
Importantly, we allow the $d$-dimensional mass matrix to become degenerate inside the $(d+1)$-dimensional bulk $Y_+$.
Manifolds $X$ together with the background fields that do not admit such $Y_+$ even when we allow the mass matrix to become degenerate inside the bulk are needed only for the purpose of detecting ordinary invertible theories that do not see the parameter space.
Because we are not interested in these ordinary invertible theories in this paper, we can safely assume that $Y_+$ always exists.

In this setup, we can state our main result:
\begin{equation} \label{eq:main_result}
    \left.\frac{\mathcal{Z}_{d}(X)}{\mathcal{Z}^{(0)}_{d}(X)}\middle/  \left|\frac{\mathcal{Z}_d(X)}{\mathcal{Z}_d^{(0)}(X)}\right| \right.= \exp \left(2\pi \mathrm{i} (\eta_K (Y_+;\mathtt{APS}) - \eta_{K^{(0)}} (Y_+;\mathtt{APS})\right) \,.
\end{equation}
Here, $\mathcal{Z}_d (X)$ is the partition function of the $d$-dimensional theory with a modulated but non-degenerate mass parameters living on $X$, and $\eta_{K}(Y_+;\mathtt{APS})$ is the APS $\eta$-invariant for a hermitian operator $K$ which extends the $d$-dimensional Dirac operator on $X$ to the bulk of $Y_+$ \cite{Atiyah:1975jf}.
The quantities with the superscript $(0)$ are for the regulator fermions.
We derive this result by considering $(d+1)$-dimensional Dirac fermions living in the bulk $Y_+$.
The original $d$-dimensional theory is realized as the localized edge modes living on the boundary $\partial Y_+ = X$.
The result works even when the mass term breaks certain global symmetries for which we want to turn on the background gauge fields, by promoting the mass parameters to be an appropriate section of the associated bundle.

The equation \eqref{eq:main_result} connects two different viewpoints. 
The left-hand side is the Berry phase of the original system in $d$-dimensions.
The right-hand side is the Berry phase of the $(d+1)$-dimensional fermion system with the parameter space $\widetilde{P} = P \cup D$ for which the original $d$-dimensional system behaves as the boundary anomalous theory.
Here, $D$ is the space of degenerate mass matrices (which again is determined by $d$ and $N_f$), and in general $\widetilde{P} = \mathbb{R}^N$ for some integer $N$. 
We will provide an interpretation of the equation from the generalized cohomology viewpoint in Section~\ref{sec:gencoh}.

The consequence of the result \eqref{eq:main_result} is that we can apply the index theorem for superconnections to calculate the right-hand side \cite{Cordova:2019jnf,Cordova:2019uob,Kanno:2021bze}.
The definition of a superconnection will be reviewed in Section \ref{Sec:superconnection}.
For the math literature regarding the superconnection and the relevant index theorem, see \cite{quillen1985superconnections,kahle2011superconnections,Gomi:2021bhy}.
In particular, this approch enables us to write down an explicit formula for the Berry curvature $\mathcal{B}$:
\begin{equation} \label{eq:berry_curvature}
    \mathcal{B} = \int_{\widetilde{M}_0 = 0}^{\widetilde{M}_0 = \infty} \left[ \ch (\mathcal{F}) - \ch (\mathcal{F}^{(0)})
    \right] \wedge \hat{A} \,,
\end{equation}
where $\mathcal{F}$ is the field strength for the superconnection, and $\ch (\mathcal{F})$ is the corresponding Chern character.\footnote{$\ch (\mathcal{F}^{(0)})$ is the contribution from the regulator fermions.}
$\hat{A}$ is the $\hat{A}$-genus responsible for the usual gravitational contribution, and
$\widetilde{M}_0$ is the overall scale of the (renormalized) mass matrix.

We also show some examples where the resulting parameterized invertible theory matches global anomalies in the parameter space in one lower dimensions.
In these cases, the Berry curvature vanishes and the partition function defines a bordism invariant.
We call such a bordism invariant a \emph{torsional Berry phase}.

\subsection{Organization} \label{Sec:organization}

This paper is organized as follows.
In Section \ref{Sec2}, we derive Eq. (\ref{eq:main_result}) in two steps, by relating the partition function of the $d$-dimensional fermion system of interest to that of an auxiliary $(d+1)$-dimensional system.
We first show that the phase of the partition function of the $(d+1)$-dimensional system is given by the APS $\eta$-invariant of a hermitian operator $K$ that we will define, and then show that the regularized $(d+1)$-dimensional partition function is equal to the regularized $d$-dimensional partition function that we are after in a certain limit. 
The essential idea is based on earlier works by Yonekura \cite{Yonekura:2016wuc}, and Yonekura and Witten \cite{Witten:2019bou}.
Then, we interpret Eq. (\ref{eq:main_result}) from the perspective of a generalized cohomology theory.
In Section \ref{Sec3}, we derive Eq.~(\ref{eq:berry_curvature}) and demonstrate it in several examples.
In Section \ref{Sec4}, we show examples of parameterized invertible theories with vanishing Berry curvatures, where the partition function can be computed using Eq.~(\ref{eq:main_result}).
In Appendix \ref{appind}, we derive the APS index theorem in the presence of the modulated mass parameters, by closely following the derivation given in \cite{Kobayashi:2021jbn} for the case of massless Dirac operators. 
It is a generalization of the theorem derived in \cite{Kanno:2021bze} to the case of manifolds with boundary,\footnote{A special case of this was also treated in \cite{Kanno:2021bze}.} and we use as an input their calculation result which was based on the Fujikawa method \cite{Fujikawa:1979ay,Fujikawa:1983bg}.

\section{Massive fermion partition functions} \label{Sec2}

Our goal in this section is to express (the phase of) the partition function of massive free Dirac fermions in $d$ spacetime dimensions with the spacetime-dependent mass parameters in terms of $\eta$-invariants (see Eq. \eqref{eq:main_result}).
We denote the number of flavors as $N_f$ which is arbitrary.
We would like to achieve this possibly in the presence of other background gauge fields for some global symmetries and in general on a curved spacetime manifold.
The global symmetries we consider are (subgroups of) $U(N_f) \times U(N_f)$ if $d$ is even, or $U(N_f)$ if $d$ is odd, as well as the fermion parity symmetry.
Usually, we will not explicitly specify for which subgroup of $U(N_f) \times U(N_f)$ or $U(N_f)$ the background gauge fields are turned on, and the general computations are valid for any choice of the subgroup unless otherwise mentioned.
The $d$-dimensional spacetime manifold will be denoted as $X$, which is assumed to be closed and oriented.
All the discussions will be in the Euclidean signature.

The Lagrangian of the theory is
\begin{equation} \label{Eq:d_dim_lagrangian}
    \mathcal{L}_d = \bar{\psi}(\slashed{D} + M)\psi \,.
\end{equation}
The flavor and spinor indices are implicit in $\psi$.
When $d$ is odd, $M$ takes values in $N_f \times N_f$ hermitian matrices.
When $d$ is even, the mass matrix has the form of $M = M^\prime + \mathrm{i} \gamma M^{\prime \prime}$ where $M^\prime$ and $M^{\prime \prime}$ are valued in $N_f \times N_f$ hermitian matrices and $\gamma$ is the chirality matrix.
We let the mass matrix be a nontrivial function on the spacetime manifold, or a section of a bundle acted by a global symmetry in case the mass term is not invariant under the chosen symmetry.
However, to ensure that the partition function is nonzero and thus we can define its phase, we demand that we do not cross the massless points, i.e., $\mathrm{det}(M)$ is nonzero everywhere on the spacetime manifold $X$.

We are in particular interested in the long-distance limit of the fermion theory.
This limit can be achieved by sending the mass eigenvalues to infinity, instead of sending the volume of the spacetime to infinity.
In this limit the fermion theory becomes invertible.
The invertible theory at the IR fixed point naturally couples with the scalar background $M$, which moves in the space of non-degenerate mass matrices. 
The partition function of the invertible theory records the long-distance reaction to a modulation of the mass $M$ along the spacetime.

To achieve our goal, it helps to prepare the fermion system by a domain-wall or interface construction \cite{jackiw1976solitons}.\footnote{We will mostly refer to it as an interface instead of the commonly-used terminology domain wall, since this interface is formed by adjusting the background field configurations by hand and is not a dynamical object of the theory.}
That is, we will realize our $d$-dimensional fermion system of interest on an interface of an auxiliary $(d+1)$-dimensional fermion system by varing the $(d+1)$-dimensional mass parameters appropriately.
A similar construction for massive fermions was considered in \cite{Fukaya:2017tsq,Fukaya:2020ddp,Fukaya:2020tjk,Onogi:2021slv}.

\begin{figure}[t]
    \centering
    \raisebox{-.17\height}{\includegraphics[width=0.5\textwidth]{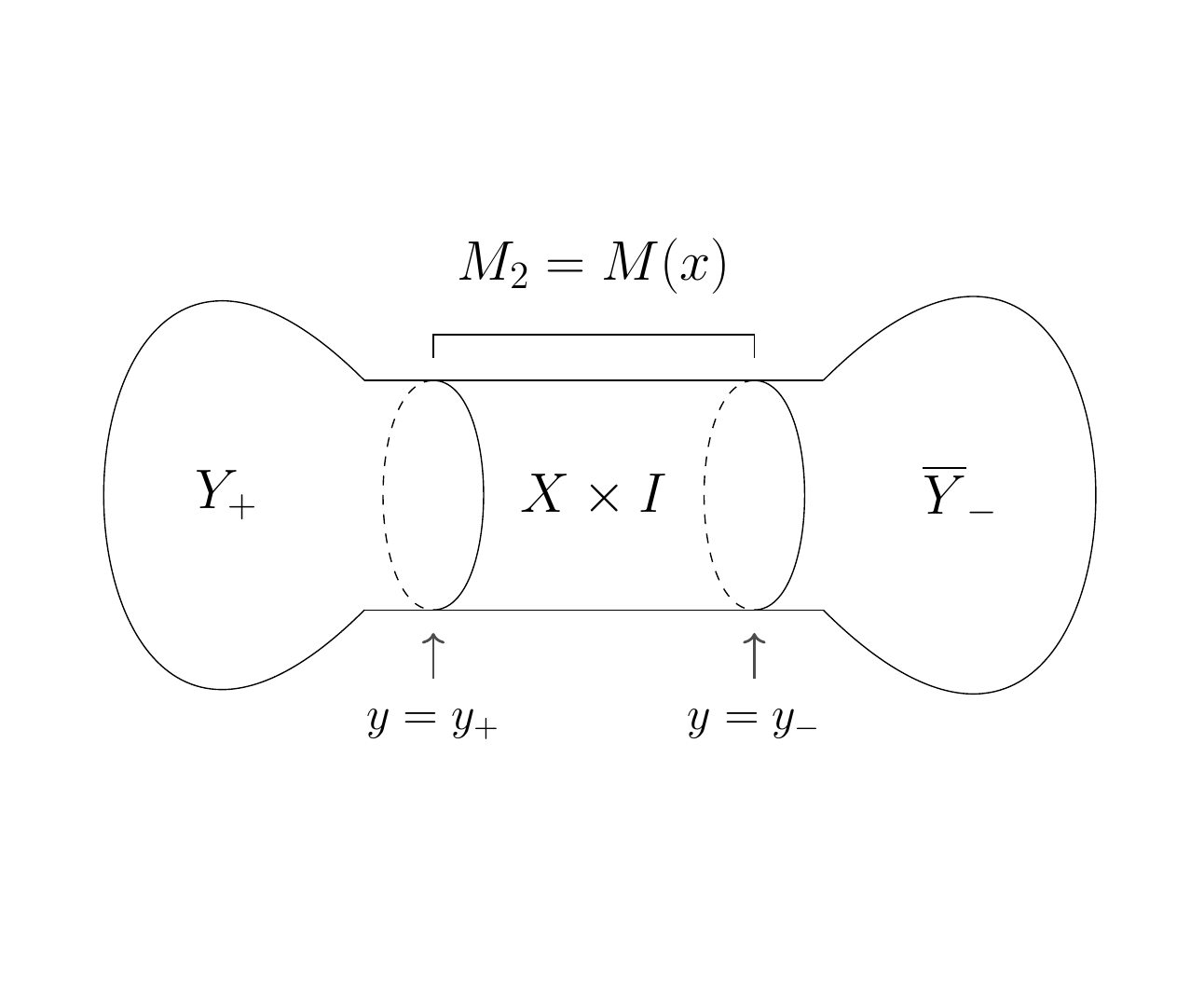}}
    \hspace{0.5cm}
    \includegraphics[width=0.4\textwidth]{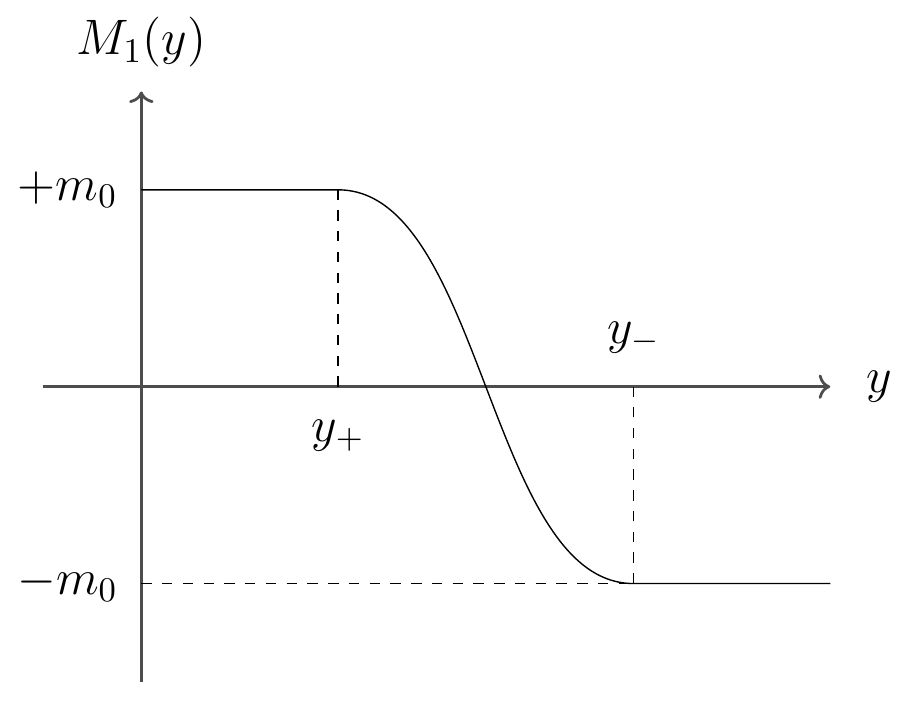}
    \caption{(Left) The $(d+1)$-dimensional closed manifold $Y = \overline{Y}_ - \cup_{\overline{X}} (X \times I) \cup_{X} Y_+$ on which the auxiliary $(d+1)$-dimensional theory is defined. (Right) The profile of $M_1(y)$ along the collar neighborhood of $X$ in $Y$ forms an interface. On the interface, the original $d$-dimensional theory of interest appears as a collection of localized modes.}
    \label{Fig:Y+Y-}
\end{figure}

Concretely, we consider a $(d+1)$-dimensional theory of $N_f$ flavors of free massive Dirac fermions if $d$ is odd, or $2N_f$ flavors of them if $d$ is even.
We put this theory on a closed manifold $Y = \overline{Y}_ - \cup_{\overline{X}} (X \times I) \cup_{X} Y_+$ which is shown in the left panel of Figure \ref{Fig:Y+Y-}.
The bar over a manifold indicates the orientation reversal of the manifold.
$Y_+ = Y_-$ are the same $(d+1)$-dimensional manifolds, and they have the boundary $\partial Y_\pm = X$.
All the background fields on $X$ extend into $Y_\pm$ in the same way.\footnote{We assume that the manifold $X$ equipped with various background gauge fields and the mass parameters is null-bordant if we allow $\text{det}(M)$ to vanish inside the null-bordism.
$\cup_{X}$ means gluing along the common boundary $X$ by the identity map, and similarly for $\cup_{\overline{X}}$.
Since we are not interested in detecting ordinary symmetry protected topological phases that do not depend on the mass parameters, this suffices for our purposes.}
$Y_\pm$ are assumed to have collar neighborhoods near the boundary, i.e., near the boundary they look like a product of $X$ and an interval.
The union of the collar neighborhoods of $X$ in $Y_\pm$ and the middle region $X \times I$ forms the collar neighborhood of $X$ in $Y$.
We denote the coordinate transverse to $X$ inside this collar neighborhood as $y$, and this will be viewed as the Euclidean time direction. 
Along the collar neighborhood, the metric factorizes as $ds^2_Y = dy^2 + ds^2_X$ where $ds^2_X$ is the metric on $X$, and any background gauge fields are required to be independent of $y$ and to have no $y$ component.
That is, locally we can write $A = A_i(x) dx^i$ for the background gauge field $A$ where $x^i$ are the coordinates along $X$ for $i=1, \dots, d$.

We will specialize to the case of odd $d$ here, and leave the similar calculations for the even $d$ case to Appendix \ref{appodddim}.
The final conclusion equally applies to both cases. 
The Lagrangian for the $(d+1)$-dimensional theory is given by
\begin{equation} \label{Eq:d+1_dim_lagrangian}
    \mathcal{L}_{d+1} = \bar{\Psi}(\slashed{D} + M_1 + \mathrm{i}\Gamma M_2)\Psi
\end{equation}
where $\Gamma$ is the $(d+1)$-dimensional chirality matrix.
We have $N_f$ flavors of $(d+1)$-dimensional Dirac fermions, where the flavor indices are again implicit.
A single Dirac fermion in $(d+1)$-dimensions has $2^{\lfloor d+1 \rfloor}$ components, whereas in $d$-dimensions it has $2^{\lfloor d \rfloor}$ components.
Since $d+1$ is even, we have $2 \times 2^{\lfloor d \rfloor} = 2^{\lfloor d+1 \rfloor}$.
The mass matrices $M_1$ and $M_2$ can in general be any $N_f \times N_f$ hermitian matrices.
We take the mass parameter $M_1$ to be $M_1 = M_1(y) \mathds{1}$ where $\mathds{1}$ is the $N_f \times N_f$ identity matrix, and $M_1(y)$ is a smooth function varying from $+m_0$ to $-m_0$ along the middle region $X \times I$ where $y \in I = [y_+ , y_-]$, with $m_0$ being a positive constant.
Inside $Y_\pm$ we will let $M_1 = \pm m_0 \mathds{1}$ to be constant.
That is, the configuration of $M_1(y)$ forms a smooth interface from $M_1(y_+) = +m_0$ to $M_1(y_-) = -m_0$, see the right panel of Figure \ref{Fig:Y+Y-}.
$M_2 = M(x)$ is taken to be equal to the mass matrix for the original $d$-dimensional fermions and is independent of $y$ inside the collar neighborhood.
$M_2$ should be suitably extended into the bulk of $Y_\pm$, but the precise form of the extension is not important for us.
The only assumption is that $M_2=M$ as well as all the background gauge fields on $X$ are extended to $Y_+$ and $Y_-$ in the same way, as mentioned earlier.

With this setup, there will appear localized $d$-dimensional massive Dirac fermions living on the interface $X$ with the Lagrangian given by Eq. (\ref{Eq:d_dim_lagrangian}).
This can be justified by considering the Dirac equaion \cite{jackiw1976solitons}, and we will make it more precise later.
Intuitively, if we take the limit $m_0 \to \infty$ while keeping $M$ finite, all the $(d+1)$-dimensional bulk modes decouple and we will be left with the $d$-dimensional fermions that we are interested in.
Below we will see how this idea is concretely realized and why this procedure of realizing the $d$-dimensional fermion on an interface of an auxiliary $(d+1)$-dimensional system is useful.

\subsection{Phase of the partition function as \texorpdfstring{$\eta$}{eta}-invariant} \label{subsec2.1}

First, we would like to show that
\begin{equation} \label{phaseeta}
    \mathcal{Z}_{d+1}(Y) = |\mathcal{Z}_d(X)|\exp \left(2\pi \mathrm{i} \eta_K (Y_+;\mathtt{APS})\right)
\end{equation}
when $m_0 \to \infty$, by applying a similar analysis as in \cite{Yonekura:2016wuc,Witten:2019bou}.
Here, $\mathcal{Z}_d (X)$ is the partition function of the $d$-dimensional theory \eqref{Eq:d_dim_lagrangian} on $X$, and similarly $\mathcal{Z}_{d+1}(Y)$ is the partition function of the $(d+1)$-dimensional theory \eqref{Eq:d+1_dim_lagrangian} on $Y$.
$\eta_K (Y_+;\mathtt{APS})$ is the APS $\eta$-invariant of a herimitian operator $K$ on $Y_+$ which we will describe below, with the APS boundary condition imposed on the boundary $\partial Y_+ = X$.

Let $Y_L \equiv X \times I$ for short. 
The partition function of the $(d+1)$-dimensional theory on the manifold $Y$ computes the following amplitude:
\begin{equation}
    \mathcal{Z}_{d+1}(Y) = \bra{Y_-} \hat{Y}_L \ket{Y_+} \,.
\end{equation}
The states $\ket{Y_\pm}$ are elements in the Hilbert space $\mathcal{H}_X$ of the spatial slice $X$, defined by the path integral over the region $Y_\pm$. The operator $\hat{Y}_L : \mathcal{H}_X \rightarrow \mathcal{H}_X$ corresponds to the Euclidean time evolution given by the cylindrical region $Y_L = X \times I$.

We define the $(d+1)$-dimensional massive Dirac operator $\sD_{d+1}$ as
\begin{equation}
    \sD_{d+1} = \Gamma^\mu D_\mu + M_1 + \mathrm{i} \Gamma M_2 \,.
\end{equation}
We pick a basis for the gamma matrices such that
\begin{align} \label{gammamatrix}
\begin{split}
    \Gamma^i &= \gamma^i \otimes \sigma^2 \,, \\
    \Gamma^y &= \mathbb{I} \otimes \sigma^3 \,, \\
    \Gamma &= \mathbb{I} \otimes \sigma^1 \,,
\end{split}
\end{align}
where $\sigma$'s are the Pauli matrices, $\gamma^i$'s are the $d$-dimensional gamma matrices, and $\mathbb{I}$ is the $2^{\lfloor d \rfloor} \times 2^{\lfloor d \rfloor}$ identity matrix. 
In this basis, inside the collar neighborhood of $X$ in $Y$, we have
\begin{align}
\begin{split}
    \sD_{d+1} &= \Gamma^y \partial_y + \Gamma^i D_i + M_1 +\mathrm{i}\Gamma M_2 \\
    &= \begin{bmatrix*}
        \partial_y +M_1 & \mathrm{i}\sD_d^\dagger \\
        \mathrm{i}\sD_d & -\partial_y + M_1
    \end{bmatrix*}
\end{split}
\end{align}
where 
\begin{equation}
    \sD_d = \gamma^i D_i + M 
\end{equation}
is the $d$-dimensional massive Dirac operator (recall $M_2 = M$ along the collar neighborhood). 
On the other hand, inside $Y_\pm$ where we have constant $M_1 = \pm m_0$, we can write
\begin{equation} \label{defK}
    \sD_{d+1} = -\mathrm{i}K \pm m_0
\end{equation}
where $K = \mathrm{i}(\Gamma^\mu D_\mu + i\Gamma M_2)$ is a hermitian operator on $Y$.

The Hamiltonian on $X$ is given by
\begin{equation}
    \hat{H}(y) = \int_X \Psi^\dagger \begin{bmatrix*}
        M_1(y) & \mathrm{i}\sD_d^\dagger \\
        -\mathrm{i}\sD_d & -M_1(y)
    \end{bmatrix*} \Psi
\end{equation}
which is dependent on the Euclidean time $y$.
Accordingly, the Euclidean time evolution along $Y_L$ is given by
\begin{equation}
    \hat{Y}_L = T\!\exp\left(-\int_{y_+}^{y_-} \hat{H}(y) dy\right) \,,
\end{equation}
where $T\!\exp$ denotes the time-ordered exponential.

For large $m_0$, the Euclidean path integral over the collar regions of $Y_\pm$ projects the states $\ket{Y_\pm}$ onto the ground states $\ket{\Omega_\pm}$ at $y=y_\pm$, respectively, as the energies of the excited states will scale with $m_0$. 
Thus, $\ket{Y_\pm} \propto \ket{\Omega_\pm}$ and the proportionality constants are just phases if we normalize all the states properly.

Combining, we can rewrite the $(d+1)$-dimensional partition function as
\begin{align} \label{bulkpart}
\begin{split}
    \mathcal{Z}_{d+1}(Y) &= \bra{Y_-} \hat{Y}_L \ket{Y_+}  \\
    &= \braket{Y_-|\Omega_-} \bra{\Omega_-} \hat{Y}_L \ket{\Omega_+} \braket{\Omega_+ | Y_+} \\
    &= \bra{\Omega_-}
     T\!\exp\left(-\int_{y_+}^{y_-} \hat{H}(y) dy\right) 
     \ket{\Omega_+} 
     \times \frac{\braket{\Omega_+ | Y_+}}{\braket{\Omega_- | Y_-}} \,,
\end{split}
\end{align}
where in the last step we used the fact that $\braket{Y_-|\Omega_-}$ is a pure phase.

To compute various state overlaps, we perform a mode expansion of $\Psi$ on the spatial slice $X$. 
Following Fujikawa \cite{Fujikawa:1983bg}, consider two hermitian operators on $X$ given by $\sH^1_d = \sD_d^\dagger \sD_d$ and $\sH^2_d = \sD_d \sD_d^\dagger$, and let the eigenvalues and the orthonormal eigenfunctions of $\sH^1_d$ be
\begin{equation}
    \sH^1_d \psi_a = \lambda_a^2 \psi_a \,.
\end{equation}
The eigenvalues are all positive, $\lambda_a > 0$ for all $a$, due to the assumption that the $d$-dimensional mass matrix is non-degenerate.
In this case, we can choose the orthonormal set of eigenfunctions of $\sH^2_d$ to be
\begin{equation} \label{fujikawa1}
    \tilde{\psi}_a = \frac{1}{\lambda_a} \sD_d \psi_a
\end{equation}
so that they satisfy $\sH^2_d \tilde{\psi}_a = \lambda_a^2 \tilde{\psi}_a$. 
The magnitude of the partition function of the $d$-dimensional fermion \eqref{Eq:d_dim_lagrangian} on $X$ is formally given by $|\mathcal{Z}_d (X)| = \prod_a \lambda_a$ \cite{Fujikawa:1983bg}.

Now, define
\begin{equation}
    \Psi_{+,a} = \begin{bmatrix*}
        \psi_a \\ 0 \end{bmatrix*} \,, \quad
        \Psi_{-,a} = \begin{bmatrix*}
            0 \\ -i\tilde{\psi}_a \end{bmatrix*} \,,
\end{equation}
and perform a mode expansion
\begin{equation}
    \Psi = \sum_a \left(A_{+,a}\Psi_{+,a} + A_{-,a}\Psi_{-,a} \right) \,.
\end{equation}
Using this expansion, we can first identify the ground state $\ket{\Omega(y)}$ for the Hamiltonian $\hat{H}(y)$ at fixed $y$.
In terms of the modes $A_{\pm,a}$, the Hamiltonian reads
\begin{equation} \label{hamiltonian}
    \hat{H}(y) = \sum_a A_a^\dagger \begin{bmatrix*}
        M_1(y) & \lambda_a \\
        \lambda_a & -M_1(y)
    \end{bmatrix*} A_a
\end{equation}
where $A = (A_{+,a},A_{-,a})^T$.
Following the conventions in \cite{Yonekura:2016wuc,Witten:2019bou}, we can diagonalize the Hamiltonian as follows. 
Define the angle $0<\theta_a(y) \leq \pi/2$ by
\begin{equation} \label{Eq:angle}
    \left(\text{cos}2\theta_a(y), \text{sin}2\theta_a(y) \right)
    = \frac{(M_1(y),\lambda_a)}{\sqrt{M_1(y)^2 + \lambda_a^2}} \,,
\end{equation}
and let
\begin{equation}
    B_a(y) = \begin{bmatrix*}
        B_{1,a}(y) \\
        B_{2,a}(y) \end{bmatrix*} = \begin{bmatrix*}
            \text{cos}\theta_a(y) & \text{sin} \theta_a(y) \\
            -\text{sin}\theta_a(y) & \text{cos} \theta_a(y)
        \end{bmatrix*} A_a \,.
\end{equation}
In terms of these new modes, the Hamiltonian is diagonalized,
\begin{equation}
    \hat{H}(y) = \sum_a \sqrt{\lambda_a^2 + M_1(y)^2}\left(B_{1,a}^\dagger(y)B_{1,a}(y)- B_{2,a}^\dagger(y)B_{2,a}(y)\right) \,.
\end{equation}
We can now easily see that the ground state $\ket{\Omega(y)}$ at fixed $y$ is the state that is annihilated by all $B_{1,a}(y)$ and all $B_{2,a}^\dagger(y)$.
In particular, $\ket{\Omega(y_\pm)} \equiv \ket{\Omega_\pm}$.

Going back to Eq. \eqref{bulkpart}, we will now argue that the fraction $\frac{\braket{\Omega_+ | Y_+}}{\braket{\Omega_- | Y_-}}$ gives us the exponentiated $\eta$-invariant of the operator $K$.
The operator $K$ that we defined in Eq.~(\ref{defK}) is a hermitian operator on a closed manifold $Y$, but when restricted to $Y_+$, it requires an appropriate choice of boundary condition on $\partial Y_+ = X$ to maintain the hermiticity.
One such boundary condition is given by the APS boundary condition. 
In our conventions, the condition is stated as follows. 
We can write $K = \mathrm{i}\Gamma^y (\partial_y + \mathcal{D}_X)$ inside the collar neighborhood of $X$, where
\begin{equation}
    \mathcal{D}_X = \begin{bmatrix*}
        0 & \mathrm{i}\sD_d^\dagger \\
        -\mathrm{i}\sD_d & 0
    \end{bmatrix*}
\end{equation}
is a hermitian operator on $X$. 
The eigenvlues of $\mathcal{D}_X$ come in pairs, $\pm\lambda_a$, since $\{\Gamma^y,\mathcal{D}_X\}=0$.
The corresponding eigenfunctions are $\Psi_{+,a}\pm\Psi_{-,a}$. 
The APS boundary condition on $Y_+$ is given by
\begin{equation} \label{APSbc}
    \mathtt{APS} : \mathcal{P}_{< 0}^{\mathcal{D}_X} \Psi |_{\partial Y_+ = X} = 0
\end{equation}
where $\mathcal{P}_{< 0}^{\mathcal{D}_X}$ is the projection operator onto the subspace spanned by the eigenfunctions with negative eigenvalues of $\mathcal{D}_X$ on $X$. 
This means in the mode expansion
\begin{align} \label{modeexpansion}
\begin{split}
    \Psi &= \sum_a (A_{+,a}\Psi_{+,a} + A_{-,a}\Psi_{-,a}) \\
    &= \sum_a \left[\frac{1}{2}(A_{+,a}+A_{-,a})(\Psi_{+,a}+\Psi_{-,a})
       +\frac{1}{2}(A_{+,a}-A_{-,a})(\Psi_{+,a}-\Psi_{-,a}) \right]\,,
\end{split}
\end{align}
the coefficients $\frac{1}{2}(A_{+,a}-A_{-,a})$ are constrained to be $0$, whereas the other coefficients $\frac{1}{2}(A_{+,a}+A_{-,a})$ are unconstrained. 
The quantum state corresponding to such a boundary condition is characterized by \cite{Yonekura:2016wuc,Witten:2019bou}
\begin{equation}
    \bra{\mathtt{APS}} (A_{+,a}-A_{-,a}) = \bra{\mathtt{APS}} (A_{+,a}+A_{-,a})^\dagger = 0
\end{equation}
for all $a$.

Following \cite{Yonekura:2016wuc,Witten:2019bou}, let $\ket{E}$ be a state defined by 
\begin{equation} \label{Eq:state_E}
    A_{+,a} \ket{E} = A_{-,a} \ket{E} = 0
\end{equation}
for all $a$.
The states $\ket{\Omega(y)}$ and $\ket{\mathtt{APS}}$ can be realized as
\begin{align} \label{Eq:state_gs_aps}
\begin{split}
    \ket{\Omega(y)} &= \prod_a B_{2,a}^\dagger \ket{E} \,, \\
    \ket{\mathtt{APS}} &= \prod_a \frac{(A_{-,a}-A_{+,a})^\dagger}{\sqrt{2}} \ket{E} \,.
\end{split}
\end{align}
Then, we claim
\begin{equation} \label{Eq:state_overlap_ratio}
    \frac{\braket{\mathtt{APS} | \Omega_-}}{\braket{\mathtt{APS} | \Omega_+}} = \frac{\prod\limits_a \left(\text{cos}\theta_a(y_-) + \text{sin}\theta_a(y_-) \right) }{\prod\limits_a \left(\text{cos}\theta_a(y_+) + \text{sin}\theta_a(y_+) \right) } = 1 \,.
\end{equation}
To see this, first notice
\begin{align}
\begin{split}
    \braket{\mathtt{APS} | \Omega(y)} &= \frac{1}{\sqrt{2}} \prod_{a,a'}
    \bra{E}\left( A_{-,a} - A_{+,a}\right) B_{2,a'}^\dagger \ket{\Omega(y)} \\
    &=\frac{1}{\sqrt{2}} \prod_{a,a'}
    \bra{E}\left( A_{-,a} - A_{+,a}\right)\left(-\text{sin}\theta_{a'}(y) A_{+,a'}^\dagger + \text{cos}\theta_{a'}(y) A_{-,a'}^\dagger  \right) \ket{\Omega(y)} \\
    &= \frac{1}{\sqrt{2}} \prod_a \left(\text{sin}\theta_a(y) + \text{cos}\theta_a(y) \right)  \,,
\end{split}
\end{align}
where the last step follows from the usual anticommutation relations for the fermionic modes $A_{\pm,a}$ and Eq. \eqref{Eq:state_E} (we normalize the state $\ket{E}$ so that $\braket{E|E}=1$).
This verifies the first equality sign in (\ref{Eq:state_overlap_ratio}).
Then, at $y = y_\pm$, we have $M(y_\pm) = \pm m_0$, and from Eq. (\ref{Eq:angle}) we see that $\theta_a(y_+) + \theta_a (y_-) = \pi/2 $ for all $a$.
From this the second equality sign of (\ref{Eq:state_overlap_ratio}) follows.

Using the fact that $\ket{Y_\pm} \propto \ket{\Omega_\pm}$ and Eq. (\ref{Eq:state_overlap_ratio}), we can rewrite the fraction $\frac{\braket{\Omega_+ | Y_+}}{\braket{\Omega_- | Y_-}}$ in Eq.~(\ref{bulkpart}) as 
\begin{equation} \label{eq:ground_state_APS}
    \frac{\braket{\Omega_+ | Y_+}}{\braket{\Omega_- | Y_-}} =
    \frac{\braket{\mathtt{APS} | Y_+}}{\braket{\mathtt{APS} | Y_-}}\frac{\braket{\mathtt{APS} | \Omega_-}}{\braket{\mathtt{APS} | \Omega_+}} = \frac{\braket{\mathtt{APS} | Y_+}}{\braket{\mathtt{APS} | Y_-}} \,.
\end{equation}
In the limit $m_0 \to \infty$, this last expression is nothing but the exponentiated $\eta$-invariant for the operator $K$ on $Y_+$ with the APS boundary condition imposed, which can be seen as follows.
Since $K$ is hermitian on $Y_+$ with this boundary condition, we may write its complete set of orthonormal eigenfunctions and the corresponding eigenvalues as
\begin{equation}
    K\Psi_j = \xi_j \Psi_j \,.
\end{equation}
Each $\Psi_j$ satisfies the APS boundary condition, $\mathcal{P}_{ < 0}^{\mathcal{D}_X} \Psi_j |_X = 0$, and we can expand the fermion field in the path integral as
\begin{equation} \label{expansion1}
    \Psi = \sum_j a_j \Psi_j
\end{equation}
with grassmannian coefficients $a_j$.
To know what is the valid expansion for the conjugate field $\bar{\Psi}$, we need to know the boundary condition imposed on it.
If we perform a mode expansion for $\bar{\Psi}$ at the boundary as
\begin{equation}
    \bar{\Psi}|_X = \sum_a \left[ B_{1,a} (\Psi_{+,a}+\Psi_{-,a})^\dagger + B_{2,a} (\Psi_{+,a}-\Psi_{-,a})^\dagger \right] \,,
\end{equation}
then we find that $B_{1,a}$ and $B_{2,a}$ are canonically conjugate to $\frac{1}{2} (A_{+,a}-A_{-,a})$ and $\frac{1}{2} (A_{+,a}+A_{-,a})$, respectively.
This is because $\Gamma^y$ from $\bar{\Psi}\Gamma^y \partial_y \Psi$ in the kinetic term permutes $\Psi_{+,a}+\Psi_{-,a}$ and $\Psi_{+,a}-\Psi_{-,a}$, see Eq.~(\ref{modeexpansion}).

As we explained above, the boundary condition on $\Psi$ is such that at the boundary $A_{+,a} - A_{-,a}$ vanishes for all $a$, whereas $A_{+,a} + A_{-,a}$ is unconstrained for all $a$.
Correspondingly, at the boundary we have to require $B_{2,a}$ which is conjugate to $A_{+,a} - A_{-,a}$ to be unconstrained, whereas $B_{1,a}$ which is conjugate to $A_{+,a} + A_{-,a}$ to vanish.
This is the boundary condition imposed on $\bar{\Psi}$.

This boundary condition is precisely satisfied by $\Psi_j^\dagger$, thus we can expand inside the path integral
\begin{equation}
    \bar{\Psi} = \sum_j b_j \Psi_j^\dagger
\end{equation}
with the grassmannian coefficients $b_j$.
Combined with Eq. \eqref{expansion1}, these expansions diagonalize the action, $S = \int_{Y_\pm} \bar{\Psi} (-\mathrm{i}K \pm m_0) \Psi = \sum\limits_j (-\mathrm{i}\xi_j \pm m_0) b_j a_j$.
Thus, the path integral gives us
\begin{equation}
    \frac{\braket{\mathtt{APS} | Y_+}}{\braket{\mathtt{APS} | Y_-}} = \prod_j \frac{-\mathrm{i}\xi_j + m_0}{-\mathrm{i}\xi_j - m_0} \to \text{exp}(2\pi \mathrm{i} \eta_K (Y_+;\mathtt{APS}))
\end{equation}
in the limit $m_0 \to \infty$.
Here,
\begin{equation} \label{Eq:eta_invariant}
\eta_K(Y_+;\mathtt{APS}) = \frac{1}{2} \sum_j \text{sgn}(\xi_j) 
\end{equation}
is the desired $\eta$-invariant of $K$ with the APS boundary condition imposed on the boundary $\partial Y_+ = X$. 
We have defined $\text{sgn}(0) \equiv +1$.
In general, the infinite sum in Eq.~(\ref{Eq:eta_invariant}) should be appropriately regularized. 
For instance, a common choice is the heat kernel regularization which is given by
\begin{equation} \label{eq:eta_invariant_reg}
    \eta_K(Y_+;\mathtt{APS}) = \lim_{\mu \to \infty}
    \left[
    \frac{1}{2}
    \sum_{j} \text{sgn}(\xi_j) \text{exp} \left(-\frac{\xi_j^2}{\mu^2} \right) 
    \right]\,.
\end{equation}

\begin{figure}[t]
    \centering
    \includegraphics[width=0.8\textwidth]{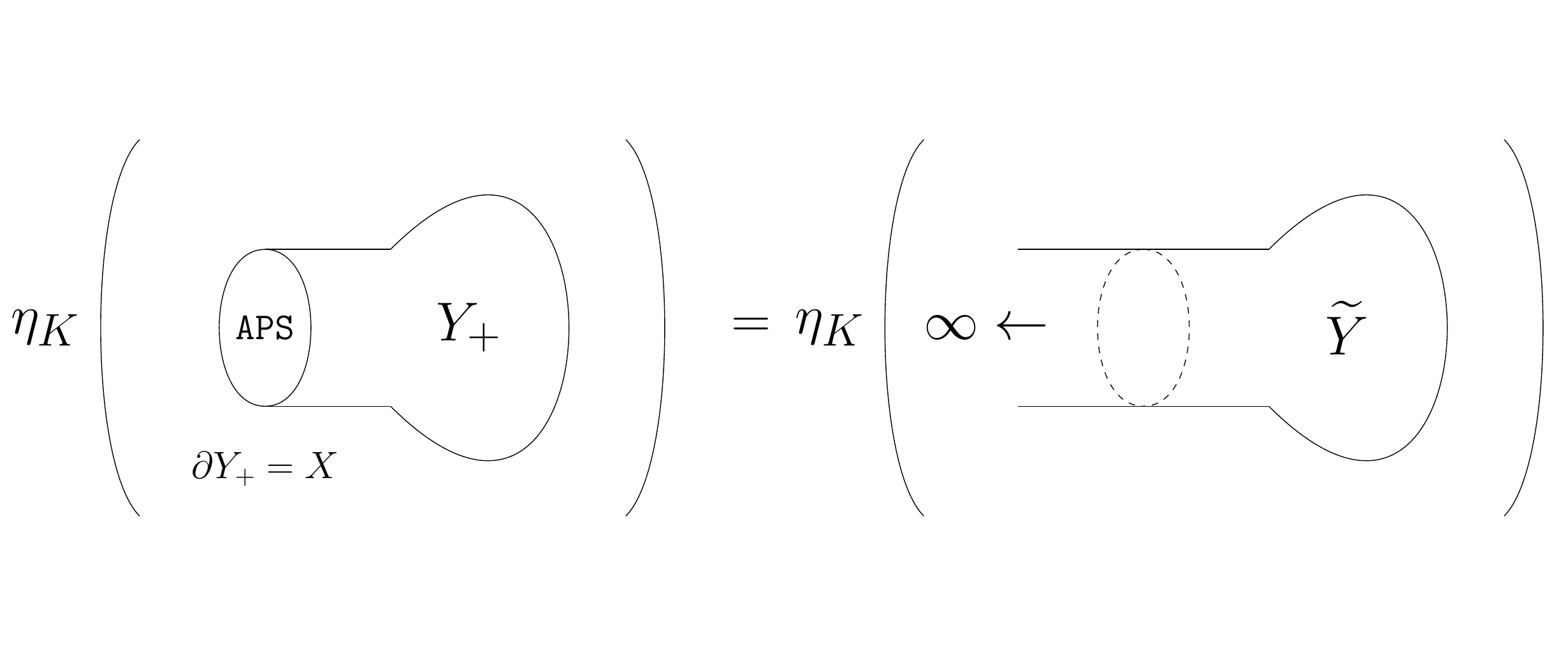}
    \caption{The APS boundary condition can be replaced by an infinite half cylinder. All the mass parameters and the background gauge fields are independent of the longitudinal direction along the half cylinder.}
    \label{Fig:APScylinder}
\end{figure}

An important property of the APS boundary condition that we will exploit in the next section is that it can be replaced by an infinitely long half cylinder attached to the boundary.
That is, if we consider the manifold $\widetilde{Y} \equiv Y_+ \cup_{X} \left( X \times [0,\infty) \right)$, we have
\begin{equation} \label{Eq:eta_cylinder}
    \eta_K(Y_+;\mathtt{APS}) = \eta_K(\widetilde{Y}) \,.
\end{equation}
All the background gauge fields are continuously glued along the boundary $X$, and they must be extended into the half-infinite cylinder in such a way they are independent of the direction transverse to $X$.
The manifold $\widetilde{Y}$ does not have a boundary (and thus we do not need to worry about boundary conditions) but is now non-compact, see Figure \ref{Fig:APScylinder}.
Eq. (\ref{Eq:eta_cylinder}) holds due to the fact that the spectrum of normalizable eigenfunctions of $K$ on 
$\widetilde{Y}$ is the same as that on $Y_+$ with the APS boundary condition imposed \cite{Atiyah:1975jf}.
Alternatively, this can be understood from Eq. (\ref{eq:ground_state_APS}), which tells us that the APS boundary condition can be replaced by the ground states $\ket{\Omega_\pm}$.
Doing so is equivalent to attaching an infinitely long half cylinder since the Euclidean path integral along the cylinder prepares the ground states.

Next, we need to compute the amplitude $\bra{\Omega_-} T[{\exp\{-\int_{y_+}^{y_-} \hat{H}(y) dy\}}] \ket{\Omega_+}$.
For our purposes, it suffices to take the limit where the length of the middle region $X \times I$ becomes infinitesimal (while maintaining $|y_+ - y_-| m_0 \gg 1$), in which case we can simply approximate
\begin{align} \label{ratio1}
\begin{split}
    \bra{\Omega_-} T[{\text{exp}\{-\int_{y_+}^{y_-} \hat{H}(y) dy\}}] \ket{\Omega_+} &\approx \braket{\Omega_- | \Omega_+}  \\
    &= \prod_a \left(\text{cos}\theta_a(y_-)\text{cos}\theta_a(y_+) + \text{sin}\theta_a(y_-)\text{sin}\theta_a(y_+)\right)  \\
    &= \prod_a \frac{\lambda_a}{\sqrt{m_0^2 + \lambda_a^2}} = |\mathcal{Z}_d (X)|_{\text{Reg.}} \,.
\end{split}
\end{align}
Here, the overlap $\braket{\Omega_- | \Omega_+}$ is computed using the presentation of the ground states given in Eq. (\ref{Eq:state_gs_aps}).
The resulting expression is essentially the product of all $\lambda_a$'s, but the eigenvalues much larger than $m_0$ are regularized and do not contribute. 
This corresponds to the regularized magnitude of the $d$-dimensional partition function \cite{Witten:2019bou}, which we denote as $|\mathcal{Z}_d (X)|_{\text{Reg.}}$.

Now, we combine Eqs. (\ref{bulkpart}), (\ref{eq:ground_state_APS}), and (\ref{ratio1}).
This gives us Eq.(\ref{phaseeta}) as desired.
Moreover, we may also consider the ratio
\begin{equation} \label{phaseeta2}
    \frac{\mathcal{Z}_{d+1}(Y)}{\mathcal{Z}_{d+1}^{(0)}(Y)} = 
    \left| \frac{\mathcal{Z}_d(X)}{\mathcal{Z}_d^{(0)}(X)}\right|
    \times
    \text{exp}(2\pi \mathrm{i} (\eta_K (Y_+;\mathtt{APS}) - \eta_{K^{(0)}}(Y_+;\mathtt{APS})) 
\end{equation}
where the superscript $(0)$ means setting $M$ to be a nonzero constant, which acts as the Pauli-Villars regulator for the $d$-dimensional fermions.
The reason for doing this is because in the end we want to obtain a purely $d$-dimensional statement, whereas the regulator $m_0$ in Eq.~(\ref{ratio1}) has the $(d+1)$-dimensional origin.

\subsection{Reducing the partition function to an interface} \label{subsec2.2}

Here, we will argue that
\begin{equation} \label{reducing}
\frac{\mathcal{Z}_{d+1}(Y)}{\mathcal{Z}^{(0)}_{d+1}(Y)} = \frac{\mathcal{Z}_d(X)}{\mathcal{Z}_d^{(0)}(X)}
\end{equation}
when $m_0 \to \infty$. 
Define two hermitian operators on $Y$, $\sH_{d+1}^1 = \sD^\dagger_{d+1} \sD_{d+1}$ and $\sH_{d+1}^2 = \sD_{d+1} \sD_{d+1}^\dagger$. 
Let the eigenvalues and eigenfunctions of $\sH_{d+1}^1$ be\footnote{We use the greek letters $\alpha, \beta, \cdots$ to label $(d+1)$-dimensional eigenfunctions $\Psi_\alpha$, and the latin letters $a, b, \cdots$ to label $d$-dimensional eigenfunctions $\psi_a$.}
\begin{equation}
    \sH_{d+1}^1 \Psi_\alpha = \Lambda_\alpha^2 \Psi_\alpha \,.
\end{equation}  
Again, $\Lambda_\alpha > 0$ for all $\alpha$ due to the fact that $\text{det}(M_1)$ and $\text{det}(M_2)$ are never simultaneously zero on $Y$ in our construction.
Then, as eigenfunctions of $\sH_{d+1}^2$, we can take
\begin{equation}
    \tilde{\Psi}_\alpha = \frac{1}{\Lambda_\alpha} \sD_{d+1} \Psi_\alpha \,,
\end{equation}
similar to Eq. \eqref{fujikawa1}. 
Now, the (regularized) fermion partition function can be written in terms of these data following \cite{Fujikawa:1983bg},
\begin{equation} \label{fujikawapartition1}
    \frac{\mathcal{Z}_{d+1}(Y)}{\mathcal{Z}^{(0)}_{d+1}(Y)} = \text{Det}(\sJ) \text{Det}(\widetilde{\sJ}^\dagger) \frac{\prod\limits_\alpha \Lambda_\alpha}{\prod\limits_\alpha \Lambda_\alpha^{(0)}}\,.
\end{equation}
Here, $\sJ$ and $\widetilde{\sJ}$ are formally unitary infinite-dimensional Jacobian matrices with entries
\begin{equation}
    (\sJ)_{\alpha\beta} = \int_Y \Psi^\dagger_\alpha \Psi^{(0)}_\beta \,,\quad (\widetilde{\sJ})_{\alpha\beta} = \int_Y \tilde{\Psi}^\dagger_\alpha \tilde{\Psi}^{(0)}_\beta \,.
\end{equation}
The quantities with superscript $(0)$ are defined in the same way as the corresponding quantities without the superscript, except for that $M$ is fixed to be a nonzero constant. 
Similarly, the $d$-dimensional (regularized) fermion partition function can be written as
\begin{equation} \label{fujikawapartition2}
    \frac{\mathcal{Z}_{d}(X)}{\mathcal{Z}^{(0)}_{d}(X)} = \text{Det}(\sK) \text{Det}(\widetilde{\sK}^\dagger) \frac{\prod\limits_a \lambda_a}{\prod\limits_a \lambda_a^{(0)}}\,,
\end{equation}
with
\begin{equation}
    (\sK)_{ab} = \int_X \psi^\dagger_a \psi^{(0)}_b \,,\quad (\widetilde{\sK})_{ab} = \int_X \tilde{\psi}^\dagger_a \tilde{\psi}^{(0)}_b \,.
\end{equation}

The spectrum of the operators $\sH_{d+1}^{1,2}$ contains that of the operators $\sH_{d}^{1,2}$ when $m_0$ is much larger than the inverse length of the collar neighborhood of $X$ in $Y$, due to the existence of the exponentially localized modes. 
Explicitly, consider an ansatz, labeled by the index $a$,
\begin{equation} \label{localizedmodes}
    \Psi_a = \mathcal{N} (\psi_a \otimes v^+) \exp \left(-\int_{y^*}^{y} M_1(y^\prime) dy^\prime \right)
\end{equation}
where $y^*$ is a point inside the interval $[y_+, y_-]$, say the midpoint, and $v^+$ is the eigenvector $\sigma^3 v^+ = + v^+$ (recall our conventions for the gamma matrices (\ref{gammamatrix})). 
$\mathcal{N}$ is a normalization constant. 
The ansatz satisfies
\begin{align}
\begin{split}
    \sH_{d+1}^1 \Psi_a &= (\sH_d^1 \psi_a) \otimes v^+ \exp \left(-\int_{y^*}^{y} M_1(y^\prime) dy^\prime \right) \\
    &= \lambda_a^2 \Psi_a \,.
\end{split}
\end{align}
The normalization constant $\mathcal{N}$ needs to be fixed such that the ansatzes satisfy the orthonormality condition
\begin{equation}
    \int_Y \Psi_a^\dagger \Psi_b = \delta_{ab} \, .
\end{equation}
This can be explicitly done if $m_0 \gg |y_- - y_+|^{-1}$, by choosing
\begin{equation}
    \mathcal{N} = \left[\int_{y_+}^{y_-} dy \, \exp \left(-2\int_{y^*}^{y} M_1(y^\prime) dy^\prime \right) \right]^{-1/2}
\end{equation}
with which the orthonormality condition is satisfied modulo corrections that are exponentially suppressed by $m_0|y_- - y_+| \gg 1$. 
In the same manner we define the ansatz $\tilde{\Psi}_a$ satisfying $\sH_{d+1}^2 \tilde{\Psi}_a = \lambda_a^2 \tilde{\Psi}_a$. 
Note that $\mathcal{N}$ is a number which depends only on the fixed interface configuration of $M_1(y)$ and does not depend on the configuration of $M_2$. 
Indeed, modulo exponentially small corrections, we have
\begin{equation} \label{norminner}
    \int_Y \Psi_a^\dagger \Psi^{(0)}_b = \int_X \psi_a^\dagger \psi^{(0)}_b \,,\quad \int_Y \tilde{\Psi}_a^\dagger \tilde{\Psi}^{(0)}_b = \int_X \tilde{\psi}_a^\dagger \tilde{\psi}^{(0)}_b \,.
\end{equation}

The expression (\ref{fujikawapartition1}) is obtained by expanding the fermion fields $\Psi$ and $\bar{\Psi}$ in terms of the eigenfunctions $\Psi_\alpha$ and $\tilde{\Psi}_\alpha$, respectively. 
Instead, we can perform a partial expansion with respect to the localized ansatzes, 
\begin{equation} \label{partialexpansion}
    \Psi = \sum_a ( c_a \Psi_a + \Psi^\prime ) \,, \quad
    \bar{\Psi} = \sum_a ( d_a \tilde{\Psi}_a^\dagger + \bar{\Psi}^\prime ) \,,
\end{equation}
where $c_a$ and $d_a$ are expansion coefficients which are grassmannian. 
$\Psi^\prime$ belongs to the orthogonal complement of the subspace spanned by the localized modes $\Psi_a$. 
Similarly, $\bar{\Psi}^\prime$ is in the orthogonal complement of the subspace spanned by $\tilde{\Psi}_a^\dagger$.

Inserting the partial expansions (\ref{partialexpansion}), the fermion path integral becomes
\begin{align}
    \frac{\mathcal{Z}_{d+1}(Y)}{\mathcal{Z}^{(0)}_{d+1}(Y)}
    &= \text{Det}(\sK) \text{Det}(\widetilde{\sK}^\dagger) \frac{\prod_a \lambda_a}{\prod_a \lambda_a^{(0)}} \times \frac{\int [D\bar{\Psi}^\prime][D\Psi^\prime] \text{exp}(-\int_Y \bar{\Psi}^\prime \sD_{d+1} \Psi^\prime)}{\int [D\bar{\Psi}^\prime][D\Psi^\prime] \text{exp}(-\int_Y \bar{\Psi}^\prime \sD_{d+1}^{(0)} \Psi^\prime)} \nonumber \\
    &= \frac{\mathcal{Z}_{d}(X)}{\mathcal{Z}^{(0)}_{d}(X)} \times \frac{\int [D\bar{\Psi}^\prime][D\Psi^\prime] \text{exp}(-\int_Y \bar{\Psi}^\prime \sD_{d+1} \Psi^\prime)}{\int [D\bar{\Psi}^\prime][D\Psi^\prime] \text{exp}(-\int_Y \bar{\Psi}^\prime \sD_{d+1}^{(0)} \Psi^\prime)}
\end{align}
where in the first line, we used Eq.~(\ref{norminner}), and in the second line, we substituted Eq.~(\ref{fujikawapartition2}).
Then, for $m_0 \to \infty$, we have 
\begin{equation} \label{bulkmodesaction}
    \int_Y \bar{\Psi}^\prime \sD_{d+1} \Psi^\prime \approx \int_Y \bar{\Psi}^\prime M_1(y) \Psi^\prime \,.
\end{equation}
Recall that $M_1 = \pm m_0$ inside $Y_\pm$, whereas in the middle region $Y_L$ it is a smooth interpolating function between the two values $\pm m_0$. 
Eq. (\ref{bulkmodesaction}) is valid because we first factored out the localized modes. 
For a localized mode, its variation along the $y$ direction will cancel the $m_0$ mass term and we cannot ignore other terms in the action even in the limit $m_0 \to \infty$.

Now, since Eq.~(\ref{bulkmodesaction}) holds independent of the configuration of the other mass parameter $M_2$, we can deduce that
\begin{equation}
    \frac{\int [D\bar{\Psi}^\prime][D\Psi^\prime] \text{exp}(-\int_Y \bar{\Psi}^\prime \sD_{d+1} \Psi^\prime)}{\int [D\bar{\Psi}^\prime][D\Psi^\prime] \text{exp}(-\int_Y \bar{\Psi}^\prime \sD_{d+1}^{(0)} \Psi^\prime)} \rightarrow 1
\end{equation}
as $m_0 \to \infty$. 
Thus, in this limit, we conclude
\begin{equation} \label{reducingtointerface}
    \frac{\mathcal{Z}_{d+1}(Y)}{\mathcal{Z}^{(0)}_{d+1}(Y)} =
    \frac{\mathcal{Z}_{d}(X)}{\mathcal{Z}^{(0)}_{d}(X)}
\end{equation}
as desired. 
Combining Eq.~(\ref{reducingtointerface}) with Eq.~(\ref{phaseeta2}) from the previous subsection, we arrive at our central claim:
\begin{equation} \label{centralclaim}
    \left.\frac{\mathcal{Z}_{d}(X)}{\mathcal{Z}^{(0)}_{d}(X)} \middle/ \left|\frac{\mathcal{Z}_d(X)}{\mathcal{Z}_d^{(0)}(X)}\right|\right.  = \text{exp}(2\pi \mathrm{i} (\eta_K (Y_+;\mathtt{APS}) - \eta_{K^{(0)}} (Y_+;\mathtt{APS})) \,.
\end{equation}
In words, we claim that the phase of the Euclidean partition function of free fermions coupled to arbitrary background gauge fields and spacetime-dependent mass parameters is in general given by the APS $\eta$-invariant of the hermitian operator $K$ that extends the $d$-dimensional Dirac operator to a $(d+1)$-dimensional manifold $Y_+$ which bounds the $d$-dimensional spacetime manifold $X$.
If the $d$-dimensional mass matrix $M$ becomes infinitely large, one expects to have $ \left|\frac{\mathcal{Z}_d(X)}{\mathcal{Z}_d^{(0)}(X)}\right|  \rightarrow 1$.
Then, Eq. (\ref{centralclaim}) defines the partition function of a \emph{parameterized} invertible field theory given by the low-energy limit of the $d$-dimensional fermions with modulated mass parameters.

As a remark, we mention that both the actual fermion system with the partition function $\mathcal{Z}_d(X)$ that we are interested in and the regulator fermions given by the partition function $\mathcal{Z}_d^{(0)}(X)$ are on the same footing at the technical level.
Although we have required the mass parameters for the regulator fermions to be constant-valued for simplicity, everything that we said is still valid even if we allow complicated mass configurations for the regulator fermions.
What this tells us is that precisely speaking, only the difference between two fermion systems is a well-defined invertible theory, and in that sense, the regulator fermions are there to specify a ``reference point.''
This indeed reflects a familiar fact that a QFT usually has an ambiguity in how one fixes the counterterms that depend only on the backgrounds fields, where different choices of the counterterms correspond to different regularization schemes.
Still, we will keep making a conceptual distinction between the physical system that we are interested in and the regulator system, while exploiting the freedom to choose a convenient ``reference point'' when needed.

\subsection{Generalized cohomology perspectives}\label{sec:gencoh}

We have related the $d$-dimensional fermion system with modulated mass parameters with a $(d+1)$-dimensional fermion system. 
It would be useful to rephrase the statement in a more general perspective using the language of the generalized cohomology theory.

The invertible field theories, which are the IR fixed points of trivially gapped QFTs, define a generalized cohomology theory \cite{Kit13,Kit15,Freed:2016rqq,Gaiotto:2017zba,Yamashita:2021cao,Yamashita:2021fkd}.
Let $\mathfrak{M}_d$ be the space of $d$-dimensional invertible theories.
Then the statement, or conjecture, is that the set of spaces 
$\mathfrak{M}_d$ is equipped with a particular structure, called the $\Omega$-spectrum, to define a generalized cohomology theory. 
The generalized cohomology theory on a topological space $P$, which we call $\Inv^d(P)$, is defined by
\begin{equation}
    \Inv^d(P) = [P,\mathfrak{M}_d] \,,
\end{equation}
where $[P,Q]$ is the set of homotopy classes of continuous maps from $P$ to $Q$.
Moreover, this cohomology theory $\Inv$ is identified with a cobordism theory \cite{Kapustin:2014tfa,Kapustin:2014dxa,Freed:2016rqq,Yonekura:2018ufj,Yamashita:2021cao,Yamashita:2021fkd}.\footnote{
When we focus on the invertible theories that are IR fixed points of free fermion theories, the space $\mathfrak{M}_d$ is homotopic to the space $P_d$ of non-degenerate mass matrices for (infinitely many) $d$-dimensional fermions.
The corresponding cohomology theory is isomorphic to the $K$-theory or $KO$-theory, when the fermions are Dirac or Majorana, respectively\cite{Kitaev:2009mg,ryu2010topological}.}
We can regard a map $\mathcal{T}:P \to \mathfrak{M}_d$ as a family of invertible QFTs parameterized by elements of $P$.
In other words, the cohomology group $\Inv^d(P)$ classifies the deformation classes of parameterized invertible QFTs with parameter space $P$ \cite{Freed:2017rlk,Cordova:2019jnf,Cordova:2019uob}.
In our context, the massive free fermion theory defines an element of $\Inv^d(P)$, where $P$ is the space of non-degenerate mass matrices for a specified number $N_f$ of flavors.\footnote{
    As said in the previous footnote, the space $P_d$ is the spectrum for the $K$ (or $KO$) theory itself when $N_f = \infty$. In this case the relevant cohomology group is $K^d(P_d) = [P_d,P_d]$, and the fermion theory corresponds to the identity map.
}

The fact that $\Inv$ is a generalized cohomology theory means that we can apply the axioms that such a theory satisfies.
In particular, it is natural to consider a \emph{relative} cohomology group $\Inv(P,A)$ where $A$ is a subset of $P$.
Physically, the relative cohomology $\Inv(P,A)$ classifies (the deformation classes) of invertible QFTs parameterized by $P$, such that it is trivial when the parameter resides in $A$. 
Given such a parameterized invertible QFT $\mathcal{I}$, and a manifold with boundary $(M,\partial M)$,
it is natural to consider the partition function $Z^\mathcal{I}(M,p)$, where $p$ is a map $p: M \to P$ whose restriction on $\partial M$ is contained in $A$ .
In particular, for the torsion part of $\Inv$, the partition function is the invariant of the relative bordism class of the pair $(M,\partial M)$ in $(P,A)$.

Recall that in Eq. (\ref{centralclaim}), the mass matrix has to be non-degenerate on the boundary of $Y_+$ but does not need to be so inside the bulk.
This means that we were considering the pair $(\widetilde{P},P)$, where $\widetilde{P}$ is the space of all (possibly degenerate) mass matrices, and the exponentiated $\eta$-invariant which appears on the right-hand side of Eq. \eqref{centralclaim} is understood as the partition function of the invertible QFT parameterized by the pair $(\widetilde{P},P)$.
Therefore, \eqref{centralclaim} is a correspondence between $\Inv^d(P)$ and $\Inv^{d+1}(\widetilde{P},P)$. 
Indeed, because $\widetilde{P}$ is contractible, the general axiom of cohomology tells us that
\begin{equation} \label{Eq:gen_coh}
    \widetilde{\Inv}^d(P) \cong \Inv^{d+1}(\widetilde{P},P),
\end{equation}
where $\widetilde{\Inv}^d(P) = \Inv^d(P,\text{pt})$ is the reduced cohomology group.\footnote{
    We have the long exact sequence $\widetilde{\Inv}^{d}(\widetilde{P})\to\widetilde{\Inv}^d(P) \to \Inv^{d+1}(\widetilde{P},P) \to \widetilde{\Inv}^{d+1}(\widetilde{P})$ and $\widetilde{\Inv}^\star(\widetilde{P})=0$ because $\widetilde{P}$ is contractible.
}
The $d$-dimensional parameterized invertibe theory is an element of the cohomology group $\widetilde{\Inv}^d(P)$ (since we are not interested in purely gravitational invertible theories, taking the reduced comology group suffices).
Then, Eq. \eqref{centralclaim} really states that the correspondence \eqref{Eq:gen_coh} in the generalized cohomology theory holds at the level of partition functions of parameterized invertible QFTs coming from free fermions.

\section{Berry curvature} \label{Sec3}

The group of ordinary invertible field theories without any parameters are known to be generated by two classes of invertible theories \cite{Freed:2016rqq,Yamashita:2021cao,Yamashita:2021fkd}.
On the one hand, there are invertible theories which match the perturbative anomalies of the boundary theories.
The partition functions of such invertible theories are given by an integral of the Chern-Simons forms for the background gauge fields as well as the curvature of the spacetime manifold, which locally are differential forms.
Such invertible theories are classified by the free part of the suitable generalized cohomology group.
On the other hand, there are invertible theories which correspond to the non-perturbative anomalies of the boundary theories.
Their partition functions define bordism invariants, and these topological invertible theories are classified by the torsion part of the same generalized cohomology group \cite{Kapustin:2014tfa,Kapustin:2014dxa,Yonekura:2018ufj}.
A general invertible theory is a mixture of these two parts.

This understanding continues to hold for the parameterized invertible field theories as well.
In particular, our fermion system in the infinite mass limit defines a parameterized invertible theory as we explained, and thus its partition function will take the following general form:
\begin{equation} \label{Eq:general_inv_partfnc}
    \left.\frac{\mathcal{Z}_{d}(X)}{\mathcal{Z}^{(0)}_{d}(X)} \middle/ \left|\frac{\mathcal{Z}_d(X)}{\mathcal{Z}_d^{(0)}(X)}\right|\right.
    \rightarrow \phi(X) \exp \left(
        2\pi \mathrm{i} \int_X \mathcal{W}
    \right) \in U(1)
\end{equation}
as mass goes to infinity, $|M| \to \infty$.
Here, $\phi(X) \in U(1)$ is the value of a bordism invariant $\phi$ evaluated on the spacetime manifold $X$ (equipped with the background gauge fields and the modulated mass parameters), and $\mathcal{W}$ is locally a (real-valued) differential form made out of the background gauge fields (as well as the curvature of the spacetime manifold) and the modulated mass parameters.

Following \cite{Kapustin:2020eby,Kapustin:2020mkl,Hsin:2020cgg}, we call the right-hand side of Eq. \eqref{Eq:general_inv_partfnc} a higher Berry phase.
The differential form $\mathcal{W}$ can then be thought of as a higher analogue of the Berry connection in quantum mechanics, and it is convenient to take the formal derivative of it to define the (higher) Berry curvature:
\begin{equation}
    \mathcal{B} =  \mathrm{d} \mathcal{W} \,.
\end{equation}
As was advocated in the Introduction, the Berry curvature is the ``anomaly polynomial'' for the theories living on the boundary of the $d$-dimensional fermion system, which will have the anomaly in the space of coupling constants (i.e., the mass parameter space) \cite{Cordova:2019jnf,Cordova:2019uob}.

In this section, we will focus on this Berry curvature, and the bordism invariant part will be discussed in Section \ref{Sec4}.
For the purpose of studying only the Berry curvature part, it will suffice to consider the spacetime manifold $X$ which is null-bordant even while keeping the mass to be non-degenerate inside the null-bordism.
That is, we will assume the existence of a $(d+1)$-dimensional manifold $Y$ whose boundary is $\partial Y = X$,\footnote{This $Y$ is not to be confused with the specific manifold $Y$ that we constructed in Section \ref{Sec2}.} where the mass matrix is extended into $Y$ in such a way that it is everywhere non-degenerate.\footnote{Note that this is different from the general discussion in Section \ref{Sec2}, where inside the manifold $Y_+$ with $\partial Y_+ = X$ the mass matrix was allowed to become degenerate.}
In this case, the fermion partition function in the infinite mass limit is given as
\begin{equation} \label{Eq:berry_curvature}
    \left.\frac{\mathcal{Z}_{d}(X)}{\mathcal{Z}^{(0)}_{d}(X)} \middle/ \left|\frac{\mathcal{Z}_d(X)}{\mathcal{Z}_d^{(0)}(X)}\right|\right.
    \rightarrow \exp \left(
        2\pi \mathrm{i} \int_Y \mathcal{B}
        \right) \,.
\end{equation}
We will provide a universal formula for the Berry curvature $\mathcal{B}$.

In view of Eq. (\ref{centralclaim}), our strategy to compute the Berry curvature will be by using the APS index theorem which relates the $\eta$-invariant to other quantities.
In the more familiar case where there are no modulated mass parameters, it is well-known that the $\eta$-invariant is partly given by the integral of the Chern-Simons form, which is an immediate consequence of the APS index theorem.
To take into account the modulated mass parameters, one needs the version of the APS index theorem which treats Dirac operators coupled to the \emph{superconnection}, which combines the ordinary connections (background gauge fields) and the mass parameters.
We will review its definition in the following subsection.
The index theorem for Dirac operators coupled to the superconnection is discussed in Appendix \ref{appind}.

\subsection{Superconnection} \label{Sec:superconnection}

The spacetime-dependent modulated mass parameters and the background gauge fields for the global symmetries can be combined into what is known as the superconnection \cite{quillen1985superconnections,kahle2011superconnections} (see also \cite{berline1992heat}).
In physics, superconnection appeared in the context of string theory \cite{Kennedy:1999nn,Kraus:2000nj,Takayanagi:2000rz,Alishahiha:2000du}, and recently in the study of massive fermions with spacetime-dependent mass terms \cite{Cordova:2019jnf,Kanno:2021bze,Gomi:2021bhy}.
Our definitions and conventions directly follow the ones given in \cite{Kanno:2021bze}, and we repeat them here for completeness.

\paragraph{Superconnection of even type}

When $d$ is even and there are $N_f$ Dirac fermions, the most general mass matrix $M$ is valued in complex $N_f \times N_f$ matrices.
In addition, one can also couple the system to the $U(N_f) \times U(N_f)$ background gauge fields which we denote as $A_+$ and $A_-$.\footnote{If we want to consider only a subgroup of $U(N_f)\times U(N_f)$, then some of the components of $A_\pm$ need to be appropriately set to zero.}
Locally, they are one-forms taking values in hermitian $N_f \times N_f$ matrices.
The background gauge invariance can be maintained by allowing the mass term to transform in the bifundamental representation of $U(N_f) \times U(N_f)$. 

This set of background fields can be packaged into a \emph{superconnection} $\mathcal{A}$, which is defined as
\begin{equation} \label{Eq:superconnection_even}
    \mathcal{A} = \begin{pmatrix*}
        -\mathrm{i} A_+ & \mathrm{i} \widetilde{M}^\dagger \\
        \mathrm{i} \widetilde{M} & -\mathrm{i} A_-
    \end{pmatrix*} \,.
\end{equation}
Here, $\widetilde{M} \equiv M/\Lambda$ is a dimensionless mass matrix where $\Lambda$ is a cutoff scale, which is to be taken to infinity at the end of calculations \cite{Kanno:2021bze} (see also the discussion around Eq. (\ref{heatkernelanompoly})). 
We call $\widetilde{M}$ the renormalized mass matrix.
The off-diagonal blocks in (\ref{Eq:superconnection_even}) anticommute with odd degree differential forms.
The field strength for the superconnection is given by
\begin{equation} \label{Eq:field_strength_even}
    \mathcal{F} = \mathrm{d}\mathcal{A} + \mathcal{A}^2
    = \begin{pmatrix*}
        -\mathrm{i} F_+ - \widetilde{M}^\dagger \widetilde{M} & \mathrm{i} D\widetilde{M}^\dagger \\
        \mathrm{i} D \widetilde{M} & -\mathrm{i} F_- - \widetilde{M} \widetilde{M}^\dagger
    \end{pmatrix*}
\end{equation}
where $F_\pm = \mathrm{d}A_\pm - \mathrm{i} A^2_\pm$ are $U(N_f)$ field strengths, and $D\widetilde{M} = \mathrm{d}\widetilde{M} - \mathrm{i} A_- \widetilde{M} + \mathrm{i} \widetilde{M} A_+$ and $D\widetilde{M}^\dagger = \mathrm{d}\widetilde{M}^\dagger -\mathrm{i} A_+ \widetilde{M}^\dagger +\mathrm{i} \widetilde{M}^\dagger A_-$ are the covariant derivatives acting on the renormalized mass matrix which is in the bifundamental representation.

Similar to the ordinary connections, we can define the Chern character for the superconnection:
\begin{equation} \label{Eq:super_chern}
    \ch (\mathcal{F}) = \sum_{k=0}^\infty (\frac{\mathrm{i}}{2\pi})^{k/2} [\text{Str}(e^\mathcal{F})]_k \,.   
\end{equation}
Here, $[\cdots]_k$ means that we take the degree $k$ part of the differential form.
The supertrace $\text{Str}$ is defined by
\begin{equation} \label{Eq:supertrace_even}
    \text{Str} \begin{pmatrix*}
        a & b \\
        c & d
    \end{pmatrix*}
    = \text{Tr}(a) - \text{Tr}(d) \,.
\end{equation}
If the mass parameters are much smaller than the cutoff scale $\Lambda$, then we simply have $\ch (\mathcal{F}) = \ch(F_+) - \ch(F_-)$ where $\ch(F_\pm)$ is the usual Chern character for the $U(N_f)$ gauge field.
Importantly, we will see that to relate the Chern character of the superconnection to the Berry curvature, we need to consider the mass parameters that are much bigger than the cutoff scale in which case the mass parameters contribute nontrivially to the Berry curvature.

\paragraph{Superconnection of odd type}

When $d$ is odd and we have $N_f$ Dirac fermions, the mass matrix $M$ takes values in $N_f \times N_f$ hermitian matrices.
We can further turn on the background gauge field $A$ for the $U(N_f)$ global symmetry, if we allow the mass matrix to transform in the adjoint representation.\footnote{Again, if we want to consider only a subgroup of $U(N_f)$, then appropriate components of $A$ should be set to zero.}

We can combine the gauge field $A$ and the mass matrix $M$ into the following superconnection:
\begin{equation} \label{Eq:superconnection_odd}
    \mathcal{A} = \begin{pmatrix*}
        -\mathrm{i}A & \mathrm{i}\widetilde{M} \\
        \mathrm{i}\widetilde{M} & -\mathrm{i}A
    \end{pmatrix*} \,.
\end{equation}
Again, the off-diagonal blocks anticommute with odd degree differential forms.
The field strength is similarly defined as
\begin{equation}
    \mathcal{F} = \mathrm{d}\mathcal{A} + \mathcal{A}^2
    = \begin{pmatrix*}
        -\mathrm{i}F - \widetilde{M}^2 & \mathrm{i}D\widetilde{M} \\
        \mathrm{i}D\widetilde{M} & -\mathrm{i}F - \widetilde{M}^2
    \end{pmatrix*}
\end{equation}
where $F = \mathrm{d}A -\mathrm{i} A^2$ and $D \widetilde{M} = \mathrm{d} \widetilde{M} - \mathrm{i} [A,\widetilde{M}]$.

The Chern character of the superconnection is defined by the same equation \eqref{Eq:super_chern} as in the even case, with the supertrace now given as
\begin{equation} \label{Eq:supertrace_odd}
    \text{Str} \begin{pmatrix*}
        a & b \\
        c & d
    \end{pmatrix*}
    =  \sqrt{2} \mathrm{i}^{-3/2}  \text{Tr}(b) \,.
\end{equation}

\subsection{Berry curvature from Chern character of superconnection}

Having defined the superconnection and the corresponding Chern character, our goal is now to compute the Berry curvature $\mathcal{B}$ in Eq. (\ref{Eq:berry_curvature}) for the $d$-dimensional fermions living on the spacetime manifold $X$.
For this, we will use the results from Section \ref{Sec2}, in particular Eqs. \eqref{Eq:eta_cylinder} and \eqref{centralclaim}, and also the APS index theorem explained in Appendix \ref{appind}:
\begin{equation} \label{eq:index}
    \eta_K(\partial W) = -\Ind Q + \lim_{\Lambda\to\infty}\int_W \ch (\mathcal{F}(\widetilde{M}=M/\Lambda)) \hat{A}(TW)  \,.
\end{equation}
On the left-hand side, we have the $(d+1)$-dimensional $\eta$-invariant of the hermitian operator $K$ that we discussed in Section \ref{Sec2} and it depends on the background gauge fields as well as the modulated mass parameters.
Here, $W$ is a possibly non-compact $(d+2)$-dimensional manifold, and the operator $K$ is defined on its boundary $\partial W$ (which does not have a boundary itself, $\partial \partial W = \emptyset$, and so we do not have to worry about the boundary condition for $K$).
On the right-hand side, $Q$ is a hermitian operator on $W$ which is defined in Eq. (\ref{Eq:operator_Q}).
The precise form of this operator $Q$ will not be important for the computation of the Berry curvature, and the details are given in Appendix \ref{appind}, including the definition of the index $\Ind Q$.
The only important part is that $Q$ needs to have a discrete spectrum so that the integer-valued index can be defined; $\Ind Q \in \mathbb{Z}$.
For this to hold, the absolute value of mass eigenvalues should become infinitely large in the asymptotic infinity of $W$ when $W$ is non-compact.

Next, $\ch (\mathcal{F}(\widetilde{M}=M/\Lambda))$ is the Chern character for the superconnection that we defined in the last subsection, and here we emphasized that the field strength of the superconnection depends on the mass parameters only through the combination $\widetilde{M} = M/\Lambda$.
$\hat{A}(TW)$ is the $\hat{A}$-genus for the tangent bundle of $W$ which will be responsible for the gravitational part of the Berry curvature in the end.
The product $\ch(\mathcal{F})\hat{A}(TW)$ is a formal polynomial of differential forms, and in the above equation we should take the degree-$(d+2)$ part of $\ch(\mathcal{F})\hat{A}(TW)$ and then integrate over $W$.
The limit $\Lambda\to\infty$ is supposed to be taken after the integral is evaluated \cite{Kanno:2021bze}.

\begin{figure}[t]
    \centering
    \includegraphics[width=0.9\textwidth]{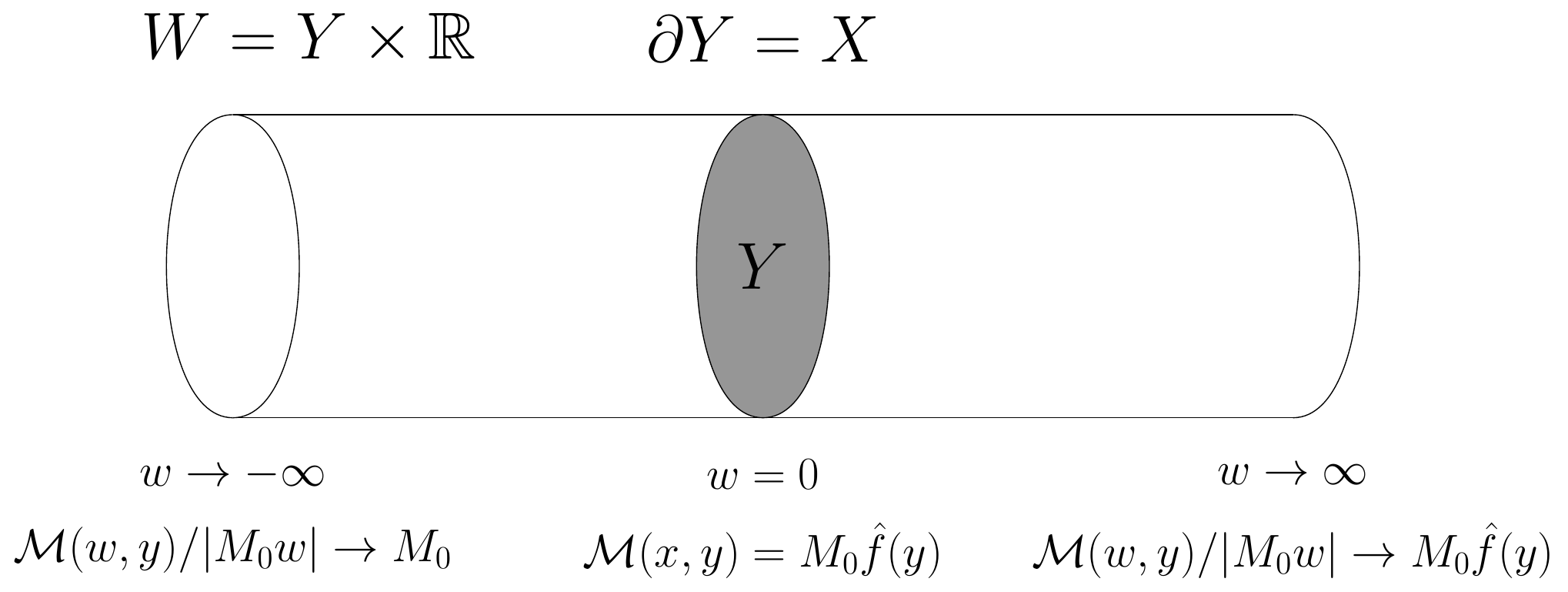}
    \caption{We define a $(d+2)$-dimensional non-compact manifold $W$ as $W = Y \times \mathbb{R}$. On $W$, the $d$-dimensional mass matrix $M(x) = M_0 f(x)$ on $X$ is extended to $\mathcal{M}(w,y)$, which has the specifically chosen asymptotic behavior as $w \to \pm\infty$.}
    \label{Fig:W}
\end{figure}

To relate this to Eq. \eqref{centralclaim}, we consider the following concrete setup.
Recall first that we assume in this section that there exists a $(d+1)$-dimensional manifold $Y$ with $\partial Y = X$ where the $d$-dimensional mass matrix is extended into $Y$ while being non-degenarate everywhere.
Take such a manifold $Y$, and we will let the $(d+2)$-dimensional manifold on which we will apply the index theorem \eqref{eq:index} to be $W = Y \times \mathbb{R}$, whose boundary is $\partial W = X \times \mathbb{R}$.\footnote{To be very precise, we want to have a collar neighborhood of $\partial W$ in $W$ to apply the derivation of the index theorem given in Appendix \ref{appind}. This can be achieved by appropriately deforming $W$ and the background field configurations, which does not affect the dicussions in this section.}
See Figure \ref{Fig:W}.

Let $w$ be the cooridinate along the $\mathbb{R}$ direction on $W$, and at the boundary of the $w=0$ slice our $d$-dimensional system of interest lives.
We will write the modulated $d$-dimensional mass matrix on $X$ as $M(x) = M_0 f(x)$ where $x$ denotes the coordinates on $X$, $M_0 >0$ is a constant mass, and $f(x)$ is a dimensionless non-degenerate matrix-valued function.
The function $f(x)$ is extended into $Y$ by a non-degenerate matrix-valued function $\hat{f}(y)$ where $y$ denotes the coordinates on $Y$.\footnote{The background gauge fields on $X$ also should be suitably extended into $Y$ as well as into $W$. We will let them extend into $Y$ in an arbitrary fashion and be independent of $w$.}
Now, we let this mass matrix to be extended into the whole $W$ (away from $w=0$) in the following way:\footnote{In Appendix \ref{appind} and other parts of the paper, we usually denote both the $d$-dimensional mass matrix and its extension to higher dimensions by the same symbol $M$, whereas here we use different symbols to be more explicit. We hope this does not cause any confusion.}
\begin{equation} \label{Eq:mass_W}
    \mathcal{M}(w,y) = \begin{cases}
    M_0^2 w
    \hat{f}(y) & \quad \text{for $w \geq w^*$} \,, \\
    M_0 \hat{f}(y) & \quad \text{at $w=0$} \,, \\
    -M_0^2 w & \quad \text{for $w \leq -w^*$} \,,
    \end{cases}
\end{equation}
where $w^* >0$ is some fixed position.
The precise way the mass parameters interpolate in the intermediate region $w \in (-w^*,w^*)$ as well as the actual value of $w^*$ will not be important as we will see.
We assume that $\mathcal{M}$ is finite for finite $|w|$, and otherwise only the asymptotic behavior of $\mathcal{M}(w,y)$ as $w\to\pm\infty$ will be relevant.
As mentioned above, to apply the index theorem, all of the absolute values of eigenvalues of $\mathcal{M}(w,y)$ should become infinite as $w\to \pm\infty$, which holds because $\hat{f}(y)$ is non-degenerate.
See Figure \ref{Fig:W} for the illustration.

We are then ready to apply the index theorem \eqref{eq:index} to this configuration.
First, we look at the integral on the right-hand side of Eq. \eqref{eq:index}.
The integral over $W$ factorizes as $\int_W = \int_Y \int_{w=-\infty}^{w=+\infty}$, and we will examine the integral over $w$.
The $\hat{A}$-genus does not depend on $w$ in this geometry: $\hat{A}(TW) = \hat{A}(T(Y\times \mathbb{R})) = \hat{A}(TY)$.
We can thus pull out the $\hat{A}$-genus out of the $w$ integral. 
We will devide the $w$ integral into three parts,
\begin{align}
\begin{split}
    & \int_{w=-\infty}^{w=\infty} \ch (\mathcal{F} (\widetilde{\mathcal{M}} = \mathcal{M}/\Lambda)) \\
    = & \int_{w=w^*}^{w=\infty} \ch (\mathcal{F} (\widetilde{\mathcal{M}}= \mathcal{M}/\Lambda)) 
    + \int_{w=-\infty}^{w=-w^*} \ch (\mathcal{F} (\widetilde{\mathcal{M}}= \mathcal{M}/\Lambda)) 
    + \int_{w=-w^*}^{w=w^*} \ch (\mathcal{F} (\widetilde{\mathcal{M}}= \mathcal{M}/\Lambda)) \,.
\end{split}
\end{align}
Consider a change of variable: $\widetilde{M}_0 (w) = M_0^2 |w| / \Lambda$.
$\widetilde{M}_0$ is dimensionless, and it determines the overall scale of the $(d+2)$-dimensional renormalized mass matrix $\widetilde{\mathcal{M}}(w,y)$.
Note that $\widetilde{\mathcal{M}}(w,y)$ depends on $w$ only through the dimensionless combination $\widetilde{M}_0$, at least outside of the intermediate region $w \in (-w^*,w^*)$.
There, the change of variable gives us
\begin{align}
\begin{split}
    \int_{w=w^*}^{w=\infty} \ch (\mathcal{F} (\widetilde{\mathcal{M}}= \mathcal{M}/\Lambda)) &= \int_{\widetilde{M}_0 = M_0^2 w^*/\Lambda}^{\widetilde{M}_0=\infty} \ch (\mathcal{F} (\widetilde{M})) \,, \\
    \int_{w=-\infty}^{w=-w^*} \ch (\mathcal{F} (\widetilde{\mathcal{M}}= \mathcal{M}/\Lambda)) &= - \int_{\widetilde{M}_0 = M_0^2 w^*/\Lambda}^{\widetilde{M}_0=\infty}
    \ch (\mathcal{F} (\widetilde{M}_0)) \,,
\end{split}
\end{align}
where $\widetilde{M} = \widetilde{M}_0 \hat{f}$.
On the right-hand side, now the $\Lambda$ dependence is only in the lower limit of the integrals, and it is easy to take the $\Lambda\to\infty$ limit.\footnote{Although we put a tilde on $\widetilde{M} = \widetilde{M}_0 \hat{f}$, there is no $\Lambda$ dependence here anymore since $\widetilde{M}_0$ is simply a dummy variable to be integrated over.}
What we are effectively doing here is to simply rewrite the integral on the manifold $W$ along the $w$ direction on the left-hand side as the integral inside the parameter space on the right-hand side.
Moreover, we can deduce that the integral over the intermediate region $w \in (-w^*,w^*)$ should vanish in the limit $\Lambda\to\infty$, since this integral will correspond to an integral in the parameter space from $\widetilde{\mathcal{M}} = 0$ to $\widetilde{\mathcal{M}}=0$ as $\mathcal{M}$ is finite for $w \in (-w^*,w^*)$.
We see here that the precise interpolating configuration of $\mathcal{M}$ in the intermediate region is irrelevant in the $\Lambda\to\infty$ limit as long as $\mathcal{M}$ remains finite as we claimed.

Combining, we obtain
\begin{equation} \label{Eq:remove_lambda}
    \lim_{\Lambda\to\infty} \int_{w=-\infty}^{w=\infty} \ch (\mathcal{F} (\widetilde{\mathcal{M}} = \mathcal{M}/\Lambda))
    = \int_{\widetilde{M}_0 = 0}^{\widetilde{M}_0=\infty} \left[\ch (\mathcal{F} (\widetilde{M}))
    - 
    \ch (\mathcal{F} (\widetilde{M}_0))\right] \,.
\end{equation}
From now on, we will simply write $\ch (\mathcal{F}) \equiv \ch (\mathcal{F}(\widetilde{M}))$ and $\ch (\mathcal{F}^{(0)}) \equiv \ch (\mathcal{F}(\widetilde{M}_0))$ for brevity.
Intuitively, $\ch (\mathcal{F}^{(0)})$ is the contribution from the regulator fermions living at $w \to -\infty$, for which we have chosen the mass parameters to be constant.
The integral from $\widetilde{M}_0 = 0$ to $\widetilde{M}_0 = \infty$ can be thought of as an integral in the parameter space along the ``radial'' direction, that is, we integrate over the overall scale of the mass matrix from zero to infinity.

If we now insert Eq. (\ref{Eq:remove_lambda}) into Eq. (\ref{eq:index}), we obtain
\begin{equation} \label{Eq:radial_integral_eta}
    \eta_K(X \times \mathbb{R}) = \int_Y \int_{\widetilde{M}_0 = 0}^{\widetilde{M}_0 = \infty} \left[ \ch (\mathcal{F}) - \ch (\mathcal{F}^{(0)})
    \right] \wedge \hat{A}(TY) \quad \text{mod $\mathbb{Z}$}
\end{equation}
where we have inserted $\partial W = X \times \mathbb{R}$ on the left-hand side.
The mod $\mathbb{Z}$ difference is from the index $\Ind Q$, and this difference can be ignored for our purpose since in the end we will exponentiate both sides.

\begin{figure}[t]
    \centering
    \includegraphics[width=0.65\textwidth]{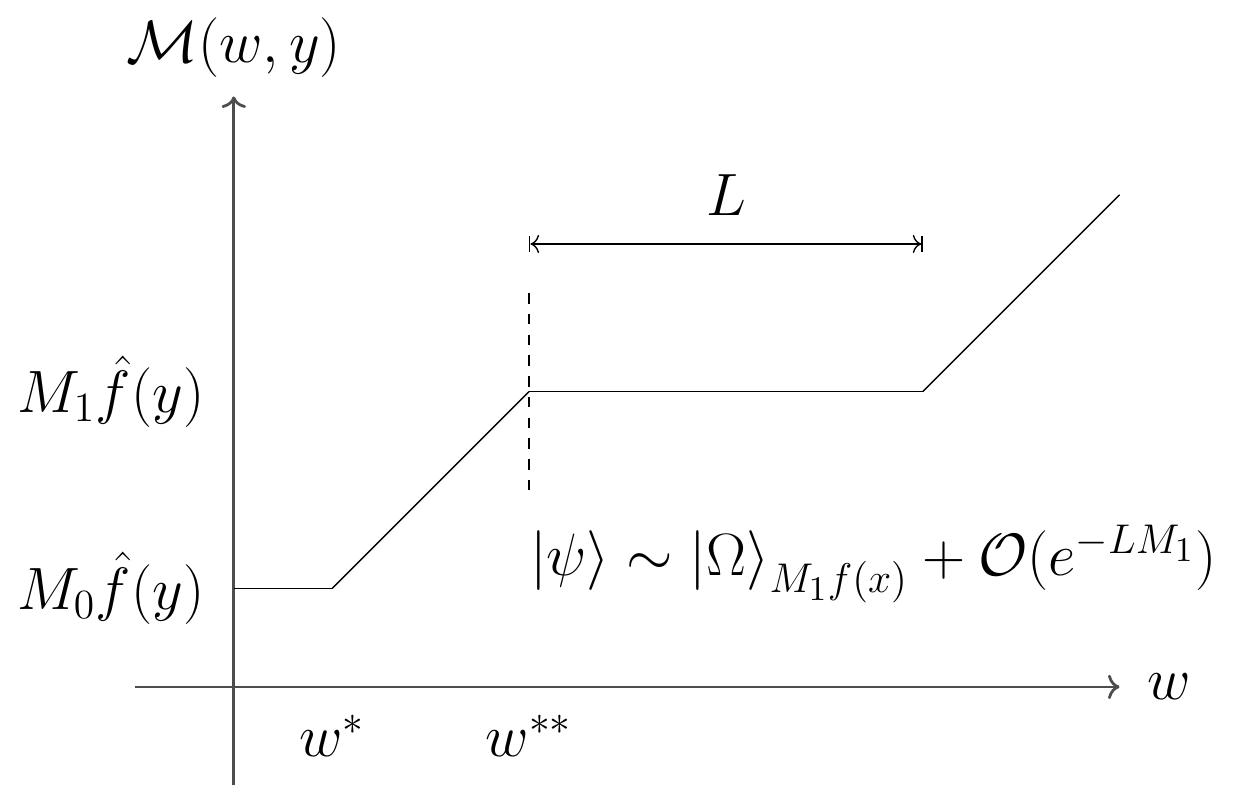}
    \caption{The modified profile of the $(d+2)$-dimensional mass matrix $\mathcal{M}(w,y)$. Only the $w>0$ region is shown. The asymptotic behavior at $w\to\infty$ is unchanged, but there is a plateau of length $L$ along which $\mathcal{M}(w,y) = M_1 \hat{f}(y)$ does not depend on $w$. Due to the plateau, the quantum state prepared by the half infinite cylinder given by $X \times [w^{**}+\epsilon,\infty)$ is exponentially close to the ground state $\ket{\Omega}_{M_1 f(x)}$ on $X$ with the mass configuration $M_1 f(x)$.}
    \label{Fig:mass_profile}
\end{figure}

Our remaining task is to relate the left-hand side of Eq. (\ref{Eq:radial_integral_eta}) to the partition function of the $d$-dimensional fermion system of interest, using the results from Section \ref{Sec2}.
To this end, first, using the fact that the precise form of the $(d+2)$-dimensional mass matrix $\mathcal{M}(w,y)$ for finite $w$ is not important, we will deform the configuration \eqref{Eq:mass_W} into the following new mass profile:
\begin{equation} \label{Eq:new_mass_W}
    \mathcal{M}(w,y) = \begin{cases}
        M_0^2 w \hat{f}(y) & \quad \text{for $w \to +\infty$} \,, \\
        M_1 \hat{f}(y) & \quad \text{for $w^{**} \leq w \leq w^{**} + L$} \,, \\
        M_1 & \quad \text{for $- w^{**} -L \leq w \leq -w^{**}$} \,, \\
        -M_0^2 w & \quad \text{for $w \to -\infty$} \,,
        \end{cases}
\end{equation}
where $w^{**} > w^{*}$ is some fixed position, $M_1 > M_0$ is a positive constant, and $L>0$ is an arbitrary length scale.
Compared to Eq. \eqref{Eq:mass_W}, we are not changing the asymptotic behavior of the mass profile at $w \to \pm \infty$, but making plateaus of length $L$ at $w^{**} \leq w \leq w^{**} + L$ and $- w^{**} -L \leq w \leq -w^{**}$ along which the mass matrix is independent of $w$.
See Figure \ref{Fig:mass_profile} for the illustration.
Since the asymptotic behaviors are unchanged, Eq. \eqref{Eq:radial_integral_eta} is still valid with the new mass profile.
Importantly, this is independent of the values of $M_1$, $w^{**}$, and $L$.

Now, consider cutting the infinitely long cylinder $\partial W = X \times \mathbb{R}$ into three pieces, $X \times (-\infty,-w^{**}-\epsilon]$, $X \times (-w^{**}-\epsilon,w^{**}+\epsilon)$, and $X \times [w^{**}+\epsilon,\infty)$, where $\epsilon>0$ is a small positive constant.
We claim
\begin{equation} \label{Eq:eta_cut}
    \eta_K(X \times \mathbb{R}) = \eta_K \left(X \times (-w^{**}-\epsilon,w^{**}+\epsilon);\mathtt{APS}\right) + \mathcal{O}(e^{-LM_1} ) \,.
\end{equation}
This follows from Eq. \eqref{Eq:eta_cylinder} from Section \ref{Sec2} (see also Figure \ref{Fig:APScylinder}), which says that for the purpose of computing an $\eta$-invariant on a manifold with boundary, the APS boundary condition can be replaced by an infinitely long half cylinder.
From the QFT point of view, this was equivalent to saying that the quantum state characterizing the APS boundary condition can be replaced by the ground state on the boundary.
To use this trick to compute $\eta_K \left(X \times (-w^{**}-\epsilon,w^{**}+\epsilon);\mathtt{APS}\right)$ on the right-hand side of \eqref{Eq:eta_cut}, note that the manifold $X \times (-w^{**}-\epsilon,w^{**}+\epsilon)$ has two disconnected boundary components (if $X$ is connected), and attaching half-infinite cylinders to the two boundary components yields the original manifold $\partial W = X \times \mathbb{R}$.

However, to replace the APS boundary condition, along the attached half-infinite cylinder, all the background field configurations including the mass parameters should remain fixed. 
With such a configuration, the path integral over the cylinder prepares the desired ground state at the original APS boundary.
Our deformed mass profile \eqref{Eq:new_mass_W} almost fulfills this purpose due to the existence of the plateaus of finite length $L$.
For instance, if we let $\ket{\psi}$ denote the state defined by the path integral over the region $w > w^{**}$, $\ket{\psi}$ is almost the ground state $\ket{\Omega}_{M_1f(x)}$ up to corrections which are suppressed by the factor of $e^{-LM_1}$, as excited states will have the energies of order $M_1$.
From this, the claimed equation \eqref{Eq:eta_cut} follows.

Lastly, Eq. \eqref{centralclaim} tells us
\begin{equation}
    \left.\frac{\mathcal{Z}_{d}(X,M_1 f(x))}{\mathcal{Z}^{(0)}_{d}(X,M_1)} \middle/ \left|\frac{\mathcal{Z}_d(X,M_1 f(x))}{\mathcal{Z}_d^{(0)}(X,M_1)}\right|\right.
    = \exp \left(2\pi \mathrm{i}
        \eta_K \left(X \times (-w^{**}-\epsilon,w^{**}+\epsilon);\mathtt{APS}\right)
    \right) \,.
\end{equation}
On the right-hand side, we do not need to include additional regulator contribution $\eta_{K^{(0)}}$ as in Eq. \eqref{centralclaim} since the configuration already includes the regulator fermions living on the boundary $w = -w^{**}-\epsilon$ where the mass matrix is fixed to be the constant $M_1$.
Now, if we take the limit $M_1 \to \infty$, then the right-hand side reduces to $\eta_K(X \times \mathbb{R})$ due to Eq. \eqref{Eq:eta_cut}.
Moreover, on the left-hand side, this limit by definition defines the (exponentiated) Berry curvature according to Eq. \eqref{Eq:berry_curvature}.
We thus have
\begin{equation}
    \int_Y \mathcal{B} = \eta_K(X \times \mathbb{R}) \quad \text{mod $\mathbb{Z}$}
\end{equation}
where again the mod $\mathbb{Z}$ difference is not important as only the exponentiated quantities are of interest.
Combined with Eq. \eqref{Eq:radial_integral_eta}, we can finally read off the form of the Berry curvature:
\begin{equation} \label{Eq:berry_curvature_formula}
    \mathcal{B} = \int_{\widetilde{M}_0 = 0}^{\widetilde{M}_0 = \infty} \left[ \ch (\mathcal{F}) - \ch (\mathcal{F}^{(0)})
    \right] \wedge \hat{A}(TY) \,.
\end{equation}
That is, we obtain the Berry curvature by integrating the Chern character (times the $\hat{A}$-genus) along the ``radial'' direction in the mass parameter space.
The resulting Berry curvature is a degree-$(d+1)$ differential form on $Y$ which bounds the $d$-dimensional spacetime $X$, that is, $\partial Y = X$.
This is the desired universal formula for the Berry curvature of Dirac fermions with modulated mass paramters.
The formula works in general spacetime dimensions and for arbitrary number of fermion species coupled to arbitrary background gauge fields and modulated mass parameters.

\subsection{Examples}

We now demonstrate how Eq. \eqref{Eq:berry_curvature_formula} is used in practice in a few examples.
A relatively simple situation is when the mass parameters are invariant under all the global symmetries for which background gauge fields are turned on.
In this case, the Chern character of the superconnection factorizes as
\begin{equation}
    \ch (\mathcal{F}) = \ch (F) \wedge \ch (\widetilde{M}) 
\end{equation}
where $\ch (F)$ is the ordinary Chern character for the background gauge field ($F$ is the field strength) and we wrote $\ch (\widetilde{M})$ to denote the Chern character of the superconnection which is obtained by setting $A=0$ in Eq. \eqref{Eq:super_chern}.
We will first discuss this type of examples, and will also give an example where the mass parameter breaks the global symmetry for which the background gauge field is turned on at the end.

\subsubsection{Single Dirac fermion in even dimensions} \label{sec:example1}

One of the simplest cases for which the mass term contributes to the Chern character is when we have a single Dirac fermion in even number of spacetime dimensions \cite{Cordova:2019jnf}.
Although these examples can be studied by the conventional application of the anomalous axial rotation of the fermions, here we explicitly see that we can reproduce them.
The general mass parameter is a complex number 
$M = M_0 e^{\mathrm{i}\alpha}$, and the Lagrangian is given by
\begin{equation}
    \mathcal{L} = \bar{\psi} \left(\slashed{D} + M_0 \exp (\mathrm{i}\alpha \gamma)\right)\psi \,,
\end{equation}
where $\gamma$ is the chirality matrix.
We will let the regulator fermions to have $\alpha=0$ and they will couple to the other background gauge fields in the same way as the physical fermions.

If we turn off the background gauge fields, then the Chern character is given by
\begin{equation}
    \ch (\widetilde{M}) = -\frac{\mathrm{i}}{2\pi}e^{-|\widetilde{M}|^2} \mathrm{d}\widetilde{M} \mathrm{d}\widetilde{M}^* \,.
\end{equation}
Integrating this along the radial direction in the mass parameter space gives us $- \mathrm{d}\alpha / 2\pi$.

From this we can read off the forms of the Berry curvature for various even dimensional Dirac fermions.
For instance, if we have a single flavor of a two-dimensional Dirac fermion with the modulated complex mass term as well as the background gauge field for the $U(1)$ vector symmetry turned on, the Berry curvature for the theory is given by  
\begin{equation}
    \mathcal{B} = -\frac{\mathrm{d}\alpha}{2\pi} \wedge \frac{F}{2\pi} \,.
\end{equation}
This implies that the effective action after we integrate out the fermion contains the topological term $-\frac{1}{2\pi}\int \alpha F$, reproducing the example we saw in Sec~\ref{sec:overview}.

As another example, we can consider four-dimensional Dirac fermions, coupled to a vector $U(N_f)$ background gauge field and with the $U(N_f)$ invariant diagonal mass term.
In four-dimensions, the background geometry also contributes to the Berry curvature through the $\hat{A}$-genus.
We can read off that the Berry curvature is given by
\begin{equation}
    \mathcal{B} = -\frac{\mathrm{d}\alpha}{2\pi} \wedge \Big( - \frac{N_f p_1}{24} + \frac{\text{tr}(F \wedge F)}{8\pi^2} \Big)
\end{equation}
where $p_1 = -\frac{1}{8\pi^2} \text{tr}(R \wedge R)$ is the first Pontryagin class of the spacetime tangent bundle and $F$ is the $U(N_f)$ background field strength.
The effective action then contains axion-coupling-like topological terms $\sim \int \alpha \,\text{tr}(R \wedge R)$ and $\sim \int \alpha \,\text{tr}(F \wedge F)$ if we integrate out the fermions.

\subsubsection{Odd dimensional examples}

In odd dimensions, the most general mass term for $N_f$ species of Dirac fermions is given by a $N_f \times N_f$ hermitian matrix.
Consider two flavors of Dirac fermions in any odd dimensions,
\begin{equation}
    \mathcal{L} = \bar{\psi}(\slashed{D} + M) \psi
\end{equation}
where $M$ is valued in $2 \times 2$ hermitian matrices.
We can write the mass matrix as $M = M_0 + M_1 \tau^1 + M_2 \tau^2 + M_3 \tau^3$ where $\tau^i$'s are Pauli matrices. 
The degenerate locus is given by the equation $M_0 ^2 = M_1^2 + M_2^2 + M_3^2$, which forms a two copies of three-dimensional cones meeting at the origin, inside the four-dimensional parameter space $\widetilde{P}= \mathbb{R}^4$ including degenerate locus.
This model was analyzed in \cite{Hsin:2020cgg}, and our results are consistent.

The Berry curvature is most concisely written if we use the following parameterization of the mass matrix.
We can locally diagonalize the mass matrix so that $M = U \Sigma U^\dagger$ where
\begin{equation}
    \Sigma = \begin{bmatrix*}
        \lambda_1 & 0 \\
        0 & \lambda_2
    \end{bmatrix*} \, , \quad
    U = \begin{bmatrix*}
        \text{cos}\frac{\theta}{2} & e^{\mathrm{i}\phi} \text{sin}\frac{\theta}{2} \\
        -e^{-\mathrm{i}\phi} \text{sin}\frac{\theta}{2} & \text{cos}\frac{\theta}{2}
    \end{bmatrix*} \, .
\end{equation}
$\lambda_1$ and $\lambda_2$ are real eigenvalues of $M$, and $\theta$ and $\phi$ parameterize a two-sphere $S^2$.
When $\lambda_1$ and $\lambda_2$ have opposite signs, this $S^2$ is non-contractible in the parameter space without crossing the degenerate locus, and it sources a nontrivial Berry curvature as we will see shortly.
When the two eigenvalues have the same sign, the $S^2$ is contractible while not hitting any degenerate locus, and it does not source any Berry curvature.
For the regulator fermions, we will set $\theta = \phi = 0$ and $\lambda_1 = \lambda_2$ to be a fixed positive constant.
They will couple to the other background gauge fields in the same way as the physical fermions.

For simplicity, consider first the case where there is no additional background gauge fields.
The integral of the Chern character along the radial direction in the parameter space in this can be obtained as
\begin{equation}
    \int_{\widetilde{M}_0=0}^{\widetilde{M}_0=\infty} \, \ch (\widetilde{M}) = \frac{1}{2}(\text{sgn}(\lambda_1) + \text{sgn}(\lambda_2)) - \frac{1}{2} (\text{sgn}(\lambda_1) - \text{sgn}(\lambda_2))\, \omega_2 \, ,
\end{equation}
where $\omega_2 = \frac{1}{4\pi} \text{sin} \theta \mathrm{d}\theta \mathrm{d}\phi$ is the volume form on $S^2$ normalized to give 1 when integrated over $S^2$.
As we anticipated, the 2-sphere $S^2$ parameterized by $\theta$ and $\phi$ leads to a nontrivial contribution proportional to the volume form on $S^2$, only when the sphere is not contractible without crossing the degenerate locus.

For a concrete example, consider a three-dimensional theory of $2N_f$ Dirac fermions. 
The mass term is given by $M \otimes I$ where $M$ is the above $2 \times 2$ hermitian matrix valued function and $I$ is the $N_f \times N_f$ identity matrix.
We couple the fermions to a background $U(N_f)$ gauge field where the mass matrix is invariant under this $U(N_f)$.
The Berry curvature 4-form of this system can then be read off,
\begin{equation}
    \mathcal{B} = \frac{1}{2}(\text{sgn}(\lambda_1) + \text{sgn}(\lambda_2)-2) \Big(- \frac{N_f p_1}{24} + \frac{\text{tr}(F \wedge F)}{8\pi^2} \Big) - \frac{1}{2} (\text{sgn}(\lambda_1) - \text{sgn}(\lambda_2)) \, \omega_2 \wedge \frac{\text{tr}(F)}{2\pi} \,.
\end{equation}
Here, the first term is nonzero even when the mass parameters are kept constant (and $-2$ is the contribution from the regulator fermions).
It reflects the familiar fact that the fermion effective action contains the (gravitational) Chern-Simons term in 3d where the sign of the level depends on the sign of the mass eigenvalues.
The second term is a gauged-WZW-like term, and its contribution is nontrivial only if the mass parameters are modulated in spacetime.

\subsubsection{Symmetry breaking mass term} \label{Sec:symm_breaking_berry}

One can also allow the mass parameters to be charged under some symmetry.
For instance, consider the case of a single flavor of a two-dimensional Dirac fermion $\psi = (\psi_L,\,\psi_R)^T$, with a complex mass parameter as before, but now turn on the background gauge field for the chiral $U(1)_L$ symmetry which acts only on the left-handed component.
The mass term explicitly breaks the $U(1)_L$ symmetry, but by promoting the mass to be a background field transforming under this symmetry, one can consistently turn on the combined background of the modulated mass term plus the gauge field.
The Lagrangian is
\begin{equation}
    \mathcal{L} = \psi_L^\dagger (D_1 + \mathrm{i}D_2) \psi_L + \psi_R^\dagger (\partial_1 - \mathrm{i}\partial_2) \psi_R +
    M_0 e^{\mathrm{i}\alpha} \psi_R^\dagger \psi_L + 
    M_0 e^{-\mathrm{i}\alpha} \psi_L^\dagger \psi_R 
\end{equation}
where $D_i = \partial_i -\mathrm{i} A_i$.
The nontrivial background gauge transformations are
\begin{equation}
    \psi_L \sim e^{\mathrm{i}\lambda} \psi_L \,, \quad
    A_i \sim A_i + \partial_i \lambda \,, \quad
    \alpha \sim \alpha - \lambda \,. 
\end{equation}
In this case, we cannot let the regulator fermions to have a constant mass term since it will not be gauge-invariant.
We will set the regulator mass to be $M_0 e^{\mathrm{i}\alpha_0}$ where $\alpha_0 \sim \alpha_0 - \lambda$.
The degree four part of the Chern character that we are after is given by
\begin{equation}
    [\ch (\mathcal{F})]_4  = \left(\frac{\mathrm{i}}{2\pi}\right)^2 e^{-\widetilde{M}_0^2}
    \left(-\frac{1}{2}F^2 + \frac{1}{3}F \wedge \widetilde{M}_0 \mathrm{d}\widetilde{M}_0 (\mathrm{d}\alpha + A)\right) \,,
\end{equation}
and after integrating from $\widetilde{M}_0 = 0$ to $\widetilde{M}_0 = \infty$, we get the following Berry curvature:
\begin{equation} \label{eq:2dchiral}
    \mathcal{B} = -\frac{1}{8\pi^2} F \wedge (d\alpha - d\alpha_0) \,.
\end{equation}
Note that the Berry curvature is invariant under the background gauge transformation.\footnote{Similarly to Section \ref{sec:example1}, the Eq. \eqref{eq:2dchiral} can also be deduced from the anomalous $U(1)_L$ rotation of $\psi_L$.}

\section{Torsional Berry phase} \label{Sec4}

In the last section we studied the (higher) Berry curvature which captures the response to the continuous deformation of parameters.
When the Berry curvature vanishes, there can still be a nontrivial response against the modulated parameters $p: X \to P$ that is not continuously deformable to (i.e., not homotopic to) a constant map.
Or, more precisely, the response can be nontrivial when the pair $(X,p)$ is not null-bordant.
Such Berry phases are classified by the torsion part of $\Inv^d(P)$, which is further identified with the Pontryagin dual of the bordism group \cite{Kapustin:2014tfa,Kapustin:2014dxa,Freed:2016rqq,Yonekura:2018ufj}:
\begin{equation}
    \Tors\Inv^d(P) = \Hom(\Tors\Omega_d^\text{spin}(P),U(1))\,.
\end{equation}
In other words, given a $d$-dimensional manifold $X$ and the modulated parameters $p:X \to P$, the Berry phase defines a bordism invariant of the pair $(X,p)$ when its Berry curvature vanishes.
We will call such a bordism invariant a torsional Berry phase, and we will calculate it in a few examples utilizing the result of Section~\ref{Sec2}.

\subsection{Reducing onto zero locus} 

To evaluate the partition function $\mathcal{Z}_d(X,M)$ when $\mathcal{B}=0$ but the pair $(X,M)$ is not null-bordant, we again use \eqref{centralclaim}.
In such cases, to extend the mass matrix $M$ into $Y_+$, we have to let the extended mass matrix to become degenerate on a locus in $Y_+$, which we will call $Y_D$.
By definition, $Y_D \cap X = \varnothing$ and in particular $Y_D$ does not have a boundary.
Then, using the assumption that $\mathcal{B}=0$, we can reduce the $\eta$-invariant on $Y_+$ to the $\eta$-invariant just for the localized modes along the degenerate mass locus $Y_D$.

Specifically, this is accomplished as follows.
If $\mathcal{B} = 0$, the value of the $\eta$-invariant on $Y_+$ is invariant under any continuous deformation of the mass parameters.
Using this freedom, we will tune the magnitude of the modulated mass term to be much larger than the regularization scale $\mu$ for the $\eta$-invariant.
That is, consider the regularized expression of the $\eta$-invariant on $Y_+$ that was given in Eq. (\ref{eq:eta_invariant_reg}),
\begin{equation} \label{Eq:eta_regularized}
    \eta_K(Y_+;\mathtt{APS}) = \lim_{\mu \to \infty}
    \left[
    \frac{1}{2}
    \sum_{j} \text{sgn}(\xi_j) \text{exp} \left(-\frac{\xi_j^2}{\mu^2} \right) 
    \right]\,.
\end{equation}
We will let $M_0/\mu \to \infty$ where $M(x) = M_0 f(x)$ as before.
When the mass becomes large, there appear exponentially localized eigenfunctions of the operator $K$ on the degenerate locus $Y_D$, and their eigenvalues are independent of the value of $M_0$.
On the other hand, all the other delocalized eigenfunctions of $K$ have eigenvalues that are proportional to the overall mass scale $|\xi|\sim M_0$.
This means that in the limit $M_0/\mu \to \infty$, the $\eta$-invariant (\ref{Eq:eta_regularized}) receives contribution only from the localized modes because of the regulator in \eqref{Eq:eta_regularized}.
If we denote the hermitian operator that acts on the localized modes on $Y_D$ as $K_D$, then according to these observations we conclude that
\begin{equation}
    \eta_K(Y_+;\mathtt{APS}) = \eta_{K_D}(Y_D)
\end{equation}
when the Berry curvature vanishes and the diabolical locus $Y_D$ is a manifold embedded in $Y_+$.\footnote{
    In general $Y_D$ might have self-intersections or other kinds of singularities, and it is not guaranteed that one can always smooth out the singularities by deforming the mass parameter configuration.
}
We list some examples of this below.

\subsection{Examples}

\subsubsection{Two-dimensional Dirac fermion} \label{Sec4.2}

Consider a single flavor of a two-dimensional Dirac fermion,
\begin{equation}
    \mathcal{L} = \bar{\psi}(\slashed{\partial} + M_0 e^{\mathrm{i}\alpha \gamma}) \psi \,.
\end{equation}
We do not turn on any additional background gauge fields for internal global symmetries (other than the spin structure).
The space of non-degenerate mass terms in this case is homotopic to a circle $S^1$, and the appropriate bordism group is
\begin{equation}
    \widetilde{\Omega}^{\text{Spin}}_{d=2} (S^1) \cong \mathbb{Z}_2 \,.
\end{equation}
As a representative of the generator, we can pick a square torus $T^2 = S^1_a \times S^1_b$.
Along the $a$-cycle the mass term winds around once, i.e., $\oint_a d\alpha = 2\pi$, and we impose the antiperiodic (NS) boundary condition for the fermion.
Along the $b$-cycle, the mass term remains constant, and we impose the periodic (R) boundary condition for the fermion.

The regularized partition function of the Dirac fermion in the infinite mass limit in this case will determine an element in the Pontryagin dual of the reduced bordism group which is again $\mathbb{Z}_2$.
If the theory corresponds to the trivial (nontrivial) element, then the partition function on the $T^2$ that we constructed will evaluate to 1 ($-1$).

Now, for the auxiliary bulk $Y_+$, we can take $D^2_a \times S^1_b$ where $D^2_a$ is a 2-disk bounding the $a$-cycle, and the mass term vanishes at the origin of the disk.
The zero locus of the mass term will then be $Y_D = \{0\} \times S^1_b$ where $0$ denotes the origin of the disk.
According to Eq. (\ref{centralclaim}) and Appendix \ref{appodddim}, the 3d hermitian operator acting on two flavors of Dirac fermions on $Y_+ = D^2_a \times S^1_b$ is
\begin{equation}
    K = \mathrm{i}(\tau^3 \Gamma^\mu \partial_\mu + \mathrm{i}\tau^2 M_0 \text{cos}\alpha - \mathrm{i}\tau^1 M_0 \text{sin}\alpha) \,.
\end{equation}
On the other hand, for the localized zero modes on the codimension two circle $Y_D = \{0\} \times S^1_b$, we simply have
\begin{equation}
    K_D = \mathrm{i}\partial \,,
\end{equation}
which is deduced by a simple counting of the number of components in the Dirac spinors.\footnote{The number of components of the codimension-$p$ localized fermion is $2^{-p}$ times the number of components for the bulk fermion.}
That is, we know that starting from two flavors of Dirac fermions in 3d, we get a single complex fermion localized on the circle.

It is easy to compute that the exponentiated $\eta$-invariant coming from this one-dimensional fermion evaluated on the circle with the periodic spin structure gives us -1.
All the nonzero eigenvalues appear in pairs of positive and negative numbers with the same magnitude, and there is a singe zero mode given by the constant mode which results in $\eta = 1/2$.
Thus, we conclude that the regularized partition function for the original 2d system is
\begin{equation} \label{eq:torus_partition_function}
    \lim_{M_0 \to \infty} \frac{\mathcal{Z}_2(T^2)}{\mathcal{Z}^{(0)}_2(T^2)} = -1 \,,
\end{equation}
indicating that this theory defines a nontrivial parameterized invertible field theory.

We demonstrate that the same result can be obtained by an explicit computation of the 2d partition functions in Appendix \ref{apptorus}.

\subsubsection{Effective action from symmetry breaking mass}
As a bit more substantial example,
we consider the cases where fermions with a symmetry breaking mass generates a nontrivial effective action in terms of the backgrounds for the remaining symmetries.
The specific model is what is considered in \cite{Thorngren:2014pza,Wang:2018qoy,Lee:2021crt}.

In the model, we have Weyl fermions in the $(\text{vector, fundamental})$ representation of $SO(2n_1)\times USp(2n_2)$ symmetry group, and consider the mass terms breaking the $USp(2n_2)$ group into $U(n_2)$, so that the fundamental representation of the former splits into the direct sum of fundamental and anti-fundamental representation of the latter.
Becuase this model has Weyl fermions, we cannot apply the result in this paper regarding Dirac fermions.
However, here we would like to pretend that Eq. \eqref{centralclaim} can be extended to more general fermion cases, perhaps by utilizing the result of \cite{Gomi:2021bhy}. 
In addition, the result here is already reported in the references mentioned above.
Their argument is from anomaly matching with the zero modes along a 't Hooft-Polyakov monopole configuration, while our method clarifies that how one would deduce such effective action directly from the fermion partition function.

The effective action we are aiming for is 
\begin{equation}\label{effS}
    \frac12 w_2(SO(2n_1)) c_1(U(n_2)),
\end{equation}
where $w_2(SO(2n_1))$ and $c_1(U(n_2))$ are the second Stiefel-Whiteney class and the first Chern class (modulo 2) for the specified symmetry bundles, respectively. 
To detect this action, we consider a $d$-dimensional manifold $X$ with constant mass $M$ breaking $USp(2n_2)$ to $U(n_2)$, and a $(SO(2n_1)\times U(n_2))/\mathbb{Z}_2$ bundle $A$ with nonzero $w_2(SO(2n_1)) c_1(U(n_2))$.
To apply an analog of \eqref{centralclaim} that should hold for Weyl fermions, we find an extension of $M$ and $A$ into $Y_+$.
As the characteristic class $w_2(SO(2n_1)) c_1(U(n_2))$ is in particular a bordism invariant, this is impossible with a constant mass $M$ over $Y_+$. 

To find an extension, we allow $M$ to vary on $Y_+$ (but it remains to be constant on $X = \partial Y_+$) and hit $0$ on a locus $Y_D$.
To be explicit, we take $X = S^2_{(1)}\times S^2_{(2)}$ (subscripts are to distinguish the two factors) and the bundle to be such that $w_2(SO(2n_1))$ is dual to the fundamental class of $S^2_{(1)}$ and $c_1(U(n_2)))$ is dual to the fundamental class of $S^2_{(2)}$.
Then, we take $Y_+ = S^2_{(1)} \times D^3_{(2)}$ where $D^3_{(2)}$ is the ball bounding $S^2_{(2)}$.
To extend the bundle, we prepare the 't Hooft-Polyakov configuration in $D^3_{(2)}$.
That is, the mass matrix $M$ becomes zero at the center of $D^3_{(2)}$, and the $U(n_2))$ bundle is enhanced to a $USp(2n_2)$ bundle at the center.\footnote{
Note that in this application, even though $M$ is constant on $X$, it is important to let it vary inside $Y_+$.
The method in the literature considered modulated $M$ on $X$, which we do not need.}

Then, the evaluation of the effective action reduces to evaluating the $\eta$-invariant of the zero modes localized at the monopole.
This zero mode counting is done in \cite{Thorngren:2014pza,Wang:2018qoy,Lee:2021crt}, and it is shown that the state arising from quantizing the zero modes is in the spinor representation of $SO(2n_1)$, which means that the $\eta$-invariant of them is $\frac12 w_2(SO(2n_1))$.
Combining all considerations, we conclude that indeed the fermion with the symmetry breaking mass term generates the effective action \eqref{effS}.

\section*{Acknowledgments}
The authors are grateful to Y.-A.~Chen, S.-J.~Huang, J.~Kaidi, Z.~Komargodski, S.~Seifnashri, S.-H.~Shao, M.~Yamashita, and G.~Zafrir for helpful conversations. KO is supported in part by JSPS KAKENHI Grant-in-Aid, No.22K13969 and in part by Simons Collaboration on Global Categorical Symmetry.

\appendix

\section{When \texorpdfstring{$d$}{d} is even} \label{appodddim}

Here, we repeat the calculations in Section \ref{Sec2} for the case of even $d$. 
The goal is again to relate the phase of the partition function of an arbitrary number $N_f$ species of Dirac fermions on a closed $d$-dimensional manifold $X$ with modulated mass parameters to $(d+1)$-dimensional $\eta$-invariants.
Since $d$ is even, we have the $d$-dimensional chirality matrix $\gamma$, and the most general $d$-dimensional mass matrix is given by $M_1 + \mathrm{i} \gamma M_2$ where $M_1$ and $M_2$ are $N_f \times N_f$ hermitian matrices.
This system can be realized on an interface, if we start from $2N_f$ species of $(d+1)$-dimensional Dirac fermions on the manifold $Y$ which we defined in Section \ref{Sec2} (depicted in Figure \ref{Fig:Y+Y-}), with the follwing Dirac operator,
\begin{equation}
    \sD_{d+1} = \Gamma^\mu D_\mu + \tau^1 M_1 + \tau^2 M_2 + \tau^3 M_3 \,,
\end{equation}
and form an interface by varying the value of $M_3$ from $+m_0$ to $-m_0$ similar to the odd $d$ case.
$\tau$'s are the $2 \times 2$ Pauli matrices. $\tau^k M_k$ is a shorthand for the tensor product of the two matrices, for $k=1,2,3$.

We define
\begin{equation}
    K = \mathrm{i}(\tau^3 \Gamma^\mu D_\mu + \mathrm{i}\tau^2 M_1 - \mathrm{i}\tau^1 M_2)
\end{equation}
which is a hermitian operator on $Y$.
When we restrict to $Y_+$, we need to choose an appropriate boundary conidition that makes $K$ hermitian.
Along the collar neighborhood of $X$ in $Y_+$, $K$ takes the form $K = \mathrm{i}\tau^3 \Gamma^y (\partial_y + \mathcal{D}_X)$ where
\begin{equation}
    \mathcal{D}_X = \Gamma^y \Gamma^i D_i + \Gamma^y (\tau^1 M_1 + \tau^2 M_2)
\end{equation}
is a hermitian operator defined on $X$.
The nonzero eigenvalues of $\mathcal{D}_X$ come in pairs because of $\{\mathcal{D}_X,\tau^3 \Gamma^y\} = 0$, and $\mathcal{D}_X$ does not have a zero mode since the $d$-dimensional mass matrix is non-degenerate.
The same APS boundary condition as in Eq. (\ref{APSbc}), $\mathcal{P}_{<0}^{\mathcal{D}_X} \Psi |_X = 0$, makes $K$ a hermitian operator on $Y_+$.

Now, the $(d+1)$-dimensional partition function on $Y$ can again be written as
\begin{align} \label{bulkpart2}
    \mathcal{Z}_{d+1}(Y) &= \bra{Y_-} \hat{Y}_L \ket{Y_+} \nonumber \\
    &= \bra{\Omega_-} T[{\text{exp}\{-\int_{y_+}^{y_-} \hat{H}(y) dy\}}] \ket{\Omega_+} \times \frac{\braket{\mathtt{APS} | Y_+}}{\braket{\mathtt{APS} | Y_-}} \times \frac{\braket{\mathtt{APS} | \Omega_-}}{\braket{\mathtt{APS} | \Omega_+}}
\end{align}
as in Eq. (\ref{bulkpart}).

Define $K^\prime = \mathrm{i}(\tau^3 \Gamma^\mu D_\mu - \mathrm{i}\tau^2 M_1 - \mathrm{i}\tau^1 M_2)$ and $\mathcal{D}_X^\prime = \Gamma^y \Gamma^i D_i - \Gamma^y (\tau^1 M_1 + \tau^2 M_2)$. $\mathcal{D}_X^\prime$ is a hermitian operator on $X$, and under the boundary condition $\mathcal{P}_{<0}^{\mathcal{D}_X^\prime} \Psi |_X = 0$, $K^\prime$ is a hermitian operator on $Y_+$.

$\mathcal{D}_X$ and $\mathcal{D}_X^\prime$ share the same spectrum of eigenvalues, because of the relation $\tau^3 \mathcal{D}_X = \mathcal{D}_X^\prime \tau^3$.
We denote their eigenfunctions and eigenvalues as $\mathcal{D}_X \psi_a = \lambda_a \psi_a$ and $\mathcal{D}_X^\prime \psi_a^\prime = \lambda_a \psi_a^\prime$, where $\psi_a^\prime = \tau^3 \psi_a$. We take the indexing convention such that $a$ runs over nonzero integers, and $\lambda_{-a} = -\lambda_{a} < 0$ for $a>0$.
We also let $\psi_{-a} = \tau^3 \Gamma^y \psi_a$. 
(Note the different notations compared to Section \ref{Sec2}.)

Along the collar neighborhood of $X$, we perform the mode expansions
\begin{equation}
    \Psi(x^i,y) = \sum_a \alpha_a(y) \psi_a (x^i) \,,\quad \bar{\Psi}(x^i,y) = \sum_a \beta_a(y) (\psi_a^\prime)^\dagger (x^i) = \sum_a \beta_a(y) \psi_a^\dagger (x^i) \tau^3 \,.
\end{equation}
The APS boundary condition imposed on $\Psi$, $\mathcal{P}_{<0}^{\mathcal{D}_X} \Psi |_X = 0$, is such that at the boundary we require all $\alpha_{a<0}$ to be zero, whereas all $\alpha_{a>0}$ are unconstrained.

If we insert the mode expansions into the action $S = \int_Y \bar{\Psi} \sD_{d+1} \Psi =  \int_Y \bar{\Psi} \tau^3 (-\mathrm{i}K + m_0) \Psi$, we obtain
\begin{equation} \label{modeexpansion2}
    S = \int dy \sum_a \left[ \beta_{-a}(y) (\partial_y + \lambda_a) \alpha_a (y) + M_3(y) \beta_a (y) \alpha_a (y) \right] \,.
\end{equation}
From this, we see that the conjugate variable of $\alpha_a$ is $\beta_{-a}$ for each $a$.
Thus, the boundary condition to be imposed on $\bar{\Psi}$ is such that at the boundary all $\beta_{a>0}$ are unconstrained, whereas all $\beta_{a<0}$ are set to be zero.

$K$ and $K^\prime$ also share the same spectrum of eigenvalues because of $\tau^3 K = K^\prime \tau^3$ which is coherent with their respective boundary conditions we explained above.
We denote the eigenfunctions and eigenvalues of $K$ and $K^\prime$ as $K \Psi_j = \xi_j \Psi_j$ and $K^\prime \Psi_j^\prime = \xi_j \Psi_j^\prime$, respectively, where $\Psi_j^\prime = \tau^3 \Psi_j$.
Inside the path integral, we will expand $\Psi = \sum\limits_j a_j \Psi_j$ with grassmannian coefficients $a_j$. 
Since $(\Psi_j^\prime)^\dagger$ satisfies the boundary condition for $\bar{\Psi}$ (but the naive choice $\Psi_j^\dagger$ does not), we will expand the conjugate field as $\bar{\Psi} = \sum\limits_j b_j (\Psi_j^\prime)^\dagger = \sum\limits_j b_j \Psi_j^\dagger \tau^3$ with grassmannian coefficients $b_j$.

This diagonalizes the action, $S = \int_{Y_\pm} \bar{\Psi} \tau^3 (-\mathrm{i}K \pm m_0) \Psi = \sum\limits_j (-\mathrm{i}\xi_j \pm m_0) b_j a_j$.
Thus, we again obtain
\begin{equation}
    \frac{\braket{\mathtt{APS} | Y_+}}{\braket{\mathtt{APS} | Y_-}} = \frac{\prod\limits_j (-\mathrm{i}\xi_j + m_0)}{\prod\limits_j (-\mathrm{i}\xi_j - m_0)} = \text{exp}(2\pi \mathrm{i} \eta_K (Y_+;\mathtt{APS}))
\end{equation}
in the limit $m_0 \to \infty$, as desired.

To compute the remaining factors in Eq. (\ref{bulkpart2}) as in Section {\ref{Sec2}}, we need to write down the Hamiltonian of the system on $X$.
From Eq. (\ref{modeexpansion2}), we read off the Hamiltonian
\begin{equation}
    \hat{H}(y) = \sum_a \left(\lambda_a \beta_{-a} \alpha_a + M_3(y) \beta_{a} \alpha_a \right)
\end{equation}
where $\beta_{-a} = \alpha_a^\dagger$.
Equivalently, we can write
\begin{equation}
    \hat{H}(y) = \sum_{a>0} [\alpha_a^\dagger \,\,\, \alpha_{-a}^\dagger] \begin{bmatrix*}
        \lambda_a & M_3(y) \\
        M_3(y) & -\lambda_a
    \end{bmatrix*}
    \begin{bmatrix*}
        \alpha_a \\
        \alpha_{-a}
    \end{bmatrix*} \,.
\end{equation}
Let
\begin{equation}
    (\text{cos}2\theta_a(y) \,, \text{sin}2\theta_a(y)) = \frac{(\lambda_a\,,M_3(y))}{\sqrt{\lambda_a^2 + M_3(y)^2}}
\end{equation}
where $-\pi/4 < \theta_a < \pi/4$ for each $a$. 
If we define, for each $a>0$,
\begin{equation}
    \begin{bmatrix*}
        A_{1,a} \\
        A_{2,a}
    \end{bmatrix*} =
    \begin{bmatrix*}
        \text{cos}\theta_a & \text{sin}\theta_a \\
        -\text{sin}\theta_a & \text{cos}\theta_a
    \end{bmatrix*}
    \begin{bmatrix*}
        \alpha_a \\
        \alpha_{-a}
    \end{bmatrix*} \,,
\end{equation}
then the modes $A_a$ diagonalize the Hamiltonian,
\begin{equation}
    \hat{H} = \sum_{a>0} \sqrt{\lambda_a^2 + m_0^2} \left(A_{1,a}^\dagger A_{1,a} - A_{2,a}^\dagger A_{2,a}\right) \,.
\end{equation}

The state corresponding to the APS boundary condition is characterized by
\begin{equation}
    \bra{\mathtt{APS}} \alpha_{a<0} = \bra{\mathtt{APS}} \alpha_{a>0}^\dagger = 0 \,.
\end{equation}
On the other hand, the ground state is characterized by
\begin{equation}
    A_{1,a}(y) \ket{\Omega(y)} = A_{2,a}^\dagger(y) \ket{\Omega(y)} = 0  
\end{equation}
for all $a>0$.

Let $\ket{E}$ be a state satisfying $\alpha_a \ket{E} = 0$ for all $a$.
Then, the states of interest can be constructed as
\begin{align}
    \bra{\mathtt{APS}} &= \bra{E} \prod_{a<0} \alpha_a \,, \nonumber \\
    \ket{\Omega(y)} &= \prod_{a>0} A_{2,a}^\dagger(y) \ket{E}\,.
\end{align}
This gives us $\braket{\mathtt{APS}|\Omega(y)} = \prod\limits_{a<0} \text{cos}\theta_a(y)$, and in particular 
\begin{equation}
\frac{\braket{\mathtt{APS}|\Omega_-}}{\braket{\mathtt{APS}|\Omega_+}} = 1 \,.
\end{equation}
Furthermore,
\begin{equation} 
\braket{\Omega_-|\Omega_+} = \prod_{a>0} \frac{\lambda_a}{\sqrt{\lambda_a^2 + m_0^2}} = \left[\prod_{a>0} \lambda_a\right]_{\text{Reg.}}\,.
\end{equation}

Recall on $X$, we have the hermitian operator $\mathcal{D}_X = \Gamma^y \Gamma^i D_i + \Gamma^y(\tau^1 M_1 + \tau^2 M_2)$ which satisfies $\{\mathcal{D}_X, \tau^3 \Gamma^y\} = 0$.
$\tau^3 \Gamma^y$ provides a $\bZ_2$ grading on the space of spinors on $X$ on which $\mathcal{D}_X$ acts.
Let $V_\pm$ be the subspace of spinors on $X$ having the eigenvalue $\tau^3 \Gamma^y = \pm 1$.
Consider the restriction of $\mathcal{D}_X$ to the subspace $V_+$, $\mathcal{D}_X |_{V_+} : V_+ \rightarrow V_-$.
$\mathcal{D}_X |_{V_+}$ can be written as a composition of two maps, $\mathcal{D}_X |_{V_+} = (\Gamma^y \tau^1)|_{V_+} \circ (\Gamma^i \tau^1 D_i + M_1 + i\tau^3 M_2)|_{V_+} \equiv (\Gamma^y \tau^1)|_{V_+} \circ \sD_d$, where $\sD_d : V_+ \rightarrow V_+$ and $(\Gamma^y \tau^1)|_{V_+} : V_+ \rightarrow V_-$.
Also, within the $V_+$ subspace, the matrices $(\Gamma^y \tau^1)|_{V_+}$ satisfy the algebra of $d$-dimensional gamma matrices, and all of them anticommute with $(\tau^3)|_{V_+}$.
Thus, it makes sense to define the $d$-dimensional gamma matrices $\gamma^i \equiv (\Gamma^y \tau^1)|_{V_+}$ and the chirality matrix $\gamma \equiv (\tau^3)|_{V_+}$.
Then, $\sD_d = \gamma^i D_i + M_1 + i\gamma M_2$ is the $d$-dimensional Dirac operator acting on the $N_f$ species of $d$-dimensional Dirac fermions living on $X$, which are precisely the elements of $V_+$.

As before, we define two hermitian operators on $X$, $\sH_d^1 \equiv \sD_d^\dagger \sD_d$ and $\sH_d^2 \equiv \sD_d \sD_d^\dagger$.
We then see $\mathcal{D}_X^2 |_{V_+} = \sH_d^1$, and $\lambda_a^2$ for $a>0$ are the eigenvalues of $\sH_d^1$.
We conclude $\prod\limits_{a>0} \lambda_a = |\mathcal{Z}_d (X)|$, and Eq. (\ref{phaseeta2}) holds for even $d$ as well.

Finally, the arguments in Section \ref{subsec2.2} are also still valid. 
The localized interface modes analogous to Eq. (\ref{localizedmodes}) are given by
\begin{equation}
    \Psi = \mathcal{N} (\psi_1 \otimes v^+_1 + \psi_2 \otimes v^+_2) \exp \left(-\int_{y^*}^{y} M_3(y^\prime) dy^\prime\right) \,,
\end{equation} 
where $\tau^3 \Gamma^y v^+_{1,2}  = + v^+_{1,2}$. $\psi_1$ and $\psi_2$ are chiral fermions in $d$ dimensions, and combined they form the $d$-dimensional Dirac fermion $\psi$.
If we apply $\sH_{d+1}^1 \equiv \sD_{d+1}^\dagger \sD_{d+1}$ to this ansatz, we obtain
\begin{equation}
    \sH_{d+1}^1 \Psi = \left(\sH_d^1 \psi\right) \exp \left(-\int_{y^*}^{y} M_3(y^\prime) dy^\prime\right)
\end{equation}
and then the remaining steps leading to Eq. (\ref{reducing}) are unchanged.
We conclude that Eq. (\ref{centralclaim}) is valid in any dimensions.

\section{APS index theorem and superconnection} \label{appind}

In this appendix, we give a derivation of the APS index theorem for a Dirac operator coupled to a superconnection (that is, modulated mass parameters and background gauge fields) on a manifold with boundary.
The derivation of the APS index theorem for a Dirac operator coupled to an ordinary connection using path integral was provided in \cite{Kobayashi:2021jbn}, and we directly generalize their method.
The case without boundary and also a special case of the APS index theorem for a cylindrical geometry were discussed in \cite{Kanno:2021bze}, and we also use their result as an input.
See \cite{quillen1985superconnections,kahle2011superconnections,Gomi:2021bhy} for the mathematical literature.

The aim is to relate the $\eta$-invariant of the hermitian operator
\begin{equation} \label{opK}
    K = \begin{cases}
        \mathrm{i}(\Gamma^\mu D_\mu + \mathrm{i}\Gamma M) & \text{for odd $d$} \,, \\
        \mathrm{i}(\tau^3 \Gamma^\mu D_\mu + \mathrm{i}\tau^2 M_1 - \mathrm{i}\tau^1 M_2) & \text{for even $d$} \,,
    \end{cases}
\end{equation}
defined on a closed and null-bordant $(d+1)$-dimensional manifold $Y$,\footnote{Inside the null-bordism here the mass matrix is allowed to be degenerate.} to the index of a Dirac-like operator $Q$ defined on a $(d+2)$-dimensional manifold $W$ with boundary $\partial W = Y$, with a suitable choice of the boundary condition.
We will define the operator $Q$ and explain its relation to the actual $(d+2)$-dimensional Dirac operator below.
We assume there is a collar neighborhood near the boundary $Y$ of $W$, and $w \in (-\epsilon,0]$ will denote the coordinate transverse to the boundary in the collar neighborhood, where the boundary is at $w=0$ and $\epsilon > 0$.
$w$ is to be interpreted as the Euclidean time direction near the boundary.\footnote{Note that the existence of the collar neighborhood near boundary was also assumed even in the original mathematical proof in \cite{Atiyah:1975jf,Atiyah:1975jg,Atiyah:1976jh}.}

For the APS index theorem to be applicable, we need a $\mathbb{Z}_2$ grading in the space of spinors, which anticommutes with the operator $Q$.
As will be explained in more detail, the operator $Q$ of our interest takes the following form:
\begin{equation} \label{Eq:operator_Q}
    Q = \begin{cases}
        \mathrm{i}\tau^1\hat{\Gamma}^{\hat{\mu}} D_{\hat{\mu}} + \tau^2 M & \text{for odd $d$} \,, \\
        \mathrm{i}\hat{\Gamma}^{\hat{\mu}} D_{\hat{\mu}} - \hat{\Gamma}(M_1 \tau^1 + M_2 \tau^2) & \text{for even $d$} \,.
    \end{cases}
\end{equation}
In both cases, the operator $Q$ acts on $2N_f$ flavors of $(d+2)$-dimensional fermions living on $W$.
Here, $\hat{\mu}= 1, \cdots, d+2$, and $\hat{\Gamma}^{\hat{\mu}}$ are $(d+2)$-dimensional gamma matrices.
For the case of even $d$, $\hat{\Gamma}$ denotes the $(d+2)$-dimensional chirality matrix.
The $\mathbb{Z}_2$-grading that anticommutes with the operator $Q$ is given by $\tau^3$ if $d$ is odd, and $\tau^3 \hat{\Gamma}$ if $d$ is even.
In the zero eigenvalue subspace of the operator $Q$, $Q$ commutes with the $\mathbb{Z}_2$-grading matrix, and they can be simultaneously diagonalized.
The eigenvalue of the $\mathbb{Z}_2$-grading matrix can be either $+1$ or $-1$, and we will call it ``chirality.''
Then, the index of the operator is defined as the number of positive chirality zero modes minus the number of negative chirality zero modes as usual. 
Equivalently, we may define an operator $Q_+$ to be the operator $Q$ restricted to the subspace of positive chirality spinors, and then the definition of the index is 
$\Ind Q \equiv \Ind Q_+ = \text{dim}\, \text{ker}\, Q_+ - \text{dim}\, \text{coker}\, Q_+ = \text{dim} \,\text{ker}\, Q_+ - \text{dim}\, \text{ker}\, Q_- $, where $Q_-$ is $Q$ restricted to the subspace of negative chirality spinors.
By definition, an operator $Q_+$ having finite dimensional kernel and cokernel is called Fredholm.
For us, requiring $Q$ to be hermitian will suffice to ensure that the index is well-defined.
When $Q$ is defined on a manifold with boundary, which is the case of our interest, we have to impose non-local boundary conditions on spinors to maintain the hermiticity of $Q$.
We will collectively call such boundary conditions as APS boundary conditions.

We do not require the manifold $W$ to be compact (nor the boundary $\partial W = Y$), but we do assume that the operator $Q$ with the boundary condition that we will explain below and the operator $K$ on the boundary have discrete spectra. 
This happens if the mass blows up towards the non-compact directions.

We will first give a detailed derivation of the index theorem assuming $d$ is odd.
Similar calculations can be done for the even dimensional case as well with minor changes.
Let $\Upphi$ be $2N_f$ species of $(d+2)$-dimensional Dirac fermions on $W$, with the Dirac operator given by
\begin{equation}
    \sD_{d+2} = \hat{\Gamma}^{\hat{\mu}} D_{\hat{\mu}} + \tau^1 m_1 + \tau^2 m_2 + \tau^3 M \,,
\end{equation}
where $m_1$ and $m_2$ are constant real numbers, and $\tau^i$'s are Pauli matrices.
We will write $m_1 + \mathrm{i}m_2 = m_0 e^{\mathrm{i} \alpha_0}$. 
$M$ is a general $N_f \times N_f$ hermitian matrix-valued function on the boundary $\partial W = Y$, see Eq. (\ref{opK}), and it extends into $W$ in such a way that it is independent of the coordinate $w$ inside the collar neighborhood.
The Euclidean partition function of this $(d+2)$-dimensional system is
\begin{equation}
    \mathcal{Z}_{d+2} = \int [D\bar{\Upphi}][D\Upphi] \text{exp}(-\int_W \bar{\Upphi} \sD_{d+2} \Upphi) \,.
\end{equation}
We may write $\sD_{d+2} = \tau^1 (-\mathrm{i}Q + m_0 e^{\mathrm{i}\alpha \tau^3})$ where
\begin{align} \label{opQ}
\begin{split}
    Q &= \mathrm{i}\tau^1\hat{\Gamma}^{\hat{\mu}} D_{\hat{\mu}} + \tau^2 M \\
    &= \begin{bmatrix*}
        0 & \mathrm{i}\hat{\Gamma}^{\hat{\mu}} D_{\hat{\mu}} - \mathrm{i}M \\
        \mathrm{i}\hat{\Gamma}^{\hat{\mu}} D_{\hat{\mu}} + \mathrm{i}M & 0 
    \end{bmatrix*} \,.
\end{split}
\end{align} 
We need to impose a boundary condition that makes $Q$ hermitian, and $\tau^3$ will be our $\mathbb{Z}_2$ grading in this case.
We have $\{\tau^3,Q\} = 0$.

On the $(d+1)$-dimensional boundary $\partial W = Y$, we define the $(d+1)$-dimensional gamma matrices to be
\begin{equation}
    \Gamma^\mu = \mathrm{i} \hat{\Gamma}^w \hat{\Gamma}^\mu \,\,,\quad \Gamma = \hat{\Gamma}^w 
\end{equation}
for $\mu = 1, \cdots, d+1$ and $\Gamma$ is the chirality matrix in $(d+1)$-dimensions.
The operator $Q_+ = \mathrm{i}(\hat{\Gamma}^{\hat{\mu}} D_{\hat{\mu}} + M)$ acting on positive chirality spinors (bottom left block of Eq.(\ref{opQ})) becomes near the boundary
\begin{equation}
    Q_+ = \mathrm{i}\hat{\Gamma}^w(\partial_w - K)
\end{equation} 
where $K = \mathrm{i}(\Gamma^\mu D_\mu + \mathrm{i}\Gamma M)$ is the hermitian operator on $Y$ that we are after (\ref{opK}).
If we define another hermitian operator $K^\prime = \mathrm{i}(\Gamma^\mu D_\mu - \mathrm{i}\Gamma M)$, then near the boundary we similarly have $Q_- = \mathrm{i}(\hat{\Gamma}^{\hat{\mu}} D_{\hat{\mu}} - M) = \mathrm{i}\hat{\Gamma}^w(\partial_w - K^\prime)$ for the upper right block of Eq. (\ref{opQ}).

For the operator $Q$ to be hermitian, we want the following quantity to vanish, for any two arbitrary spinors $\Upphi$ and $\Uppsi$,
\begin{equation} \label{hermiticity1}
    \int_W \Upphi^\dagger Q \Uppsi - \left(\int_W \Uppsi^\dagger Q \Upphi\right)^{*} = \int_Y \left(\Upphi_+^\dagger \hat{\Gamma}^w \Uppsi_- + \Upphi_-^\dagger \hat{\Gamma}^w \Uppsi_+ \right)
\end{equation}
where the $\pm$ subscripts indicate the positive and negative chirality parts (i.e., eigenvalues of $\tau^3$), respectively.
We integrated by parts to obtain the right-hand side.
On the boundary $Y$, note that we have the relation $K\hat{\Gamma}^w = - \hat{\Gamma}^w K^\prime$.
If we write the eigenvalues and eigenfunctions of $K$ as $K \Psi_j = \xi_j \Psi_j$, then $K^\prime$ has the exact opposite spectrum to $K$, $K^\prime \Psi_j^\prime = -\xi_j \Psi_j^\prime$, where $\Psi_j^\prime = \hat{\Gamma}^w \Psi_j$.
Using this fact, we may impose the following boundary conidition:
\begin{equation} \label{APSbc2}
    \mathcal{P}^{K}_{\geq 0}\, \Upphi_+ |_{\partial W = Y} = 0 \,\,,
    \quad \mathcal{P}^{K^\prime}_{> 0}\, \Upphi_- |_{\partial W = Y} = 0 \,.
\end{equation}
$\mathcal{P}^{K}_{\geq 0}$ is the projection operator to the subspace of spinors on $Y$ spanned by the eigenfunctions of $K$ with the eigenvalues greater than or equal to zero, and similarly $\mathcal{P}^{K^\prime}_{> 0}$ is the projection operator to the subspace spanned by the eigenfunctions of $K^\prime$ with the eigenvalues strictly greater than zero.
Because of $K\hat{\Gamma}^w = - \hat{\Gamma}^w K^\prime$, we see that for the spinors satisfying this boundary condition, the boundary term on the right-hand side of Eq. (\ref{hermiticity1}) vanishes, and $Q$ is hermitian with this boundary condition. There also exist more general choices of boundary conditions, but we will not need them here. 
See \cite{Yonekura:2016wuc,Kobayashi:2021jbn}.

To know the corresponding boundary condition to be imposed on $\bar{\Upphi}$ inside the path integral, we consider the follwong mode expansions of fermion fields near the boundary $\partial W = Y$, in terms of the eigenfunctions of $K$ and $K^\prime$:
\begin{align} \label{indexmode}
\begin{split}
    \Upphi_+ (y^\mu,w) &= \sum_j \alpha_j^+ (w) \Psi_j (y^\mu) \,, \\
    \Upphi_- (y^\mu,w) &= \sum_j \alpha_j^- (w) \Psi_j^\prime (y^\mu) = \sum_j \alpha_j^- (w) \hat{\Gamma}^w \Psi_j (y^\mu) \,,\\
    \bar{\Upphi}_+ (y^\mu,w) &= \sum_j \beta_j^+ (w) (\Psi_j^\prime)^\dagger (y^\mu) = \sum_j \beta_j^+ (w) \Psi_j^\dagger (y^\mu )\hat{\Gamma}^w \,,  \\
    \bar{\Upphi}_- (y^\mu,w) &= \sum_j \beta_j^- (w) \Psi_j^\dagger (y^\mu) \,.
\end{split}
\end{align}
In terms of $\alpha^\pm_j$ and $\beta^\pm_j$, the action near the boundary becomes
\begin{equation}
    S = \int dw \sum_j \left[ \beta_j^+(w) (\partial_w - \xi_j)\alpha_j^+(w) + \beta_j^-(w)(\partial+\xi_j)\alpha_j^-(w) + \cdots \right]
\end{equation}
where $\cdots$ contains terms proportional to $m_0$. We see that $\beta_j^\pm$ is conjugate to $\alpha_j^\pm$ for each $j$. 
The APS boundary condition (\ref{APSbc2}) imposed on $\Upphi$ corresponds to setting at the boundary $\alpha^+_j = 0$ if $\xi_j \geq 0$, $\alpha_j^- = 0$ if $\xi_j < 0$, and leaving the remaining coefficients to be unconstrained. 
The corresponding boundary condition for $\bar{\Upphi}$ can then be read off. 
At the boundary, we let $\beta^+_j = 0$ if $\xi_j < 0$, $\beta^-_j = 0$ if $\xi_j \geq 0$, and leave the remaining coefficients to be unconstrained.

The Hamiltonian near the boundary is given by
\begin{equation} \label{hamiltonian2}
    \hat{H} = \sum_j \xi_j \left( (\alpha_j^-)^\dagger \alpha_j^- - (\alpha_j^+)^\dagger \alpha_j^+ \right) + \cdots
\end{equation}
where we again omitted terms proportional to $m_0$.
The quantum state corresponding to the APS boundary condition (\ref{APSbc2}) is characterized by the condition ($\beta^\pm_j = (\alpha^\pm_j)^\dagger$)
\begin{equation}
    \begin{cases}
        \bra{\mathtt{APS}} \alpha^+_j = 0 & \text{if $\xi_j \geq  0$} \\
        \bra{\mathtt{APS}} (\alpha^+_j)^\dagger = 0 & \text{if $\xi_j < 0$} \\
        \bra{\mathtt{APS}} \alpha^-_j = 0 & \text{if $\xi_j < 0$} \\
        \bra{\mathtt{APS}} (\alpha^-_j)^\dagger = 0 & \text{if $\xi_j \geq 0$}
    \end{cases} 
\end{equation}
and we can see that $\ket{\mathtt{APS}}$ is a ground state of the Hamiltonian (\ref{hamiltonian2}) when $m_0 = 0$. 
There will be more than one ground states if there are zero modes $\xi_j = 0$, and our choice of the boundary condition corresponds to one of them.
We set the ground state energy to be zero.

To derive the index of $Q$ under this boundary condition, our strategy is to compute the following ratio of the two $(d+2)$- dimensional partition functions
\begin{equation} \label{indexratio}
    \frac{\mathcal{Z}_{d+2} [m_0 e^{\mathrm{i}\alpha_0}]}{\mathcal{Z}_{d+2} [m_0]} = \frac{\int [D\bar{\Upphi}][D\Upphi] \exp \left(-\int_W \bar{\Upphi} \tau^1(-\mathrm{i}Q+m_0 e^{\mathrm{i}\alpha_0 \tau^3}) \Upphi \right)}
    {\int [D\bar{\Upphi}][D\Upphi] \exp \left(-\int_W \bar{\Upphi} \tau^1(-\mathrm{i}Q+m_0) \Upphi \right)}
\end{equation}
in two different ways, for an arbitrary constant phase $\alpha_0$.

Define $Q^\prime = \mathrm{i}\tau^1 \hat{\Gamma}^{\hat{\mu}} D_{\hat{\mu}} - \mathrm{i}\tau^2 M$. 
For $Q^\prime$, we impose different boundary conditions, $\mathcal{P}^{K^\prime}_{>} \Upphi_+ |_Y =0$ and $\mathcal{P}^{K}_{\geq} \Upphi_- |_Y = 0$.
Under this boundary condition, $Q^\prime$ is a hermitian operator on $W$.
The two operators are related by $Q\tau^1 = -\tau^1 Q^\prime$, which respects the respective boundary conditions.
That is, if we write the eigenvalues and the eigenfunctions of $Q$ as $Q \Upphi_k = \upsilon_k \Upphi_k$, then $Q^\prime$ has the exact opposite spectrum to $Q$, $Q^\prime \Upphi_k^\prime = -\upsilon_k \Upphi_k^\prime$, with the eigenfunctions being $\Upphi_k^\prime = \tau^1 \Upphi_k$.

Boundary conditions are chosen in such a way that we can expand the fermion fields as
\begin{equation}
    \Upphi = \sum_k a_k \Upphi_k \,\,,
    \quad
    \bar{\Upphi} = \sum_k b_k (\Upphi_k^\prime)^\dagger
    = \sum_k b_k \Upphi_k^\dagger \tau^1 \,,
\end{equation}
where $a_k$ and $b_k$ are grassmannian coefficients.
The path integral measure can be defined to be
\begin{equation}
    [D\bar{\Upphi}][D\Upphi] = \prod_k db_k da_k \,.
\end{equation}
Consider the following change of variables for the path integral,
\begin{equation} \label{changevar}
    \Upphi \rightarrow \Upphi^\prime = e^{\mathrm{i}(\alpha/2) \tau^3} \Upphi
    \,\,, \quad
    \bar{\Upphi} \rightarrow \bar{\Upphi}^\prime = \bar{\Upphi} e^{-\mathrm{i}(\alpha/2) \tau^3} \,,
\end{equation}
where $\alpha$ can be any arbitrary function which is constant on the boundary, so that it does not affect the boundary conditions.
If we write $\Upphi^\prime = \sum_k a_k^\prime \Upphi_k$ and $\bar{\Upphi}^\prime = \sum_k b_k^\prime \Upphi_k^\dagger \tau^1$, then the coefficients $a_k^\prime$ and $b_k^\prime$ are given by
\begin{align}
\begin{split}
    a_k^\prime &= \int_W \Upphi_k^\dagger \Upphi^\prime
    = \sum_l \left(\int_W \Upphi_k^\dagger e^{\mathrm{i}\alpha \tau^3 /2} \Upphi_l \right) a_l \,, \\
    b_k^\prime &= \int_W \bar{\Upphi}^\prime \tau^1 \Upphi_k
    = \sum_l b_l \left(\int_W \Upphi_l^\dagger e^{\mathrm{i}\alpha \tau^3 /2} \Upphi_k \right) \,.
\end{split}
\end{align}
The path integral measure is potentially non-invariant under this change of variables,
\begin{equation}
    \prod_k db_k da_k = \left[\text{Det} (e^{\mathrm{i}\alpha \tau^3/2})\right]^2 \prod_k db_k^\prime da_k^\prime = \left[\exp(\mathrm{i}\text{Tr}(\alpha \tau^3))\right] \prod_k db_k^\prime da_k^\prime
\end{equation}
where the determinant and the trace is over the whole space of spinors on $W$ spanned by the eigenfunctions $\Upphi_k$.

The action also changes under (\ref{changevar}).
For instance, the numerator in Eq. (\ref{indexratio}) changes as
\begin{equation}
    \int_W \bar{\Upphi} \tau^1 (-\mathrm{i}Q + m_0 e^{\mathrm{i}\alpha_0 \tau^3}) \Upphi
    = \int_W \left[ \bar{\Upphi}^\prime \tau^1 (-\mathrm{i}Q + m_0 e^{\mathrm{i}(\alpha_0-\alpha) \tau^3}) \Upphi^\prime - \frac{\mathrm{i}}{2} (\partial_{\hat{\mu}}\alpha) \bar{\Upphi}^\prime \hat{\Gamma}^{\hat{\mu}} \tau^3 \Upphi^\prime \right] \,.
\end{equation}

Now, consider making this change of variables to the numerator in Eq. (\ref{indexratio}) with constant $\alpha = \alpha_0$.
Then, the path integral measure picks up the factor $\exp (\mathrm{i}\alpha_0 \text{Tr}(\tau^3)) = \exp (\mathrm{i}\alpha_0 \Ind Q)$, and the remaining path integral is identical to the denominator.
Thus, we get
\begin{equation} \label{index1}
    \frac{\mathcal{Z}_{d+2} [m_0 e^{\mathrm{i}\alpha_0}]}{\mathcal{Z}_{d+2} [m_0]} = e^{\mathrm{i}\alpha_0 \Ind Q} \,.
\end{equation}
Note that this is independent of $m_0$, as long as $m_0 \neq 0$.
If $m_0 = 0$, the partition function vanishes when there exist any zero modes, making the ratio ill-defined.

Then, following the trick from \cite{Kobayashi:2021jbn}, we consider a different change of variables again to the numerator of Eq. (\ref{indexratio}), where $\alpha$ is given by $\alpha = \alpha_0 \theta(w_0 - w)$ inside the collar neighborhood of the boundary and remains to be equal to constant $\alpha_0$ everywhere inside the bulk.
$w_0 \in [0,-\epsilon)$ is a fixed Euclidean time inside the collar neighborhood, and $\theta(w_0 - w)$ is the Heaviside step function which vanishes for $w > w_0$, i.e., near the boundary. Under this change of variables, we get
\begin{equation}
    \frac{\mathcal{Z}_{d+2} [m_0 e^{\mathrm{i}\alpha_0}]}{\mathcal{Z}_{d+2} [m_0]} = e^{\mathrm{i}\text{Tr}(\alpha_0 \theta(w_0 - w) \tau^3)} \times 
    \frac{\int [D\bar{\Upphi}][D\Upphi] e^{-\int_W \bar{\Upphi} \tau^1(-\mathrm{i}Q+m_0 e^{\mathrm{i}\alpha_0 \theta(w-w_0)\tau^3}) \Upphi} e^{-\frac{\mathrm{i}}{2} \alpha_0 \mathcal{O} (w_0) }}
    {\int [D\bar{\Upphi}][D\Upphi] e^{-\int_W \bar{\Upphi} \tau^1(-\mathrm{i}Q+m_0) \Upphi}}
\end{equation}
where $\mathcal{O}(w_0) = \int_{Y, w=w_0} \bar{\Upphi} \hat{\Gamma}^w \tau^3 \Upphi$ is an ``axial'' charge operator inserted at $w=w_0$ slice inside the collar neighborhood.

We divide the manifold $W$ into two pieces.
Let $W_1$ be a cylindrical region near the boundary $\partial W = Y$ of length $w_0 \leq w \leq 0$, and $W_0$ be the rest of $W$, such that gluing $W_1$ and $W_0$ along their common boundary gives the whole $W$.
Path integral over $W_0$ defines a state as usual, $\ket{W_0}$, and the path integral over $W_1$ gives a time evolution which we denote as $\hat{W}_1$.
Then, we have
\begin{equation}
    \frac{\int [D\bar{\Upphi}][D\Upphi] e^{-\int_W \bar{\Upphi} \tau^1(-\mathrm{i}Q+m_0 e^{\mathrm{i}\alpha_0 \theta(w-w_0)\tau^3}) \Upphi} e^{-\frac{\mathrm{i}}{2} \alpha_0 \mathcal{O}(w_0) }}
    {\int [D\bar{\Upphi}][D\Upphi] e^{-\int_W \bar{\Upphi} \tau^1(-\mathrm{i}Q+m_0) \Upphi}}
    = \frac{\braket{\mathtt{APS}|\hat{W}_1 (m_0 e^{\mathrm{i}\alpha_0})e^{-\frac{\mathrm{i}}{2}\alpha_0\hat{\mathcal{O}}(w_0)}|W_0}}{\braket{\mathtt{APS}|\hat{W}_1(m_0)|W_0}}
\end{equation}
where for the time evolution along $W_1$, we have $m_1 + \mathrm{i}m_2 = m_0 e^{\mathrm{i}\alpha_0}$ for the numerator, whereas $m_1 + \mathrm{i}m_2 = m_0$ for the denominator.
$\hat{\mathcal{O}}$ is the quantum operator corresponding to the charge $\mathcal{O}$.

Now, since $\ket{\mathtt{APS}}$ is a ground state on $Y$
when $m_0 = 0$, if we take the limit $m_0 \rightarrow 0$ in the above, we obtain
\begin{equation}
    \frac{\braket{\mathtt{APS}|\hat{W}_1 (m_0 e^{\mathrm{i}\alpha_0})e^{-\frac{\mathrm{i}}{2}\alpha_0\hat{\mathcal{O}}(w_0)}|W_0}}{\braket{\mathtt{APS}|\hat{W}_1(m_0)|W_0}} \rightarrow \frac{\braket{\mathtt{APS}|e^{-\frac{\mathrm{i}}{2}\alpha_0\hat{\mathcal{O}}(w_0)}|W_0}}{\braket{\mathtt{APS}|W_0}} \,.
\end{equation}

Next, using the mode expansions (\ref{indexmode}), the operator $\hat{\mathcal{O}}(w_0)$ can be expressed as
\begin{equation}
    \hat{\mathcal{O}}(w_0) = \sum_j \left[ (\alpha_j^+)^\dagger \alpha_j^+ - (\alpha_j^-)^\dagger \alpha_j^- \right] \,.
\end{equation}
The state $\bra{\mathtt{APS}}$ is an eigenstate of this operator, with the eigenvalue being the $\eta$-invariant of $K$ on $Y$,
\begin{align}
\begin{split}
    \bra{\mathtt{APS}} \hat{\mathcal{O}}(w_0) &= \bra{\mathtt{APS}} \left(\sum_{\xi_j \geq 0} 1 - \sum_{\xi_j < 0} 1 \right)  \\
    &= \bra{\mathtt{APS}} 2\eta_K(Y) \,.
\end{split}
\end{align}
Thus, we obtain
\begin{equation} \label{indexeta}
    \frac{\braket{\mathtt{APS}|\hat{W}_1 (m_0 e^{\mathrm{i}\alpha_0})e^{-\frac{\mathrm{i}}{2}\alpha_0\hat{\mathcal{O}}(w_0)}|W_0}}{\braket{\mathtt{APS}|\hat{W}_1(m_0)|W_0}} = \exp \left(-\mathrm{i}\alpha_0 \eta_K (Y) \right)
\end{equation}
in the limit $m_0 \to 0$.

Finally, we would like to compute $\exp \left(\mathrm{i}\text{Tr}(\alpha_0 \theta(w_0 - w)\tau^3)\right)$.
This quantity is analogous to (the exponential of) $\text{Tr}(\gamma^5)$ appearing in the Fujikawa's computation of an axial anomaly \cite{Fujikawa:1979ay,Fujikawa:1983bg}, and can be computed by a similar heat kernel regularization method.
Since $\alpha_0 \theta(w_0 -w)$ vanishes near the boundary, such a local computation is still applicable even in the precense of the boundary \cite{Kobayashi:2021jbn}.
The computation of $\exp \left(\mathrm{i}\text{Tr}(\alpha_0 \theta(w_0 - w)\tau^3)\right)$ was performed in \cite{Kanno:2021bze} using the Fujikawa method.
We will simply adopt their result here:
\begin{align} \label{heatkernelanompoly}
\begin{split}
    \exp \left(\mathrm{i}\text{Tr}(\alpha_0 \theta(w_0 - w)\tau^3)\right) &= \lim_{\Lambda\to\infty} \int_{W_0} \ch (\mathcal{F}) \hat{A} (TW) \\
    &= \lim_{\Lambda\to\infty} \int_W \ch (\mathcal{F}) \hat{A} (TW) \,.
\end{split}
\end{align}
The field strength $\mathcal{F}$ for the superconnection (of odd type since $d+2$ is odd) and the Chern character $\ch (\mathcal{F})$ were defined in Section \ref{Sec:superconnection}.
Recall the field strength depends on the mass parameters as $\mathcal{F} = \mathcal{F} (\widetilde{M}=M/\Lambda)$.
The cutoff scale $\Lambda$ is introduced through the heat kernel regularization in the Fujikawa method.
Here, it is important that on the right-hand side, we should first evaluate the integral and then take the limit $\Lambda \to \infty$.
The second equality sign follows from the fact that $\ch (\mathcal{F}) \hat{A} (TW)$ vanishes near the boundary due to the conditions imposed on the collar neighborhood of the boundary.

Combining Eq. (\ref{index1}), Eq. (\ref{indexeta}), and Eq. (\ref{heatkernelanompoly}), we see that
\begin{equation}
    \frac{\mathcal{Z}_{d+2} [m_0 e^{\mathrm{i}\alpha_0}]}{\mathcal{Z}_{d+2} [m_0]} = \exp \left(\mathrm{i}\alpha_0 \Ind Q \right)
    = \exp \left[ \mathrm{i}\alpha_0 \left(
        -  \eta_K (Y) + \lim_{\Lambda\to\infty} \int_W \ch (\mathcal{F}) \hat{A} (TW) 
        \right) \right]
\end{equation}
for $m_0 \to 0$.
Since $\alpha_0$ is arbitrary, we conclude that
\begin{equation} \label{indexthm}
    \eta_K (Y)
    = -  \Ind Q  + \lim_{\Lambda\to\infty}  \int_W \ch (\mathcal{F}) \hat{A} (TW) \,.
\end{equation}
This is the desired APS index theorem.

The same result still holds when $d$ is even, with the appropriately modified definition of $K$ as in Eq. (\ref{opK}) and using the superconnection of even type from Section \ref{Sec:superconnection}.
We will only briefly sketch the derivation for the even $d$ case since all the steps are analogous to the odd $d$ case.
Again, we have $2N_f$ species of $(d+2)$-dimensional Dirac fermions $\Upphi$ on a $(d+2)$-dimensional manifold $W$ with the boundary $\partial W = Y$.
The $(d+2)$-dimensional Dirac operator in the bulk $W$ is given by
\begin{equation}
     \sD_{d+2} = \hat{\Gamma}^{\hat{\mu}} D_{\hat{\mu}} + m_1 + \mathrm{i}\hat{\Gamma}(m_2 \tau^3 + M_1 \tau^1 + M_2 \tau^2)
\end{equation}
where $\hat{\Gamma}$ is the chirality matrix in $(d+2)$-dimensions which is now even, $m_1$ and $m_2$ are real constants, and $M_1$ and $M_2$ are $N_f \times N_f$ hermitian matrix valued modulated masses.
On the boundary $\partial W$, $M_1$ and $M_2$ reduces to those defining the operator $K$ in Eq. \eqref{opK}.
As before, we write $m_1 + im_2 = m_0 e^{i\alpha_0}$.
We may write $\sD_{d+2} = (-\mathrm{i}Q + m_0 e^{\mathrm{i}\alpha_0 \hat{\Gamma}\tau^3})$ where 
\begin{equation}
Q = \mathrm{i}\hat{\Gamma}^{\hat{\mu}} D_{\hat{\mu}} - \hat{\Gamma}(M_1 \tau^1 + M_2 \tau^2) \,.
\end{equation} 
$Q$ again is a hermitian operator on $W$ with appropriately chosen boundary conditions.
The aim is to compute the index of this $Q$.
The $\mathbb{Z}_2$-grading that is necessary to define the index in this case is now given by the matrix $\tau^3 \hat{\Gamma}$ which acts both on the flavor and the spinor indices, and it anticommutes with $Q$.
The rest of the steps are essentially the same as the odd $d$ case, and we will not repeat them here.

\section{Explicit torus partition functions} \label{apptorus}

In this appendix, we rederive Eq. (\ref{eq:torus_partition_function}) from Section \ref{Sec4.2} by explicitly computing the torus partition functions of the 2d Dirac fermion with the modulated complex mass parameter.
We will consider a slightly more general configuration where the winding number of the mass term along the $a$-cycle is an arbitrary integer $\tilde{n}$.
If we write the angular coordinate along the $a$-cycle as $\theta_a$, then we have $\alpha(\theta_a) = \tilde{n} \theta_a$.

We compute the partition function by interpreting the $b$-cycle as the compactified Euclidean time direction.
In the Minkowski signature, the canoncial quantization on $S^1_a$ can be done by the expansion
\begin{equation}
    \psi(t,x) = \sum_{n \in \mathbb{Z} + 1/2} \left[ a_n \phi_n^a (x) e^{-\mathrm{i}E_n^a t} + b_n \phi_n^b (x) e^{-\mathrm{i}E_n^b t} \right] \,,
\end{equation}
where the explicit forms of the functions $\phi_n^a,\, \phi_n^b$ are not important. 
The energies are given by\footnote{In this appendix, we let the coordinates on the torus to be dimensionless (and $2\pi$ periodic), and thus all the energies and the mass parameters are also dimensionless.}
\begin{equation} \label{t2spectrum}
    \begin{cases}
        E_n^a &= \frac{1}{2}\tilde{n} + \frac{1}{2} \sqrt{(\tilde{n} - 2n)^2 + 4M_0^2} \,, \\
        E_n^b &= \frac{1}{2}\tilde{n} - \frac{1}{2} \sqrt{(\tilde{n} - 2n)^2 + 4M_0^2} \,,
    \end{cases}
\end{equation}
and in the limit $M_0 \to \infty$, we have $|E_n^i| \to \infty$ for all $n$ and both $i = a,b$.

For a single fermionic mode with energy $E$, we pick a regularization scheme such that the thermal partition function $\text{Tr}[(-1)^F e^{-\beta H}]$ for this mode becomes 
\begin{equation}
    \lim_{|E| \to \infty} \mathcal{Z}_1[E] =
    \begin{cases}
        1 & \text{if}\,\, E>0  \,,\\
        -1 & \text{if}\,\, E<0 \,.
    \end{cases}
\end{equation}
In the infinite energy limit, the partition function receives a contribution only from the ground state, and we are setting the ground state energy to be zero.
The sign of the partition function corresponds to the assignment of the fermion number to the ground state.
The sign difference for the two cases is to account for the level crossing at $E=0$.

Then, the 2d partition function in the infinite mass limit becomes
\begin{equation}
    \lim_{M_0 \to \infty} \mathcal{Z}_{2}(T^2) = \prod_{E_n^i} \left( \lim_{|E_n^i| \to \infty} \mathcal{Z}_1[E_n^i] \right)
\end{equation}
which formally corresponds to $(-1)^\chi$ where $\chi$ is the number of negative energy eigenvalues. 
The number of negative energy eigenvalues for large $M_0$ is infinite, and the partition function is not well-defined without an additional regularization.

We perform the regularization in the following way.
Inside the parameter space $\widetilde{P} = \mathbb{C}$ including the degenerate point, we can continuously deform the mass parameter and change its winding number.
In such a process, the energy spectrum will continuously vary, and some of the energy levels cross the zero energy. 
Each time an energy level crosses the zero energy, it changes the sign of the partition function. 
Consider continuously changing the winding number from $\tilde{n}$ to zero, first by setting $M_0$ to zero and increasing it again while keeping $\alpha = 0$.
If the number of energy levels that cross zero in the process is $N$, then the regularized partition function becomes
\begin{equation} \label{2dratio2}
    \lim_{M_0 \to \infty} \frac{\mathcal{Z}_{2}(T^2)}{\mathcal{Z}^{(0)}_{2}(T^2)} = (-1)^N \,.
\end{equation}

The remaining task is to compute $N$.
At the end of the deformation process, the energy spectrum is given by
\begin{equation} \label{t2spec2}
    E = \pm \sqrt{n^2 + M_0^2}
\end{equation}
with large $M_0$, whereas at the beginning of the process we have Eq. (\ref{t2spectrum}), again with large $M_0$. 
We have to glue the two spectra across $M_0 = 0$, and count the number of energy levels which cross the zero energy.
When we start from the spectrum (\ref{t2spec2}) with sufficiently large $M_0$, all the $E_n^a$'s are positive whereas all the $E_n^b$'s are negative. 
As we decrease $M_0$, $E_n^a$'s always remain positive down to $M_0 = 0$, but some of the $E_n^b$'s which were originally negative become positive or zero at $M_0 = 0$, depending on $\tilde{n}$ and $n$. 
Now, from $M_0 = 0$, we evolve the spectrum again by increasing $M_0$, according to (\ref{t2spec2}). 
A feature of the spectrum (\ref{t2spec2}) is that, if the energy $E$ was positive (negative) at $M_0 = 0$, it always remains positive (negative) all the way to $M_0 \to \infty$.

\begin{figure}[t]
    \centering
    \includegraphics[width=0.5\textwidth]{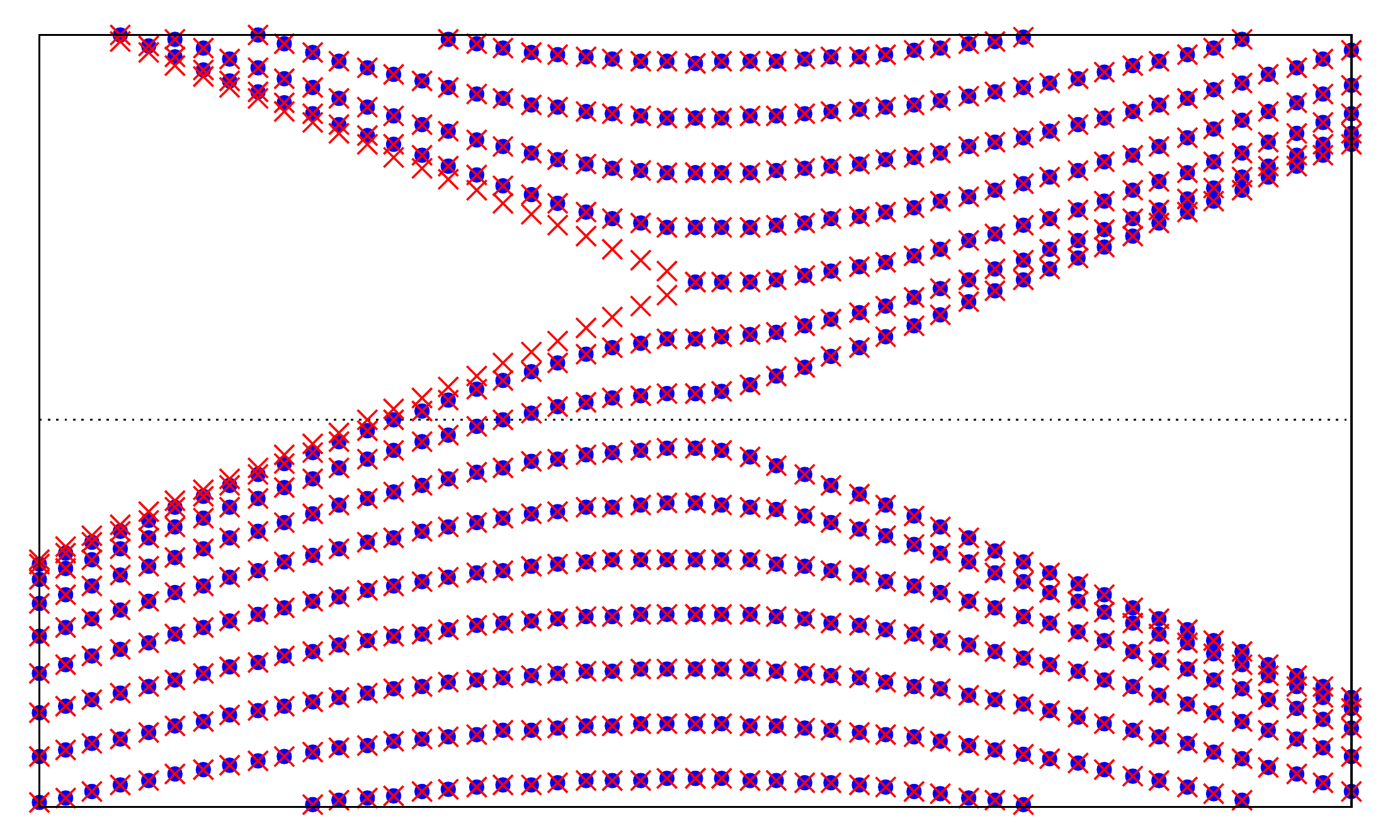}
    \caption{The spectral flow for the case of $\tilde{n}=5$.
    The vertical axis corresponds to the energy, and from the left to the right the deformation of the mass parameter occurs.
    The stand-alone red crosses indicate non-degenerate energy eigenvalues, whereas the red crosses that are overlaid with blue dots indicate doubly-degenerate energy eigenvalues.
    The horizontal black dashed line corresponds to the zero energy. We see that the number of zero energy crossings across the spectral flow is indeed equal to $\tilde{n}=5$.}
    \label{Fig:flow}
\end{figure}

From this, we deduce that the number of energy levels that cross the zero energy while following the spectral flow of energy eigenvalues across the deformation process is equal to the number of positive $E_n^b$'s at $M_0 = 0$. 
The number of $n$'s satisfying this condition is $N=\tilde{n}$. 
Thus,
\begin{equation} 
    \lim_{M_0 \to \infty} \frac{\mathcal{Z}_{2}(T^2)}{\mathcal{Z}^{(0)}_{2}(T^2)} = (-1)^{\tilde{n}} \,,
\end{equation}
and the special case $\tilde{n} =1$ confirms Eq. (\ref{eq:torus_partition_function}).
See Figure \ref{Fig:flow} for an examplary spectal flow of the energy spectrum across the deformation process.

\def\arxivfont{\rm}
\bibliographystyle{ytamsalpha}
\baselineskip=.95\baselineskip
\bibliography{ref}

\newcommand{\etalchar}[1]{$^{#1}$}
\providecommand{\bysame}{\leavevmode\hbox to3em{\hrulefill}\thinspace}
\providecommand{\MR}{\relax\ifhmode\unskip\space\fi MR }
\providecommand{\MRhref}[2]{%
  \href{http://www.ams.org/mathscinet-getitem?mr=#1}{#2}
}
\providecommand{\href}[2]{#2}
\providecommand{\doihref}[2]{\href{#1}{#2}}
\providecommand{\arxivfont}{\tt}
\begin{thebibliography}{FFM{\etalchar{+}}20b}

\bibitem[AH20]{Angelescu:2020yzf}
A.~Angelescu and P.~Huang, \emph{{Integrating Out New Fermions at One Loop}},
  \doihref{http://dx.doi.org/10.1007/JHEP01(2021)049}{JHEP \textbf{01} (2021)
  049}, \href{http://arxiv.org/abs/2006.16532}{{\arxivfont arXiv:2006.16532
  [hep-ph]}}.

\bibitem[AIO00]{Alishahiha:2000du}
M.~Alishahiha, H.~Ita, and Y.~Oz, \emph{{On superconnections and the tachyon
  effective action}},
  \doihref{http://dx.doi.org/10.1016/S0370-2693(01)00175-7}{Phys. Lett. B
  \textbf{503} (2001) 181--188},
  \href{http://arxiv.org/abs/hep-th/0012222}{{\arxivfont
  arXiv:hep-th/0012222}}.

\bibitem[APS75a]{Atiyah:1975jg}
M.~F. Atiyah, V.~K. Patodi, and I.~M. Singer, \emph{{Spectral asymmetry and
  Riemannian geometry. II}},
  \doihref{http://dx.doi.org/10.1017/S0305004100051872}{Math. Proc. Cambridge
  Phil. Soc. \textbf{78} (1975) 405}.

\bibitem[APS75b]{Atiyah:1975jf}
\bysame, \emph{{Spectral asymmetry and Riemannian Geometry. I}},
  \doihref{http://dx.doi.org/10.1017/S0305004100049410}{Math. Proc. Cambridge
  Phil. Soc. \textbf{77} (1975) 43}.

\bibitem[APS76]{Atiyah:1976jh}
\bysame, \emph{{Spectral asymmetry and Riemannian geometry. III}},
  \doihref{http://dx.doi.org/10.1017/S0305004100052105}{Math. Proc. Cambridge
  Phil. Soc. \textbf{79} (1976) 71--99}.

\bibitem[AW99]{Abanov:1999qz}
A.~G. Abanov and P.~B. Wiegmann, \emph{{Theta terms in nonlinear sigma
  models}}, \doihref{http://dx.doi.org/10.1016/S0550-3213(99)00820-2}{Nucl.
  Phys. B \textbf{570} (2000) 685--698},
  \href{http://arxiv.org/abs/hep-th/9911025}{{\arxivfont
  arXiv:hep-th/9911025}}.

\bibitem[AWH22]{Aasen:2022cdu}
D.~Aasen, Z.~Wang, and M.~B. Hastings, \emph{{Adiabatic paths of Hamiltonians,
  symmetries of topological order, and automorphism codes}},
  \href{http://arxiv.org/abs/2203.11137}{{\arxivfont arXiv:2203.11137
  [quant-ph]}}.

\bibitem[Ber84]{Berry:1984jv}
M.~V. Berry, \emph{{Quantal phase factors accompanying adiabatic changes}},
  \doihref{http://dx.doi.org/10.1098/rspa.1984.0023}{Proc. Roy. Soc. Lond. A
  \textbf{392} (1984) 45--57}.

\bibitem[BGV92]{berline1992heat}
N.~Berline, E.~Getzler, and M.~Vergne, \emph{Heat kernels and dirac operators},
  Die Grundlehren der mathematischen Wissenschaften in Einzeldarstellungen,
  Springer-Verlag, 1992. \url{https://books.google.com/books?id=vRc-QgAACAAJ}.

\bibitem[CFLS19a]{Cordova:2019jnf}
C.~C\'ordova, D.~S. Freed, H.~T. Lam, and N.~Seiberg, \emph{{Anomalies in the
  Space of Coupling Constants and Their Dynamical Applications I}},
  \doihref{http://dx.doi.org/10.21468/SciPostPhys.8.1.001}{SciPost Phys.
  \textbf{8} (2020) 001}, \href{http://arxiv.org/abs/1905.09315}{{\arxivfont
  arXiv:1905.09315 [hep-th]}}.

\bibitem[CFLS19b]{Cordova:2019uob}
\bysame, \emph{{Anomalies in the Space of Coupling Constants and Their
  Dynamical Applications II}},
  \doihref{http://dx.doi.org/10.21468/SciPostPhys.8.1.002}{SciPost Phys.
  \textbf{8} (2020) 002}, \href{http://arxiv.org/abs/1905.13361}{{\arxivfont
  arXiv:1905.13361 [hep-th]}}.

\bibitem[CJH85]{callan1985anomalies}
C.~G. Callan~Jr and J.~A. Harvey, \emph{Anomalies and fermion zero modes on
  strings and domain walls}, Nuclear Physics B \textbf{250} (1985) 427--436.

\bibitem[EQV{\etalchar{+}}20]{Ellis:2020ivx}
S.~A.~R. Ellis, J.~Quevillon, P.~N.~H. Vuong, T.~You, and Z.~Zhang, \emph{{The
  Fermionic Universal One-Loop Effective Action}},
  \doihref{http://dx.doi.org/10.1007/JHEP11(2020)078}{JHEP \textbf{11} (2020)
  078}, \href{http://arxiv.org/abs/2006.16260}{{\arxivfont arXiv:2006.16260
  [hep-ph]}}.

\bibitem[FFM{\etalchar{+}}20a]{Fukaya:2020ddp}
H.~Fukaya, M.~Furuta, S.~Matsuo, T.~Onogi, S.~Yamaguchi, and M.~Yamashita,
  \emph{{A physicist-friendly reformulation of the Atiyah-Patodi-Singer index
  and its mathematical justification}},
  \doihref{http://dx.doi.org/10.22323/1.363.0061}{PoS \textbf{LATTICE2019}
  (2019) 061}, \href{http://arxiv.org/abs/2001.01428}{{\arxivfont
  arXiv:2001.01428 [hep-lat]}}.

\bibitem[FFM{\etalchar{+}}20b]{Fukaya:2020tjk}
H.~Fukaya, M.~Furuta, Y.~Matsuki, S.~Matsuo, T.~Onogi, S.~Yamaguchi, and
  M.~Yamashita, \emph{{Mod-two APS index and domain-wall fermion}},
  \href{http://arxiv.org/abs/2012.03543}{{\arxivfont arXiv:2012.03543
  [hep-th]}}.

\bibitem[FH16]{Freed:2016rqq}
D.~S. Freed and M.~J. Hopkins, \emph{{Reflection positivity and invertible
  topological phases}},
  \doihref{http://dx.doi.org/10.2140/gt.2021.25.1165}{Geom. Topol. \textbf{25}
  (2021) 1165--1330}, \href{http://arxiv.org/abs/1604.06527}{{\arxivfont
  arXiv:1604.06527 [hep-th]}}.

\bibitem[FKS17]{Freed:2017rlk}
D.~S. Freed, Z.~Komargodski, and N.~Seiberg, \emph{{The Sum Over Topological
  Sectors and $\theta$ in the 2+1-Dimensional $\mathbb{C}\mathbb{P}^1$
  $\sigma$-Model}},
  \doihref{http://dx.doi.org/10.1007/s00220-018-3093-0}{Commun. Math. Phys.
  \textbf{362} (2018) 167--183},
  \href{http://arxiv.org/abs/1707.05448}{{\arxivfont arXiv:1707.05448
  [cond-mat.str-el]}}.

\bibitem[FOY17]{Fukaya:2017tsq}
H.~Fukaya, T.~Onogi, and S.~Yamaguchi, \emph{{Atiyah-Patodi-Singer index from
  the domain-wall fermion Dirac operator}},
  \doihref{http://dx.doi.org/10.1103/PhysRevD.96.125004}{Phys. Rev. D
  \textbf{96} (2017) 125004},
  \href{http://arxiv.org/abs/1710.03379}{{\arxivfont arXiv:1710.03379
  [hep-th]}}.

\bibitem[Fuj79]{Fujikawa:1979ay}
K.~Fujikawa, \emph{{Path Integral Measure for Gauge Invariant Fermion
  Theories}}, \doihref{http://dx.doi.org/10.1103/PhysRevLett.42.1195}{Phys.
  Rev. Lett. \textbf{42} (1979) 1195--1198}.

\bibitem[Fuj84]{Fujikawa:1983bg}
\bysame, \emph{{On the Evaluation of Chiral Anomaly in Gauge Theories with
  Gamma(5) Couplings}},
  \doihref{http://dx.doi.org/10.1103/PhysRevD.29.285}{Phys. Rev. D \textbf{29}
  (1984) 285}.

\bibitem[GJF17]{Gaiotto:2017zba}
D.~Gaiotto and T.~Johnson-Freyd, \emph{{Symmetry Protected Topological phases
  and Generalized Cohomology}},
  \doihref{http://dx.doi.org/10.1007/JHEP05(2019)007}{JHEP \textbf{05} (2019)
  007}, \href{http://arxiv.org/abs/1712.07950}{{\arxivfont arXiv:1712.07950
  [hep-th]}}.

\bibitem[GKKS17]{Gaiotto:2017yup}
D.~Gaiotto, A.~Kapustin, Z.~Komargodski, and N.~Seiberg, \emph{{Theta, Time
  Reversal, and Temperature}},
  \doihref{http://dx.doi.org/10.1007/JHEP05(2017)091}{JHEP \textbf{05} (2017)
  091}, \href{http://arxiv.org/abs/1703.00501}{{\arxivfont arXiv:1703.00501
  [hep-th]}}.

\bibitem[GKSW14]{Gaiotto:2014kfa}
D.~Gaiotto, A.~Kapustin, N.~Seiberg, and B.~Willett, \emph{{Generalized Global
  Symmetries}}, \doihref{http://dx.doi.org/10.1007/JHEP02(2015)172}{JHEP
  \textbf{02} (2015) 172}, \href{http://arxiv.org/abs/1412.5148}{{\arxivfont
  arXiv:1412.5148 [hep-th]}}.

\bibitem[GL20]{Gorantla:2020fao}
P.~Gorantla and H.~T. Lam, \emph{{Interface junctions in QCD${}_4$}},
  \doihref{http://dx.doi.org/10.21468/SciPostPhys.10.4.085}{SciPost Phys.
  \textbf{10} (2021) 085}, \href{http://arxiv.org/abs/2004.13300}{{\arxivfont
  arXiv:2004.13300 [hep-th]}}.

\bibitem[GW81]{Goldstone:1981kk}
J.~Goldstone and F.~Wilczek, \emph{{Fractional Quantum Numbers on Solitons}},
  \doihref{http://dx.doi.org/10.1103/PhysRevLett.47.986}{Phys. Rev. Lett.
  \textbf{47} (1981) 986--989}.

\bibitem[GY21]{Gomi:2021bhy}
K.~Gomi and M.~Yamashita, \emph{{Differential $KO$-theory via gradations and
  mass terms}}, \href{http://arxiv.org/abs/2111.01377}{{\arxivfont
  arXiv:2111.01377 [math.KT]}}.

\bibitem[HHN{\etalchar{+}}19]{Hidaka:2019jtv}
Y.~Hidaka, Y.~Hirono, M.~Nitta, Y.~Tanizaki, and R.~Yokokura,
  \emph{{Topological order in the color-flavor locked phase of a ( 3+1
  )-dimensional U(N) gauge-Higgs system}},
  \doihref{http://dx.doi.org/10.1103/PhysRevD.100.125016}{Phys. Rev. D
  \textbf{100} (2019) 125016},
  \href{http://arxiv.org/abs/1903.06389}{{\arxivfont arXiv:1903.06389
  [hep-th]}}.

\bibitem[HHY21]{Huang:2021ubo}
S.-J. Huang, C.-T. Hsieh, and J.~Yu, \emph{{Effective field theories of
  topological crystalline insulators and topological crystals}},
  \doihref{http://dx.doi.org/10.1103/PhysRevB.105.045112}{Phys. Rev. B
  \textbf{105} (2022) 045112},
  \href{http://arxiv.org/abs/2107.03409}{{\arxivfont arXiv:2107.03409
  [cond-mat.str-el]}}.

\bibitem[HKT20]{Hsin:2020cgg}
P.-S. Hsin, A.~Kapustin, and R.~Thorngren, \emph{{Berry Phase in Quantum Field
  Theory: Diabolical Points and Boundary Phenomena}},
  \href{http://arxiv.org/abs/2004.10758}{{\arxivfont arXiv:2004.10758
  [cond-mat.str-el]}}.

\bibitem[HT20]{Honda:2020txe}
M.~Honda and Y.~Tanizaki, \emph{{Topological aspects of 4D Abelian lattice
  gauge theories with the $\theta$ parameter}},
  \doihref{http://dx.doi.org/10.1007/JHEP12(2020)154}{JHEP \textbf{12} (2020)
  154}, \href{http://arxiv.org/abs/2009.10183}{{\arxivfont arXiv:2009.10183
  [hep-th]}}.

\bibitem[JR76]{jackiw1976solitons}
R.~Jackiw and C.~Rebbi, \emph{Solitons with fermion number $1/2$}, Physical
  Review D \textbf{13} (1976) 3398.

\bibitem[Kah11]{kahle2011superconnections}
A.~Kahle, \emph{Superconnections and index theory}, Journal of Geometry and
  Physics \textbf{61} (2011) 1601--1624.

\bibitem[Kap14]{Kapustin:2014tfa}
A.~Kapustin, \emph{{Symmetry Protected Topological Phases, Anomalies, and
  Cobordisms: Beyond Group Cohomology}},
  \href{http://arxiv.org/abs/1403.1467}{{\arxivfont arXiv:1403.1467
  [cond-mat.str-el]}}.

\bibitem[Kit09]{Kitaev:2009mg}
A.~Kitaev, \emph{{Periodic table for topological insulators and
  superconductors}}, \doihref{http://dx.doi.org/10.1063/1.3149495}{AIP Conf.
  Proc. \textbf{1134} (2009) 22--30},
  \href{http://arxiv.org/abs/0901.2686}{{\arxivfont arXiv:0901.2686
  [cond-mat.mes-hall]}}.

\bibitem[Kit13]{Kit13}
A.~Kitaev, \emph{{On the classification of short-range entangled states}},
  2013. \url{http://scgp.stonybrook.edu/video_portal/video.php?id=2010}. Talk
  at Simons Center.

\bibitem[Kit15]{Kit15}
\bysame, \emph{{Homotopy-theoretic approach to spt phases in action: Z16
  classification of three- dimensional superconductors}}, 2015.
  \url{http://www.ipam.ucla.edu/programs/workshops/symmetry-and-topology-in-quantum-matter/?tab=schedule}.
  Symmetry and Topology in Quantum Matter Workshop, Institute for Pure and
  Applied Mathematics, University of California.

\bibitem[KK19]{Karasik:2019bxn}
A.~Karasik and Z.~Komargodski, \emph{{The Bi-Fundamental Gauge Theory in 3+1
  Dimensions: The Vacuum Structure and a Cascade}},
  \doihref{http://dx.doi.org/10.1007/JHEP05(2019)144}{JHEP \textbf{05} (2019)
  144}, \href{http://arxiv.org/abs/1904.09551}{{\arxivfont arXiv:1904.09551
  [hep-th]}}.

\bibitem[KL00]{Kraus:2000nj}
P.~Kraus and F.~Larsen, \emph{{Boundary string field theory of the D anti-D
  system}}, \doihref{http://dx.doi.org/10.1103/PhysRevD.63.106004}{Phys. Rev. D
  \textbf{63} (2001) 106004},
  \href{http://arxiv.org/abs/hep-th/0012198}{{\arxivfont
  arXiv:hep-th/0012198}}.

\bibitem[KS20a]{Kapustin:2020eby}
A.~Kapustin and L.~Spodyneiko, \emph{{Higher-dimensional generalizations of
  Berry curvature}},
  \doihref{http://dx.doi.org/10.1103/PhysRevB.101.235130}{Phys. Rev. B
  \textbf{101} (2020) 235130},
  \href{http://arxiv.org/abs/2001.03454}{{\arxivfont arXiv:2001.03454
  [cond-mat.str-el]}}.

\bibitem[KS20b]{Kapustin:2020mkl}
\bysame, \emph{{Higher-dimensional generalizations of the Thouless charge
  pump}}, \href{http://arxiv.org/abs/2003.09519}{{\arxivfont arXiv:2003.09519
  [cond-mat.str-el]}}.

\bibitem[KS21]{Kanno:2021bze}
H.~Kanno and S.~Sugimoto, \emph{{Anomaly and Superconnection}},
  \href{http://arxiv.org/abs/2106.01591}{{\arxivfont arXiv:2106.01591
  [hep-th]}}.

\bibitem[KS22]{Kapustin:2022apy}
A.~Kapustin and N.~Sopenko, \emph{{Local Noether theorem for quantum lattice
  systems and topological invariants of gapped states}},
  \href{http://arxiv.org/abs/2201.01327}{{\arxivfont arXiv:2201.01327
  [math-ph]}}.

\bibitem[KSTZ17]{Komargodski:2017dmc}
Z.~Komargodski, A.~Sharon, R.~Thorngren, and X.~Zhou, \emph{{Comments on
  Abelian Higgs Models and Persistent Order}},
  \doihref{http://dx.doi.org/10.21468/SciPostPhys.6.1.003}{SciPost Phys.
  \textbf{6} (2019) 003}, \href{http://arxiv.org/abs/1705.04786}{{\arxivfont
  arXiv:1705.04786 [hep-th]}}.

\bibitem[KT17]{Kikuchi:2017pcp}
Y.~Kikuchi and Y.~Tanizaki, \emph{{Global inconsistency, \textquoteright{}t
  Hooft anomaly, and level crossing in quantum mechanics}},
  \doihref{http://dx.doi.org/10.1093/ptep/ptx148}{PTEP \textbf{2017} (2017)
  113B05}, \href{http://arxiv.org/abs/1708.01962}{{\arxivfont arXiv:1708.01962
  [hep-th]}}.

\bibitem[KTTW14]{Kapustin:2014dxa}
A.~Kapustin, R.~Thorngren, A.~Turzillo, and Z.~Wang, \emph{{Fermionic Symmetry
  Protected Topological Phases and Cobordisms}},
  \doihref{http://dx.doi.org/10.1007/JHEP12(2015)052}{JHEP \textbf{12} (2015)
  052}, \href{http://arxiv.org/abs/1406.7329}{{\arxivfont arXiv:1406.7329
  [cond-mat.str-el]}}.

\bibitem[KW99]{Kennedy:1999nn}
C.~Kennedy and A.~Wilkins, \emph{{Ramond-Ramond couplings on Brane - anti-Brane
  systems}}, \doihref{http://dx.doi.org/10.1016/S0370-2693(99)00967-3}{Phys.
  Lett. B \textbf{464} (1999) 206--212},
  \href{http://arxiv.org/abs/hep-th/9905195}{{\arxivfont
  arXiv:hep-th/9905195}}.

\bibitem[KY21]{Kobayashi:2021jbn}
S.~K. Kobayashi and K.~Yonekura, \emph{{Atiyah-Patodi-Singer index theorem from
  axial anomaly}}, \href{http://arxiv.org/abs/2103.10654}{{\arxivfont
  arXiv:2103.10654 [hep-th]}}.

\bibitem[LOT21]{Lee:2021crt}
Y.~Lee, K.~Ohmori, and Y.~Tachikawa, \emph{{Matching higher symmetries across
  Intriligator-Seiberg duality}},
  \doihref{http://dx.doi.org/10.1007/JHEP10(2021)114}{JHEP \textbf{10} (2021)
  114}, \href{http://arxiv.org/abs/2108.05369}{{\arxivfont arXiv:2108.05369
  [hep-th]}}.

\bibitem[OY21]{Onogi:2021slv}
T.~Onogi and T.~Yoda, \emph{{Comments on the Atiyah-Patodi-Singer index
  theorem, domain wall, and Berry phase}},
  \doihref{http://dx.doi.org/10.1007/JHEP12(2021)096}{JHEP \textbf{12} (2021)
  096}, \href{http://arxiv.org/abs/2109.08274}{{\arxivfont arXiv:2109.08274
  [hep-th]}}.

\bibitem[QSV21]{Quevillon:2021sfz}
J.~Quevillon, C.~Smith, and P.~N.~H. Vuong, \emph{{Axion Effective Action}},
  \href{http://arxiv.org/abs/2112.00553}{{\arxivfont arXiv:2112.00553
  [hep-ph]}}.

\bibitem[Qui85]{quillen1985superconnections}
D.~Quillen, \emph{Superconnections and the chern character}, Topology
  \textbf{24} (1985) 89--95.

\bibitem[RSFL10]{ryu2010topological}
S.~Ryu, A.~P. Schnyder, A.~Furusaki, and A.~W. Ludwig, \emph{Topological
  insulators and superconductors: tenfold way and dimensional hierarchy}, New
  Journal of Physics \textbf{12} (2010) 065010.

\bibitem[Sha20]{Sharon:2020doo}
A.~Sharon, \emph{{Global Aspects of Spaces of Vacua}},
  \doihref{http://dx.doi.org/10.1007/JHEP11(2020)083}{JHEP \textbf{11} (2020)
  083}, \href{http://arxiv.org/abs/2004.11182}{{\arxivfont arXiv:2004.11182
  [hep-th]}}.

\bibitem[Shi21]{Shiozaki:2021weu}
K.~Shiozaki, \emph{{On adiabatic cycles of quantum spin systems}},
  \href{http://arxiv.org/abs/2110.10665}{{\arxivfont arXiv:2110.10665
  [cond-mat.str-el]}}.

\bibitem[STU20]{Sulejmanpasic:2020zfs}
T.~Sulejmanpasic, Y.~Tanizaki, and M.~\"Unsal, \emph{{Universality between
  vector-like and chiral quiver gauge theories: Anomalies and domain walls}},
  \doihref{http://dx.doi.org/10.1007/JHEP06(2020)173}{JHEP \textbf{06} (2020)
  173}, \href{http://arxiv.org/abs/2004.10328}{{\arxivfont arXiv:2004.10328
  [hep-th]}}.

\bibitem[Tho14]{Thorngren:2014pza}
R.~Thorngren, \emph{{Framed Wilson Operators, Fermionic Strings, and
  Gravitational Anomaly in 4d}},
  \doihref{http://dx.doi.org/10.1007/JHEP02(2015)152}{JHEP \textbf{02} (2015)
  152}, \href{http://arxiv.org/abs/1404.4385}{{\arxivfont arXiv:1404.4385
  [hep-th]}}.

\bibitem[Tho17]{Thorngren:2017vzn}
\bysame, \emph{{Topological Terms and Phases of Sigma Models}},
  \href{http://arxiv.org/abs/1710.02545}{{\arxivfont arXiv:1710.02545
  [cond-mat.str-el]}}.

\bibitem[TK10]{Teo:2010zb}
J.~C.~Y. Teo and C.~L. Kane, \emph{{Topological Defects and Gapless Modes in
  Insulators and Superconductors}},
  \doihref{http://dx.doi.org/10.1103/PhysRevB.82.115120}{Phys. Rev. B
  \textbf{82} (2010) 115120}, \href{http://arxiv.org/abs/1006.0690}{{\arxivfont
  arXiv:1006.0690 [cond-mat.mes-hall]}}.

\bibitem[TK17]{Tanizaki:2017bam}
Y.~Tanizaki and Y.~Kikuchi, \emph{{Vacuum structure of bifundamental gauge
  theories at finite topological angles}},
  \doihref{http://dx.doi.org/10.1007/JHEP06(2017)102}{JHEP \textbf{06} (2017)
  102}, \href{http://arxiv.org/abs/1705.01949}{{\arxivfont arXiv:1705.01949
  [hep-th]}}.

\bibitem[TS18]{Tanizaki:2018xto}
Y.~Tanizaki and T.~Sulejmanpasic, \emph{{Anomaly and global inconsistency
  matching: $\theta$-angles, $SU(3)/U(1)^2$ nonlinear sigma model, $SU(3)$
  chains and its generalizations}},
  \doihref{http://dx.doi.org/10.1103/PhysRevB.98.115126}{Phys. Rev. B
  \textbf{98} (2018) 115126},
  \href{http://arxiv.org/abs/1805.11423}{{\arxivfont arXiv:1805.11423
  [cond-mat.str-el]}}.

\bibitem[TTU00]{Takayanagi:2000rz}
T.~Takayanagi, S.~Terashima, and T.~Uesugi, \emph{{Brane - anti-brane action
  from boundary string field theory}},
  \doihref{http://dx.doi.org/10.1088/1126-6708/2001/03/019}{JHEP \textbf{03}
  (2001) 019}, \href{http://arxiv.org/abs/hep-th/0012210}{{\arxivfont
  arXiv:hep-th/0012210}}.

\bibitem[TU22]{Tanizaki:2022ngt}
Y.~Tanizaki and M.~\"Unsal, \emph{{Center vortex and confinement in Yang-Mills
  theory and QCD with anomaly-preserving compactifications}},
  \doihref{http://dx.doi.org/10.1093/ptep/ptac042}{PTEP \textbf{2022} (2022)
  04}, \href{http://arxiv.org/abs/2201.06166}{{\arxivfont arXiv:2201.06166
  [hep-th]}}.

\bibitem[\"U20]{Unsal:2020yeh}
M.~\"Unsal, \emph{{Strongly coupled QFT dynamics via TQFT coupling}},
  \doihref{http://dx.doi.org/10.1007/JHEP11(2021)134}{JHEP \textbf{11} (2021)
  134}, \href{http://arxiv.org/abs/2007.03880}{{\arxivfont arXiv:2007.03880
  [hep-th]}}.

\bibitem[WQB{\etalchar{+}}21]{wen2021flow}
X.~Wen, M.~Qi, A.~Beaudry, J.~Moreno, M.~J. Pflaum, D.~Spiegel, A.~Vishwanath,
  and M.~Hermele, \emph{Flow of (higher) berry curvature and bulk-boundary
  correspondence in parametrized quantum systems}, arXiv preprint
  arXiv:2112.07748 (2021) .

\bibitem[WWW18]{Wang:2018qoy}
J.~Wang, X.-G. Wen, and E.~Witten, \emph{{A New SU(2) Anomaly}},
  \doihref{http://dx.doi.org/10.1063/1.5082852}{J. Math. Phys. \textbf{60}
  (2019) 052301}, \href{http://arxiv.org/abs/1810.00844}{{\arxivfont
  arXiv:1810.00844 [hep-th]}}.

\bibitem[WY19]{Witten:2019bou}
E.~Witten and K.~Yonekura, \emph{{Anomaly Inflow and the $\eta$-Invariant}},
  {The Shoucheng Zhang Memorial Workshop}, 9 2019.
  \href{http://arxiv.org/abs/1909.08775}{{\arxivfont arXiv:1909.08775
  [hep-th]}}.

\bibitem[Yam21]{Yamashita:2021fkd}
M.~Yamashita, \emph{{Differential models for the Anderson dual to bordism
  theories and invertible QFT's, II}},
  \href{http://arxiv.org/abs/2110.14828}{{\arxivfont arXiv:2110.14828
  [math.AT]}}.

\bibitem[Yon16]{Yonekura:2016wuc}
K.~Yonekura, \emph{{Dai-Freed theorem and topological phases of matter}},
  \doihref{http://dx.doi.org/10.1007/JHEP09(2016)022}{JHEP \textbf{09} (2016)
  022}, \href{http://arxiv.org/abs/1607.01873}{{\arxivfont arXiv:1607.01873
  [hep-th]}}.

\bibitem[Yon18]{Yonekura:2018ufj}
\bysame, \emph{{On the cobordism classification of symmetry protected
  topological phases}},
  \doihref{http://dx.doi.org/10.1007/s00220-019-03439-y}{Commun. Math. Phys.
  \textbf{368} (2019) 1121--1173},
  \href{http://arxiv.org/abs/1803.10796}{{\arxivfont arXiv:1803.10796
  [hep-th]}}.

\bibitem[YY21]{Yamashita:2021cao}
M.~Yamashita and K.~Yonekura, \emph{{Differential models for the Anderson dual
  to bordism theories and invertible QFT's, I}},
  \href{http://arxiv.org/abs/2106.09270}{{\arxivfont arXiv:2106.09270
  [math.AT]}}.

\end{thebibliography}

\end{document}